\documentclass[
  journal=pasa,
  manuscript=research-paper, %% or "review"
  year=2026,
  volume=YY,
]{cup-journal}
\usepackage{amsmath}
\usepackage{amssymb,microtype,siunitx,booktabs}
\usepackage{aas-macros}
\usepackage{graphicx}
\usepackage{float}
\usepackage{multirow}
\usepackage{tikz}
\usepackage{xspace}
\usepackage{commath}
\usepackage{mathtools}
\usepackage{xcolor}
\usepackage{nicematrix}
\usepackage{orcidlink}
\usepackage{placeins}
\hypersetup{colorlinks,citecolor=largeblue,linkcolor=largeblue,urlcolor=largeblue}
\usepackage{chngcntr}

\newcommand{\rtwo}{$\rm R_{200}$\xspace}
\newcommand{\rrtwo}{$\rm R/R_{200}$\xspace}
\newcommand{\ha}{H$\alpha$\xspace}
\newcommand{\hb}{H$\beta$\xspace}
\newcommand{\oiii}{[OIII]$\lambda$5007\xspace}
\newcommand{\nii}{[NII]$\lambda$6584\xspace}

\newcommand{\hd}{H$\delta$\xspace}
\newcommand{\lt}{$<$\xspace}
\newcommand{\gt}{$>$\xspace}
\newcommand{\logMstar}{$\rm log(M_\ast/M_\odot)$}
\newcommand{\lsim}{$\lesssim$\xspace}

\newcommand{\fraction}[3]{$#1^{+#3}_{-#2}$\%}

\newcommand{\s}{$\sim$}
\newcommand{\AAs}{\AA\xspace}
\newcommand{\arcsec}{$^{\prime\prime}$}

\title{The SAMI Galaxy Survey: Quenching of Star Formation in Clusters \\ III. Ram-Pressure-Affected Galaxy Populations}

\author{\orcidlink{0000-0002-1045-2559} Oğuzhan Çakır}
\affiliation{School of Mathematical and Physical Sciences, Macquarie University, Sydney, NSW 2109, Australia}
\alsoaffiliation{Astrophysics and Space Technologies Research Centre, Macquarie University, Sydney, NSW 2109, Australia}
\alsoaffiliation{ARC Centre of Excellence for All-Sky Astrophysics in 3 Dimensions (ASTRO 3D), Australia}
\email[Oğuzhan Çakır]{oguzhan.cakir@students.mq.edu.au}

\author{\orcidlink{0000-0002-2879-1663} Matt S. Owers}
\affiliation{School of Mathematical and Physical Sciences, Macquarie University, Sydney, NSW 2109, Australia}
\alsoaffiliation{Astrophysics and Space Technologies Research Centre, Macquarie University, Sydney, NSW 2109, Australia}
\alsoaffiliation{ARC Centre of Excellence for All-Sky Astrophysics in 3 Dimensions (ASTRO 3D), Australia}

\author{\orcidlink{0000-0002-7422-9823} Luca Cortese}
\affiliation{International Centre for Radio Astronomy Research (ICRAR), University of Western Australia, M468, 35 Stirling Highway, Crawley, WA 6009, Australia}
\alsoaffiliation{ARC Centre of Excellence for All-Sky Astrophysics in 3 Dimensions (ASTRO 3D), Australia}

\author{\orcidlink{0000-0002-5896-0034} Mina Pak}
\affiliation{School of Mathematical and Physical Sciences, Macquarie University, Sydney, NSW 2109, Australia}
\alsoaffiliation{ARC Centre of Excellence for All-Sky Astrophysics in 3 Dimensions (ASTRO 3D), Australia}
\alsoaffiliation{Korea Astronomy and Space Science Institute (KASI), 776 Daeduk-daero, Yuseong-gu, Daejeon  34055, Republic of Korea}

\author{\orcidlink{0009-0009-9074-716X} Gabriella Quattropani}
\affiliation{School of Mathematical and Physical Sciences, Macquarie University, Sydney, NSW 2109, Australia}
\alsoaffiliation{Astrophysics and Space Technologies Research Centre, Macquarie University, Sydney, NSW 2109, Australia}
\alsoaffiliation{ARC Centre of Excellence for All-Sky Astrophysics in 3 Dimensions (ASTRO 3D), Australia}

\author{\orcidlink{0000-0002-9332-5386} Stefania Barsanti}
\affiliation{Sydney Institute for Astronomy (SIfA), School of Physics, The University of Sydney, NSW 2006, Australia}
\alsoaffiliation{Research School of Astronomy and Astrophysics, Australian National University, Canberra, ACT 2611, Australia}
\alsoaffiliation{ARC Centre of Excellence for All-Sky Astrophysics in 3 Dimensions (ASTRO 3D), Australia}

\author{\orcidlink{0000-0003-1627-9301} Julia J. Bryant}
\affiliation{Sydney Institute for Astronomy (SIfA), School of Physics, The University of Sydney, NSW 2006, Australia}
\alsoaffiliation{Astralis-USydney, School of Physics, University of Sydney, NSW 2006, Australia}
\alsoaffiliation{ARC Centre of Excellence for All-Sky Astrophysics in 3 Dimensions (ASTRO 3D), Australia}

\author{\orcidlink{0000-0001-5005-3125} Warrick J. Couch}
\affiliation{Centre for Astrophysics and Supercomputing, Swinburne University of Technology, Hawthorn, Victoria 3122, Australia}

\author{\orcidlink{0000-0003-2880-9197} Scott M. Croom}
\affiliation{Sydney Institute for Astronomy (SIfA), School of Physics, The University of Sydney, NSW 2006, Australia}
\alsoaffiliation{ARC Centre of Excellence for All-Sky Astrophysics in 3 Dimensions (ASTRO 3D), Australia}

\author{\orcidlink{0000-0002-4326-8598} Pratyush K. Das}
\affiliation{School of Mathematics and Physics, University of Queensland, Brisbane, QLD 4072, Australia}

\author{\orcidlink{0000-0002-6998-6993} Jon S. Lawrence}
\affiliation{Australian Astronomical Optics, Macquarie University, NSW 2109, Australia}

\author{\orcidlink{0000-0003-3514-6280} Yifan Mai}
\affiliation{Sydney Institute for Astronomy (SIfA), School of Physics, The University of Sydney, NSW 2006, Australia}
\alsoaffiliation{ARC Centre of Excellence for All-Sky Astrophysics in 3 Dimensions (ASTRO 3D), Australia}
\alsoaffiliation{Australian Astronomical Optics, Macquarie University, NSW 2109, Australia}

\author{\orcidlink{0000-0003-2723-0810} Andrei Ristea}
\affiliation{Centre for Astrophysics and Supercomputing, Swinburne University of Technology, Hawthorn, Victoria 3122, Australia}
\alsoaffiliation{ARC Centre of Excellence in Optical Microcombs for Breakthrough Science (COMBS)}

\author{\orcidlink{0000-0001-6444-9307} Sebastian F. S\'anchez}
\affiliation{Instituto de Astronom\'ia, Universidad Nacional Autonoma de M\'exico, Circuito Exterior, Ciudad Universitaria, Ciudad de M\'exico 04510, Mexico}

\author{\orcidlink{0000-0002-1576-2505} Sarah Sweet}
\affiliation{School of Mathematics and Physics, University of Queensland, Brisbane, QLD 4072, Australia}
\alsoaffiliation{ARC Centre of Excellence for All-Sky Astrophysics in 3 Dimensions (ASTRO 3D), Australia}

\author{\orcidlink{0000-0003-2552-0021} Jesse van de Sande}
\affiliation{School of Physics, University of New South Wales, Sydney, NSW 2052, Australia}
\alsoaffiliation{ARC Centre of Excellence for All-Sky Astrophysics in 3 Dimensions (ASTRO 3D), Australia}

\author{\orcidlink{0000-0003-4546-7731} Glenn van de Ven}
\affiliation{Department of Astrophysics, University of Vienna, Türkenschanzstraße 17, 1180 Vienna, Austria}

\author{\orcidlink{0000-0002-4556-2619} Sukyoung K. Yi}
\affiliation{Department of Astronomy and Yonsei University Observatory, Yonsei University, Seoul 03722, Republic of Korea}

\received {dd Mm 2025}
\revised  {dd Mm 2025}
\accepted {dd Mm 2025}
\published{dd Mm 2025}

%% First letter not capped

\begin{document}

\begin{abstract}
Cluster environments influence galaxy evolution by curtailing star formation activity, notably through ram-pressure stripping (RPS). This process can leave observable signatures—such as gas tails and truncated gas disks—that are crucial for understanding how RPS affects galaxies. In this study, using spatially resolved spectroscopic data from the SAMI Galaxy Survey, we identify galaxies undergoing or recently affected by RPS in eight nearby clusters ($\rm 0.029<z<0.058$), through a visual classification scheme based on the ionised gas (\ha+\nii) morphologies, split into "unperturbed", "asymmetric", and "truncated". Alongside, we measure non-parametric structural parameters (concentration, asymmetry, and offset between gas and stars) to quantify the ionised gas morphologies. We find that combinations of parameters such as concentration, shape asymmetry, and stellar–ionised gas centre offsets are useful in categorising the degree of RPS in line with their ionised gas morphologies. The projected phase-space analysis shows that asymmetric galaxies are found in a narrow region in cluster-centric distance ($\rm 0.1<R/R_{200}<0.6$, where \rtwo is the characteristic cluster radius) and have a larger dispersion in line-of-sight velocity ($\sigma(|v_{pec}|)_\mathrm{Asym} = 0.71^{+0.09}_{-0.07}\  \sigma_{200}$, with $\sigma_{200}$ being the cluster velocity dispersion within \rtwo), compared to the truncated and unperturbed samples that are more broadly distributed and predominantly located at larger cluster-centric distances. This suggests that asymmetric galaxies are likely recent infallers---having crossed within 0.5 R$\mathrm{_{200}}$ in the past \s1 Gyr. In terms of star formation activity, we find that the resolved star-forming main sequence (rSFMS; $\Sigma_\mathrm{SFR} - \Sigma_\ast$) of unperturbed and RPS candidates (asymmetric and truncated) differ. RPS candidates yield a much steeper rSFMS relation compared to the unperturbed counterparts, primarily emerging from having lower $\Sigma_\mathrm{SFR}$ values for the low mass density regime (i.e., $\mathrm{log} \ \Sigma_\ast \lesssim 8 \ \mathrm{M}_\odot \ \rm kpc^{-2}$), with the steepest gradient deriving from the truncated sample. Moreover, radial specific star formation rate profiles introduce different trends for unperturbed and RPS candidates. Star formation in RPS candidates is suppressed in the outskirts relative to unperturbed galaxies and is more prominent for the truncated sample compared to the asymmetric counterparts. In contrast, central (i.e., $\rm r/r_{eff}<0.5$) star formation activity in RPS candidates is comparable with that in their unperturbed and field counterparts, suggesting no elevated activity. Taken together, this suggests an evolutionary trend linked to the RPS stage, where unperturbed galaxies likely represent recently accreted systems (pre-RPS), while asymmetric and truncated galaxies may correspond to populations undergoing RPS and post-RPS phases, respectively, favouring outside-in quenching.
\end{abstract}

\section{Introduction} \label{sec:intro}
In high-density environments, such as clusters of galaxies, gas stripping plays a crucial role in the evolution of constituent galaxies \citep{Boselli2006, Cortese2021}. One such mechanism is ram pressure stripping \citep[RPS;][]{GG1972}. As the galaxy moves through the cluster, the hot ($\rm T \sim 10^{7-8}\ K$), diffuse ($\rm 10^{-3} \ particles \ cm^{-3}$) intracluster medium (ICM) is able to remove the gas component of infalling galaxies by ram pressure \citep{GG1972}. Ram pressure can remove the hot halo gas, leading to a gradual decline in star-formation over long timescales \citep[strangulation;][]{LTC1980} or rapidly through stronger RPS that even evaporates the cold gas in the disk \citep{Roediger2006, Kapferer2009}. 
\\
\\
The signatures of ram pressure stripping manifest themselves through truncated disks, extra-planar emission, and one-sided spatial tails visible across the electromagnetic spectrum, in radio \citep{Solanes2001, Vollmer2001, Chung2009, Yoon2017, Jachym2019, Roberts2021a}, optical (narrow-band + broad-band) \citep{Gavazzi2001, Koopmann2004b, Cortese2007, Owers2012,Abramson2014, Ebeling2014, Kenney2014, Poggianti2016, Boselli2018, Owers2019, Poggianti2025}, ultraviolet \citep{Smith2010, George2018, George2025}, X-ray \citep{Sun2010, Poggianti2019b}.
\\
\\
The strength of these signatures ranges from mild to severe, and the presence and severity of these signatures are important indicators of the occurrence and degree of RPS. To identify the RPS candidates, studies typically adopt a bimodal approach, either by visual inspection of these features using multi-wavelength imaging \citep{Smith2010, Poggianti2016, Yoon2017, RP2020, Roberts2021a, Roberts2022a} or through structural measurements \citep{McPartland2016, RP2020, Roberts2021b, Bellhouse2022, Roberts2022b, Krabbe2024}. Focusing on the most extreme stripping cases --- "jellyfish" galaxies \citep{Bekki2009} where \textit{in situ} star formation occurs in the stripped gas, forming star-forming tails that can extend up to several tens of kpc \citep{Cortese2007, Sun2010, Sheen2017, Poggianti2019a}  --- \citet{Poggianti2016} classified 344 RPS candidates across 71 nearby clusters using B-band imaging and a visual classification scheme based on the degree of stripping (jellyfish class; \texttt{JClass}). \citet{Yoon2017} applied a similar approach to classify HI morphologies of 35 galaxies in the Virgo cluster from the VLA Imaging of Virgo in Atomic Gas Survey \citep[VIVA;][]{Chung2009}, while \citet{Roberts2021a} exploited the Low-Frequency Array Two-metre Sky Survey \citep[LoTSS;][]{Shimwell2019} radio continuum data to conduct a systematic search of RPS candidates, resulting in 95 galaxies with radio tails across 29 nearby clusters. 

On the quantitative side, morphological parameters --- e.g., concentration (C), asymmetry (A), Gini coefficient (G), second-order moment of light ($\rm M_{20}$), Sérsic index (n), bulge strength parameter F(G, $\rm M_{20}$) --- have been explored to identify RPS candidates. In such a study, \citet{McPartland2016} used C, A, G, $M_{20}$ and skeletal decomposition ($Sk_{0-1}$ and $Sk_{1-2}$) parameters to identify candidates at $\rm z\sim0.5$ using Hubble imaging through a semi-automated approach combining visual features and structural metrics. \citet{Krabbe2024} performed a similar analysis using Legacy Survey imaging for 600 galaxies across different environments, showing that strong jellyfish candidates (\texttt{JClass}\gt3) can be distinguished from isolated galaxies through parameter combinations such as C–A, n–A, and F(G, $\rm M_{20}$)–A. All these findings show that galaxies with more extreme visual stripping evidence emerge as outliers in certain parameter spaces.
\\
\\
In terms of star formation, RPS-affected galaxies show a wide range of activity reported by both observations and simulations. Follow-up study on the Virgo spirals with truncated disks, \citet{Crowl2008} showed that stellar populations in the outer disks are consistent with being quenched within the last 0.5 Gyr. More recently, exploiting the spatially resolved data from the Sydney-AAO Multi-object Integral-field spectrograph \citep[SAMI;][]{Croom2012} Galaxy Survey \citep[SAMI-GS;][]{Bryant2015}, \citet{Owers2019} identified galaxies with strong \hd features (i.e., \hd-strong galaxies, HDSGs), as a tracer of recently quenched star formation and found that half of the cluster HDSGs (9/17) show ongoing star formation in their centres, supporting an outside-in quenching scenario. Simulations also demonstrate that RPS is able to rapidly quench star formation in a significant fraction of galaxies in group/cluster environments during the first pericentric passage, which is prominent for low-mass galaxies \citep{Oman2016, Jung2018, Lotz2019, Rhee2024}.

Simulations indicate that RPS can also trigger an episode of intense star formation in both the disk and the stripped gas material by compression of giant molecular clouds \citep{Bekki2003} or ram-pressure-driven mass flows to the centre of the galaxy \citep{Zhu2024}. This may manifest as global and/or local elevated star formation activity. The global enhancement of star formation rates in jellyfish galaxies has been consistently reported by several observational studies \citep{Vulcani2018, Ebeling2019, Roman-Oliveira2019, RP2020, Roberts2021a, Lee2022, Vulcani2024}, who found that jellyfish galaxies occupy the upper envelope of the star formation main sequence (SFMS). 

With the emergence of integral field units, we are able to study the local/resolved ($\lesssim$ 1 kpc) properties of galaxies, such as star formation activity and metallicity, for a variety of galaxies with different morphologies, star formation stages, and even merger stages, particularly over the last decade \citep{Ryder1995, Sanchez2013, Cano-Diaz2016, Hsieh2017, Ellison2018, Medling2018, Sanchez2019, Thorp2019, Bluck2020a, Bluck2020b, Sanchez2020, Sanchez2021, Mun2024}. This effort has been extended to understand the interplay between cluster environments and star formation activity through large campaigns such as the GAs Stripping Phenomena in galaxies with MUSE \citep[GASP;][]{Poggianti2017}. To that end, \citet{Vulcani2020} reported locally elevated star formation activity within the disk and stripped tails for the jellyfish galaxies identified within the GASP survey. A similar result of enhanced star formation activity has been found at the centre of a ram-pressure-affected galaxy by \citet{Zhu2024} using tailored wind-tunnel simulations.
\\
\\
While previous studies have predominantly focused on individual clusters or specific populations (e.g., jellyfish galaxies) to identify and examine stripping candidates, these approaches may not fully represent the broader quenching context. In particular, they may miss galaxies undergoing stripping without clear visible signatures. Therefore, homogeneous sampling of galaxies in clusters is essential. To achieve this, we use the SAMI-GS, which provides data for around 900 galaxies across clusters with diverse dynamical states, without any assumption on underlying galaxy population or feature-based selection \citep{Owers2017}. Accordingly, this study is one of the few attempts to search for RPS candidates systematically without any pre-selection toward a certain galaxy population. 
\\
\\
The outline of the paper is as follows. Section~\ref{sec:Data and Selection} describes the SAMI-GS, the data products used, whether existing or generated in this study. Section~\ref{sec:Analysis and Results} introduces the visual classification scheme to identify the RPS candidates, and investigates the positions in projected phase space and star-forming properties of galaxies in different visual classes. In Section~\ref{sec: Discussion}, we discuss our findings and interpretation with respect to previous studies. Section~\ref{sec: Conclusion} highlights the key findings of the study. Throughout this paper, we assume a flat $\Lambda$CDM cosmology with $\mathrm{H_0 = 70 \ km \ s^{-1} \ Mpc^{-1}}$, $\Omega_\mathrm{m} = 0.3, \ \Omega_\Lambda = 0.7$.

\section{Data \& Sample Selection} \label{sec:Data and Selection}
\subsection{SAMI Galaxy Survey}\label{subsec:SAMI-GS}
The SAMI-GS was carried out with the SAMI instrument mounted at the prime focus of the 3.9 m Anglo-Australian Telescope (AAT). The SAMI instrument uses 13 integral-field units \citep[namely hexabundles;][]{Bland-Hawthorn2011, Bryant2014} composed of 61 fused 1.6 arcsecond-diameter fibres. These hexabundles are fed into the AAOmega spectrograph \citep{Sharp2006}. As of SAMI Data Release 3 \citep[DR3;][]{Croom2021}, the SAMI-GS observed 3068 unique galaxies sampled from the Galaxy and Mass Assembly Survey \citep[GAMA; ][]{Driver2011} equatorial regions (\texttt{G09, G12, G15}; 2151 galaxies) and eight cluster regions selected from the catalogue of \cite{dePropris2002} in the 2-degree Field Galaxy Redshift Survey, and the Cluster Infall Regions in the Sloan Digital Sky Survey \citep[\texttt{CIRS};][]{Rines2006}, with halo masses of $\rm 14.25\leq log(M_{200}/M_\odot)\leq15.19$ (917 galaxies) described in \citet{Owers2017}. The observed galaxies have stellar masses ranging between $8<$ \logMstar $<12$.
\\
\\
The details for target selection are given in \citet{Bryant2015} for GAMA regions and \citet{Owers2017} for the cluster sample. Briefly, the primary targets are selected with a step-like stellar mass function with an increasing lower limit as a function of redshift, where redshift refers to the flow-corrected value, $z_{tonry}$, for the GAMA regions \citep{Tonry2000, Baldry2012}. To define primary targets for the cluster regions, only two steps are adopted as \logMstar \gt 9.5 if the cluster redshift, $ z_{cl},<  0.045$, otherwise \logMstar \gt 10. Additionally, cluster targets are selected within $\rm R_{200}$ and $|v_{pec}|/\sigma_{200}\leq3.5$, where $v_{pec}$ is the peculiar velocity of galaxies with respect to the cluster redshift, $v_{pec} = c \times (z_{gal} - z_{cl}) / (1 + z_{cl}$) with c being the speed of light, and $\sigma_{200}$ is the velocity dispersion of the cluster measured within \rtwo defined by \citet{Owers2017}. Note that both $z_{cl}$ and $z_{gal}$ are not flow-corrected. This selection is less strict than the cluster membership allocation done by \citet{Owers2017}, which was based on caustic analysis, and it therefore includes non-bona fide members with redshifts close to the cluster.

One of the main aims of this study is to understand the environmental impact on star formation activity. To achieve this, we need a sample from a non-cluster field as a reference/control. Therefore, we use the GAMA portion of the survey as a control sample, which will enable a direct comparison to assess environmental effects. GAMA galaxies are selected through a step-like stellar mass function for $0.005<z_{tonry}<0.095$ \citep{Bryant2015}, which is shown in Figure~\ref{fig:sample_selection}.

%The SAMI-GS also observed a set of secondary targets selected by less conservative criteria, yet, exempted from this paper.

\begin{figure*}[!t]
    \centering
    \includegraphics[width=\textwidth]{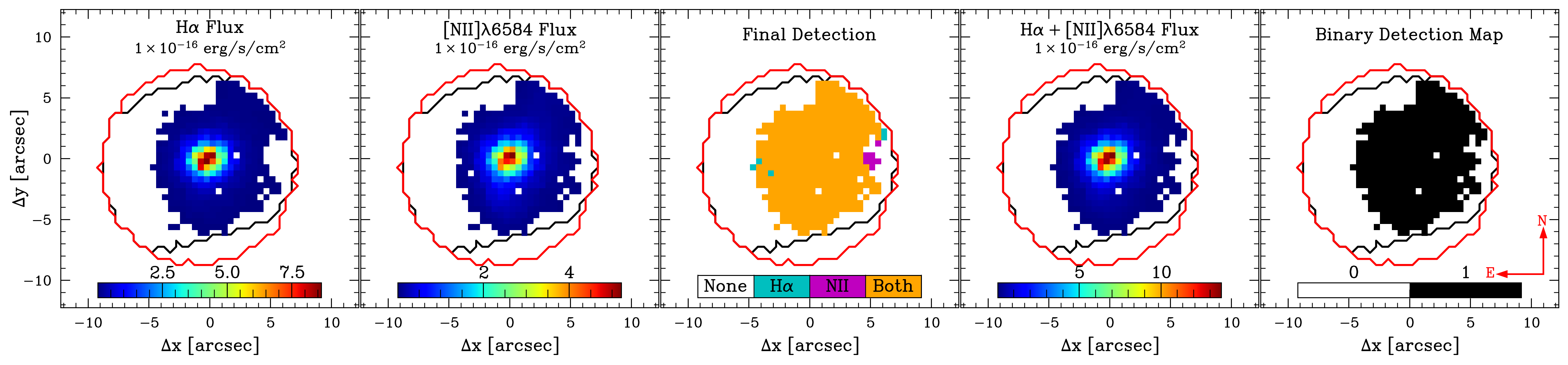}
    \caption{A visualisation of the steps involved in producing the ionised gas map for an example galaxy, \texttt{9011900084}. The two leftmost panels show the \ha and \nii flux maps generated using the detection procedure described in Section~\ref{subsec:spectral classification maps}, and the colour bar presents the flux values. The middle panel presents the final emission detection map, highlighting connected spaxels and coloured by the detected emission line, which is also shown in the colour bar.The two rightmost panels show the total emission map, derived from the final detection map using the same colour bar as the leftmost panels but applied to \ha + \nii emission. The corresponding binary detection map is also shown, where ``1" and ``0" indicate spaxels with and without detected emission, respectively. The black and red contours displayed in all panels refer to the stellar continuum defined in Section~\ref{subsec:Visual Classification}, and the SAMI field of view, respectively. The orientation is as North towards up, and East towards the left.}
    \label{fig:ionised gas detection}
\end{figure*}

\subsection{SAMI Data Products}\label{subsec:Data Products}
\subsubsection{Emission Line Measurements}\label{Emission line measurements}
The emission line fitting procedure is outlined in \citet{Green2018, Scott2018, Croom2021}. Briefly, strong emission lines (i.e. [OII]$(\lambda\lambda3726,3729)$, \hb, [OIII]$\lambda5007$, \ha, \nii, [SII]$(\lambda\lambda6716, 6731)$) are simultaneously fitted by \textsc{\texttt{Lzifu}} \citep{Ho2016} per spaxel. The underlying stellar continuum is modelled and subtracted following \citet{Owers2019}. Based on the redshifts from the SAMI input catalogues \citep{Bryant2015, Owers2017}, \textsc{\texttt{Lzifu}} then fits up to three Gaussian components to the emission lines, assuming the same gas kinematics across lines. Note that this component analysis is only applied to \ha since it usually has a higher SNR relative to other lines. In this paper, we use the total flux of \ha along with \hb, [OIII]$\lambda5007$, and \nii lines.

\subsubsection{Spectral Classification Maps}\label{subsec:spectral classification maps}
We adopt the spectral classification scheme described in \citet{Owers2019}. The scheme incorporates the emission and absorption line features simultaneously. A spectrum is classified as "emission" if a spaxel exhibits either \ha or \nii (i.e. the primary lines) and an additional line from \hb, \oiii [SII]$\lambda6716$, or [SII]$\lambda6731$ with SNR=(flux/$\sigma_{flux})>3$, where $\sigma_{flux}$ is the associated flux uncertainty estimated through a least-square fitting procedure \citep{Ho2016}. To avoid spurious detections due to template mismatches, the primary line must also have equivalent width, EW \gt 1 \AA. Otherwise, the spectrum is classified as "absorption". 

The emission line classification of \citet{Owers2019} is summarised as follows:
\begin{itemize}[leftmargin=0.1in]
    \item If \ha, \nii, and at least one of \hb or \oiii are detected, Baldwin, Phillips \& Terlevich (BPT) diagnostic \citep{BPT1981} is applied to classify the ionisation source as "star-forming (SF)", or "intermediate (INT)", or "non-star-forming (NSF)", through the demarcation lines defined by \citet{Kewley2001} (Ke01) and \citet{Kauffmann2003} (Ka03).
  
    \item If only \ha and \nii are present, WHAN diagram of \citet{CidFernandes2010} is used. SF, INT, and NSF classes are separated using $log([NII]\lambda6584/H\alpha) = -0.32$ and $log([NII]\lambda6584/H\alpha) = -0.1$ values, which approximate the Ke01 and Ke03 demarcation lines, respectively.
 
    \item In case of one of the primary lines missing, the classification is done as follows - star-forming (SF) if \ha is present, and non-SF if only \nii is detected.
    
    \item Additional sub-classes are defined based on the \ha line strength, following the recipe given in \citet{CidFernandes2010}. If EW(\ha) \lt 3 \AA, the spectrum can be divided into "wSF" or "rINT" or "rNSF". NSF spectra with EW(\ha) ranging between 3 and 6 \AAs are classified as "wNSF", whereas those with EW(\ha) \gt 6 \AAs are "sNSF". Here, \textit{"r", "w", and "s"} stand for "retired", "weak",  and "strong" sub-classes defined by \citet{CidFernandes2010}.
\end{itemize}

\noindent Absorption line measurements are carried out on absorption spectra as well as on emission spectra classified as NSF spectra, whose emission is not powered by photoionisation. The spectra with strong Balmer absorption (i.e. EW$(\overline{\mathrm{H}_{\delta\gamma\beta}})<-3$ \AAs and $|\mathrm{SNR}(\mathrm{EW}(\overline{\mathrm{H}_{\delta\gamma\beta}}))|>3$) are classified as "\hd-strong (HDS)" or "non-star-forming \hd-strong (NSF-HDS)". The absorption spectra not exhibiting strong Balmer absorption are categorised as "passive".
\\
\\
Combining emission and absorption classification, the global spectroscopic classification is performed as follows:  "passive galaxies (PASGs)" are defined as with more than 90\% of the spaxels classified as "passive", "rNSF", "rINT", "wNSF", "sNSF", or "wSF"; "star-forming galaxies (SFGs)" are those present with at least 10\% of the spaxels as "SF" or "INT"; galaxies, where "HDS" or "NSF-HDS" spaxels are making up more than 10\% of the total, are classified as  "\hd-strong galaxies (HDSGs)". Please refer to \citet{Owers2019} for the complete procedure of spectroscopic classification.

\subsubsection{Ionized Gas Maps}\label{subsec:ionized gas maps}

The primary goal of the paper is to identify galaxies with signatures of ongoing or recent ram pressure stripping and study their properties. Since it is well-established that RPS acts on the gas component \citep{Boselli2006, Cortese2021}, we therefore need to define the ionised gas distribution. The origin of the ionisation might be due to either recently formed stars (SF), as observed in jellyfish galaxies \citep{Poggianti2019a}, or non-star-forming (NSF) ionisation \citep{Fossati2016, Pedrini2022}. To account for both SF and NSF ionisation, we produce the ionised gas maps by combining \ha+\nii emission that will be used to characterise the morphological parameters of the gas distribution.
\\
\\
We determine the spaxels with reliable emission measurements by running a per-spaxel detection filter, which applies a set of criteria to exclude bad spaxels. The main criteria on emission line properties are the same as those outlined in \citet{Owers2019} to define emission spectra (also summarised in Section \ref{subsec:spectral classification maps}). Additional criteria spaxel selection criteria are: $\mathrm{|v_{gas}|}\ \& \ \sigma_{\mathrm{gas}} < 500 \rm \ km\ s^{-1}$, and their associated uncertainties must be less than 30 $\rm km\ s^{-1}$, ensuring the reliable emission fits are being used. 
%The detection maps for \ha and \nii are produced separately from these spaxels satisfying the criteria. 
\\
\\
The final step combines the \ha and \nii maps while enforcing an 8-connectivity rule, and requires a spaxel with either \ha or \nii detection to have at least three neighbouring detections. This ensures spatial coherence and excludes isolated low-SNR spaxels and/or high-SNR spurious detection.
\\
\\
Once the final detection map is produced,  for the spaxels with detections of both \ha and \nii, we sum fluxes (i.e. \ha + \nii) and propagate the uncertainties (i.e. $\sqrt{\sigma_{H\alpha}^2 + \sigma_{NII}^2}$); otherwise, we keep the flux and uncertainty as they are. Based on this final detection map, we also generate a binary detection map, in which "1" is assigned if there is a detection and "0" if none. Figure \ref{fig:ionised gas detection} illustrates the products resulting from each step of the detection procedure.

% \begin{figure*}[!t]
%     \centering
%     \includegraphics[width=\textwidth]{figures/SAMI_selection_plot.png}
%     \caption{\textbf{Left panel:} Stellar mass and redshift ($z_{tonry}$ for GAMA) distribution of the full SAMI sample \citep{Croom2021} with available spectroscopic classification from \citet{Owers2019} (N=2908). The cyan and red points represent the primary targets for GAMA and cluster regions, respectively, whereas \textbf{the blue and black open} circles show the secondary targets for the same samples. \textbf{The black solid line} shows the selection steps in stellar mass for primary GAMA targets as defined in \citet{Bryant2015}. The black dashed line marks the lower stellar mass limit we adopted in this study --- $log(M_\ast/M_\odot)=10$. The blue hatched area encloses GAMA galaxies selected as a control sample. \textbf{Right panel:} The distribution of detection fraction ($f_{det}$) of cluster galaxies, as defined \textbf{in} Section~\ref{subsec:Sample selection}, with $N_{spaxel}(Emission) \neq 0$ (N=194). The red, blue, and green histograms show the distributions for \textbf{passive galaxies (PASGs), star-forming galaxies (SFGs), and \hd-strong galaxies (HDSGs),} respectively. The black vertical line marks the cut applied as $f_{det} > 0.2$.} 
%     \label{fig:sample_selection}
% \end{figure*}

\subsubsection{SFR Maps}\label{subsec:SFR maps}
One key element of this study is to understand how RPS influences star-formation activity. Using IFU observations from SAMI, we are able to conduct a spatially resolved examination of star-forming properties of cluster galaxies. In this section, we describe the procedure to generate resolved SFR maps estimated from \ha fluxes.
\\
\\
We make use of the per-spaxel classification scheme of \citet{Owers2019} to identify star-forming spaxels. For these star-forming spaxels, we correct \ha fluxes for intrinsic dust extinction and Galactic foreground extinction similar to the recipes given in \citet{Medling2018} and \citet{Mun2024}. The dust extinction is corrected using the observed Balmer decrement (BD), $(\mathrm{H}\alpha/\mathrm{H}\beta)_\mathrm{{obs}}$. Here, BD maps are smoothed by a truncated Gaussian kernel with FWHM of 1.6 spaxels in order to account for aliasing in the differential atmospheric refraction effect (refer to \citet{Medling2018} for details). We assume an intrinsic BD, $(\mathrm{H}\alpha/\mathrm{H}\beta)_\mathrm{{int}}$, value of 2.86 based on Case B recombination line \citep{Osterbrock1989, Calzetti2000}, and Milky Way extinction curve (i.e. $R_V \simeq 3.1$) from \citet{Cardelli1989}. For spaxels without \hb detection, or with observed BD less than intrinsic value (i.e., 2.86), we fix the correction value and its error to 1 and 0, respectively, by following \citet{Medling2018}. Additionally\footnote{\url{https://docs.datacentral.org.au/sami/data-release-3/emission-line-data-products/}}, the spaxels with observed BD greater than 10, or \ha flux $>40\times 10^{-16}\ \rm erg/s/cm^{-2}$ are excluded, to eliminate spurious \ha fits to the edges of the bundles. The foreground Galactic extinction is corrected for \ha fluxes using dust maps produced by \citet{Schlegel1998, Schlafly2011} for the given coordinates. We use the observed frame wavelengths to obtain the extinction coefficients.
\\
\\
Once extinction-corrected \ha fluxes are obtained, we convert them to \ha luminosities via luminosity distances for the adopted cosmology. The redshifts of cluster galaxies are fixed to the cluster redshift to minimise the scatter due to peculiar motions relative to the cosmological redshift of the cluster, while $z_{tonry}$ values \citep{Baldry2012,Bryant2015} are used for the GAMA sample. We then determine SFRs by implementing the estimator given by \citet{Kennicutt1998} as follows:

\begin{equation}
    SFR = \frac{7.92 \times 10^{-42} \times L_{H\alpha}}{1.53} \ \ [\mathrm{M_\odot \ yr^{-1}}]
\end{equation}
Here, $L_{H\alpha}$ is \ha luminosity, and 1.53 is the conversion factor between the \citet{Salpeter1955} and \citet{Chabrier2003} initial mass functions.
\\
\\
For the given redshift range, a spaxel covers a range of physical size in kpc. Therefore, to compare SFRs within the same physical size, we determine star formation surface densities, $\Sigma_{\rm SFR}$, by normalising SFR with the physical area of each spaxel as follows:
\begin{equation}
       \Sigma_{SFR} = \frac{SFR}{(0.5 \times PS(z) )^2}\ \ [\mathrm{M_\odot \ yr^{-1} \ kpc^{-2}}]
\end{equation}
where 0.5 is the SAMI pixel scale in arcsecond/spaxel and $PS(z)$ is the physical size in kpc/arcsecond for the adopted cosmology. Redshifts are treated the same as above. Uncertainties are estimated by propagating the formal errors of \ha and \hb fluxes using the \texttt{uncertainties} Python package \citep{Lebigot2024}.

\begin{figure*}[!t]
    \centering
    \includegraphics[width=\textwidth]{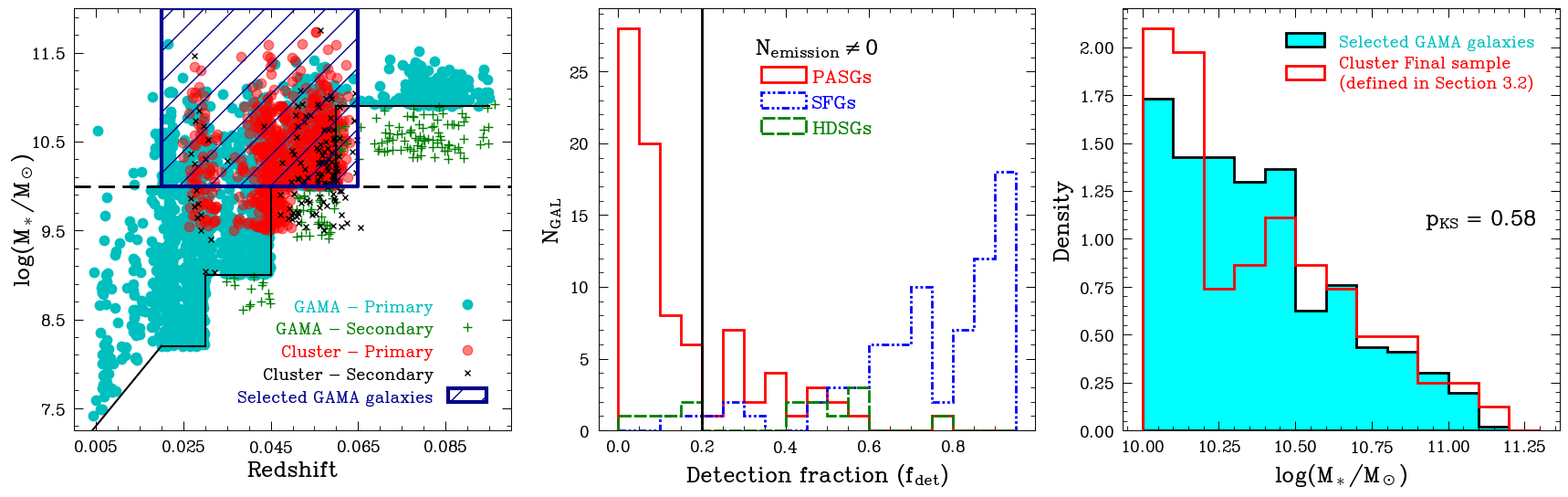}
    \caption{\textbf{Left panel:} Stellar mass and redshift ($z_{tonry}$ for GAMA) distribution of the full SAMI sample \citep{Croom2021} with available spectroscopic classification from \citet{Owers2019} (N=2908). The cyan and red points represent the primary targets for GAMA and cluster regions, respectively, whereas the green pluses and black crosses show the secondary targets for the same samples. The black solid line shows the selection steps in stellar mass for primary GAMA targets as defined in \citet{Bryant2015}. The black dashed line marks the lower stellar mass limit we adopted in this study --- $log(M_\ast/M_\odot)=10$. The blue hatched area encloses GAMA galaxies selected as a control sample. \textbf{Middle panel:} The distribution of detection fraction ($f_{det}$) of cluster galaxies, as defined in Section~\ref{subsec:Sample selection}, with $N_{spaxel}(Emission) \neq 0$ (N=194). The red, blue, and green step histograms show the distributions for passive galaxies (PASGs), star-forming galaxies (SFGs), and \hd-strong galaxies (HDSGs), respectively. The black vertical line marks the cut applied as $f_{det} > 0.2$. \textbf{Right panel:} The stellar mass distributions of the final cluster sample defined in Section~\ref{subsec:Results - Populations} (N=81) and the selected star-forming GAMA sample (N=462, the cyan histogram). P-value from the KS test for the comparison between the final cluster and the GAMA samples is shown on the right.} 
    \label{fig:sample_selection}
\end{figure*}

\subsubsection{Stellar Mass Maps}\label{subsec:stellar mass maps}
Stellar mass maps (i.e. per spaxel) are generated through the empirical relation given by \citet{Taylor2011} and adjusted by \citet{Bryant2015} and \citet{Owers2017} for the observed frame as follows

\begin{equation}
\begin{split}
     log(M_\ast/M_\odot) & = -0.4i + 2\ log(D_L(z)) -2 - log(1 + z)\\ &+ (1.2117 - 0.5893z) + (0.7106 - 0.1467z)\\ & \times (g - i)
\end{split}
\end{equation}\label{eq:stellar mass}
\noindent where $i$ is the Galactic extinction-corrected $i-$band apparent magnitude, $(g-i)$ is the Galactic extinction-free colour, $D_L$ is the luminosity distance under the assumed cosmology, and $z$ refers to redshift. Here, the luminosity distances are determined using the same approach as in Section~\ref{subsec:SFR maps} for both cluster and GAMA regions. The remaining three terms use the redshifts as described in Section~\ref{subsec:SAMI-GS}. \citet{Taylor2011} produce this relation for SDSS photometry, which is missing for the southern sky portion of the SAMI cluster sample. Therefore, we exploit $griz$ imaging from the Legacy Survey Data Release 10 \citep[LS DR10; ][]{Dey2019}, for full SAMI cluster coverage and to generate maps homogeneously across the sample. We first retrieve cutout images matching the size and the pixel resolution to SAMI cubes (i.e. 25\arcsec and 0.5\arcsec/pixel, respectively) and apply $m =  22.5 - 2.5 \ log_{10}(flux)$ for magnitude conversion. Unlike previous data releases, DR10 provides $i$-band data for the first time, incorporating additional DECam imaging, although not as complete as $grz$-bands in terms of spatial coverage. For any galaxies with missing $i$-band imaging, we generate mock $i$-band magnitudes/images via the relation, given below, for the available $riz$ magnitudes, where we use the largest aperture (i.e., 14\arcsec) and non-extinction corrected values from the \texttt{Tractor} catalogue.

\begin{equation}
    i_{LS,\ mock} = 0.318 \times (r-z)_{LS} + 0.057 + z_{LS} \ \ \text{(rms)} = 0.361
\end{equation}
We correct the magnitudes only for foreground Milky Way extinction, given that Equation 3 is not strongly impacted by the intrinsic dust attenuation \citep[see Section 5.2.3 in][]{Taylor2011}. Using transformations given by \citet{Abbott2021} for DES, we convert extinction-free Legacy magnitudes to SDSS magnitudes and then produce the stellar mass maps via Equation 3. Following the approach of \citet{Bryant2015}, only spaxels with $-0.2 < (g-i) < 2$ are included in the stellar mass maps. The empirical relation yields very precise mass measurements with an error ($1\sigma$) of \s0.1 dex independent of colour \citep{Taylor2011}.
\\
\\
Similar to SFRs, we also determine the spaxel-wise stellar mass surface densities, $\Sigma_\ast$, as follows:
\begin{equation}
        \Sigma_\ast = \frac{M_\ast}{(0.5 \times PS(z) )^2}\ [\mathrm{M_\odot \ kpc^{-2}}]
\end{equation}

\subsection{Sample Selection}\label{subsec:Sample selection}
Our main sample is constructed from the SAMI-DR3 \citep{Croom2021} primary targets with spectroscopic classification from \citet{Owers2019} - 732 cluster members that are allocated by \citet{Owers2017} and 1855 SAMI-GAMA galaxies ($\rm N_{TOT} = 2587$). Based on these galaxies, we select the sample by applying the following criteria. 

In the original selection for cluster galaxies in the SAMI-GS, two different stellar mass limits were adopted. As shown in the left panel of Figure \ref{fig:sample_selection}, for $\rm z_{cl}>0.045$, galaxies with \logMstar\lt10 were observed as secondaries. Therefore, to maintain and probe the same mass completeness across the sample, we include cluster members with \logMstar\gt10, removing 143 galaxies. Moreover, we visually exclude 24 galaxies due to miscentring or nearby objects within the hexabundle, to avoid possible contamination. Finally, we apply a cut depending on the ionised gas detection. We define a detection fraction as follows
\begin{equation}\label{eq:f_det}
    \rm f_{det} = \frac{N_{emission}}{N_{detection}}
\end{equation}
where $\rm N_{detection}$ is the total number of spaxels with either emission or continuum detection (i.e. $\rm continuum\ \cup \ emission$). The right panel in Figure \ref{fig:sample_selection} highlights $\rm f_{det}$ distribution as a function of spectral types. Here we only show galaxies with $\rm N_{emission}\neq0$ (N=194). Passive galaxies are the main occupants of the lower detection region, with their emission primarily originating from sources other than extended star formation, such as LINER/LIER emission from p-AGB stars, or a central AGN. To minimise this contamination, we exclude galaxies with $\rm f_{det}<0.2$ regardless of spectral type. Additionally, in the sample, some of the brightest cluster galaxies (BCGs) within each cluster are identified as having ionised gas emission larger than the threshold, $\rm f_{det}>0.2$. Since their emission is not related to RPS, we exclude these BCGs. These cuts result in a cluster sample of 123 members.

As already mentioned in Section~\ref{subsec:SAMI-GS}, we aim to compare cluster galaxies with their field counterparts. This is important since it will allow us to assess the environmental effects, through RPS, on galaxy properties, especially on star formation activity. For this purpose, as shown in the left panel of Figure~\ref{fig:sample_selection}, we select a subsample of star-forming GAMA galaxies with \logMstar$\gtrsim10$ and $\rm 0.02\lesssim z_{tonry} \lesssim0.065$ to probe the similar redshift and stellar mass regions with the cluster sample, returning 462 galaxies.  We compare how similar the stellar mass distributions of the final cluster sample defined in Section~\ref{subsec:Results - Populations} (N=81) and the GAMA galaxies selected as a control sample(N=462), using the two-sample Kolmogorov-Smirnov test \citep{KS}. As shown in the right panel of Figure~\ref{fig:sample_selection}, this comparison returns a p-value of 0.58, quantifying the similarity.

\begin{figure*}[!t]
    \centering
    \includegraphics[width=\textwidth]{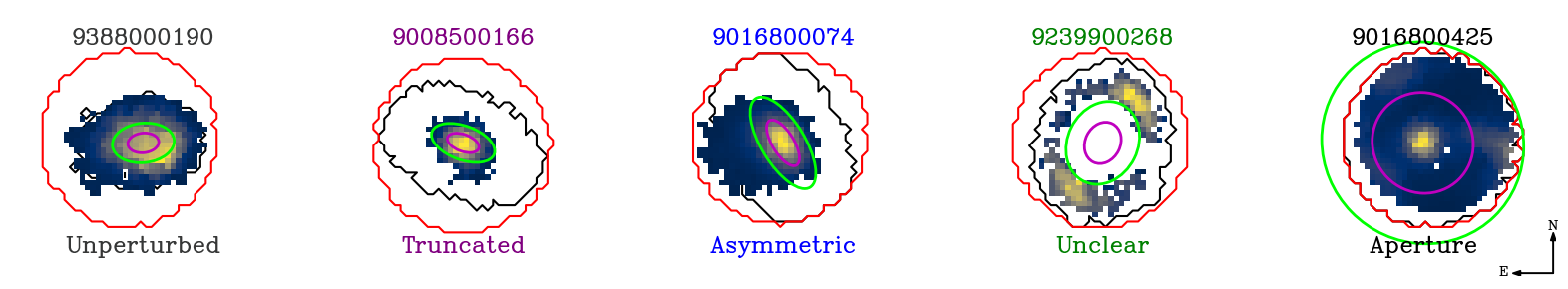}
    \caption{Examples of visual classes defined in Section~\ref{subsec:Visual Classification}. Galaxy IDs above each panel are coloured based on their classification stated below: unperturbed - grey, truncated - magenta, asymmetric - blue, unclear - green, aperture - black. The \texttt{cividis} colour map shows the total (i.e. \ha+\nii) emission, while the black and red contours represent the boundary of the stellar continuum at SNR=2, and the SAMI field of view (i.e. \s15$^{\prime \prime}$ in diameter), respectively. The magenta and lime ellipses correspond to 0.5 and 1 $\rm R_e$, respectively. North is towards the up, and East is towards the left.}
    \label{fig:example_for_visual_classes}
\end{figure*}
\section{Analysis and Results}\label{sec:Analysis and Results}
% two different identification methods explained below; 
% , and quantitative shape measurements (e.g. asymmetry, and concentration). 
\subsection{Visual Classification of Ionised Gas Distribution}\label{subsec:Visual Classification}
To understand how RPS alters the properties of cluster galaxies, it is essential to identify those undergoing RPS. To achieve this, we employ a visual classification scheme of the ionised gas morphology. Alongside optical imaging, visual classification has also been adapted by other wavelength regimes to examine different sub-components of galaxies, such as ionised gas and neutral gas components  \citep{Chung2009, Smith2010, Poggianti2016, Yoon2017, Jaffe2018, Roberts2021a}, which better trace and are used to identify ram-pressure affected galaxies in cluster environments. This is what motivates our approach of basing our visual classification scheme on the ionised gas distribution.
\\
\\
To do this, for the cluster galaxies, four visual classes are used to describe how the emission is distributed with respect to the stellar continuum. We define the continuum by taking the median value of the blue ($6400 < \lambda < 6540$ \AA) and red ($6590 < \lambda < 6700$ \AA) sidebands around the \ha and \nii emission lines, considering only spaxels with continuum $\mathrm{S/N} > 2$. The description for each class is given below. 
%We highlight each class with a different colour: unperturbed - grey; truncated - magenta; asymmetric - blue, and unclear - green. 
\begin{itemize}
    \item \textbf{\textcolor{darkgray}{Unperturbed (N)}:} Ionised gas is evenly distributed across the galaxy (i.e. follows the continuum) without any clear truncation and/or asymmetry with respect to the continuum.
    
    \item \textbf{\textcolor{magenta}{Truncated (T)}:} Relatively symmetric ionised emission distribution at the centre of the galaxy with marginal to clear truncation at sides. 
    
    \item \textbf{\textcolor{blue}{Asymmetric (A)}:} Ionised gas exhibits clear asymmetric features such as one-sided tail, extra-planar emission, and truncation at the opposite side of these asymmetric features.
    
    \item \textbf{\textcolor{green}{Unclear (U)}:} Irregular emission patterns and/or clear asymmetry which might not be explained by RPS or other features (e.g. spiral arms, etc.), or cases where other visual classes are not sufficient to describe.

    % \item \textbf{Aperture:} \textbf{The SAMI field of view corresponds to \s7.5 arcseconds in radius and \s80\% of primary targets have \textbf{been} imaged out to at least 1$R_e$ \citep{Croom2012, Bryant2015}. Still, there are galaxies having $R_e$ larger than the bundle radius, indicating that we only cover the central regions. We visually classify these cases as "aperture", where the ionised gas fills the entire bundle, preventing us from having a clear classification. Additionally, if a galaxy is partially covered by the bundle due to it is not well-centred, we also classify it as aperture.} 
\end{itemize}
\noindent Figure \ref{fig:example_for_visual_classes} shows examples of each of the visual classes defined above. 
The ionised gas maps were classified by four of us (OC, MSO, MP, and GQ) using the scheme above. Each person assessed the maps by giving a primary classification and a comment, such as "Asymmetric" and "Mild Asymmetry". Considering the primary classification, we assign a final classification for each galaxy as the visual class with the maximum count. If there is an equality between votes, such as (2N - 2T), we then take the comments into account. Here, we adopt a basic weighting scheme, shown below, as a tiebreaker: 

$\mathrm{w =              \begin{cases}
        1, & \text{No comment}\\
        0.75, & \text{Mild Asymmetry}\\
        0.75, & \text{Mild Truncation}\\
        0.5, & \text{Another class (e.g., N, T, A, U)}       
        \end{cases}}$ 
\\
\\
The minimum value we set for \textit{w} is 0.5, indicating that the person voting is indecisive between two classes. If there is no comment, no correction is needed (i.e. $w=1$). The remaining comments are set to 0.75 as the mid-point. Accordingly, we multiply each count of given votes by the weights defined per comments (i.e. $1 \times w$). The class with the maximum weighted count is then assigned as the final class. Lastly, in case of no consensus (i.e. 1 vote per class), we classified them as "Unclear".
\\
\\
With the final classifications established, we assessed the consistency between classifiers across all classes. For this purpose, we define a "Gold" sample, in which at least three out of four classifiers agree on the same visual class (e.g., 3/4 or 4/4 agreement). This Gold sample includes \s71\% (87/123) of the galaxies, indicating that the majority of the votes are in agreement, as an initial check. For the classifier-based assessment, we treat the final classes from the Gold sample as the "ground truth" and compare them with the votes per classifier. We use f1-scores and accuracy to quantify consistency. Three out of four classifiers achieve f1-scores over 0.80, 0.90 for "Unperturbed" and "Asymmetric" classes, respectively, and the minimum f1-score is 0.7. All f1-scores for "Truncated" galaxies exceed 0.90. For the "Unclear" class, f1-scores range between 0.36 and 0.8, which aligns with the definition of "unclear". The minimum accuracy across classifiers is 0.86. Overall, we find good agreement across the classes, particularly for the "Truncated" and "Asymmetric" galaxies, as the primary focus of this study.
\\
\\
Note that the SAMI field of view corresponds to \s7.5 arcseconds in radius and \s80\% of primary targets have been imaged out to at least 1$R_e$ \citep{Croom2012, Bryant2015}. Still, there are galaxies having $R_e$ larger than the bundle radius, indicating that we only cover the central regions. However, since $R_e$ is measured using $r$-band imaging \citep{Croom2021}, larger $R_e$ values may not necessarily mean that they are affected by aperture according to our visual classification. If the ionised gas of a galaxy fills the entire bundle, we visually classify that galaxy as "Aperture", irrespective of its $\rm R_e$. An example of an aperture-affected case is shown in Figure \ref{fig:example_for_visual_classes} (i.e. ID=\texttt{9016800425}). Additionally, if a galaxy is partially covered by the bundle due to it is not well-centred, we also classify it as "Aperture".
%Although they seem unperturbed within the SAMI field of view, we lack information from outside the bundle. Consequently, we do not classify these cases as unperturbed by defining a separate class, "Aperture"

\subsection{Populations}\label{subsec:Results - Populations}
In the previous section, we classified galaxies based on their ionised gas morphologies. We are able to examine whether these classes differ in key properties, such as structural parameters, projected phase-space distribution, and resolved star formation activities. Identifying such differences may provide insights into transformation as a function of RPS activity. The first step towards investigating the impact of RPS is to understand the distribution of classes across the sample before examining their properties in detail. 

\begin{figure}[!t]
    \centering
    \includegraphics[width=\linewidth]{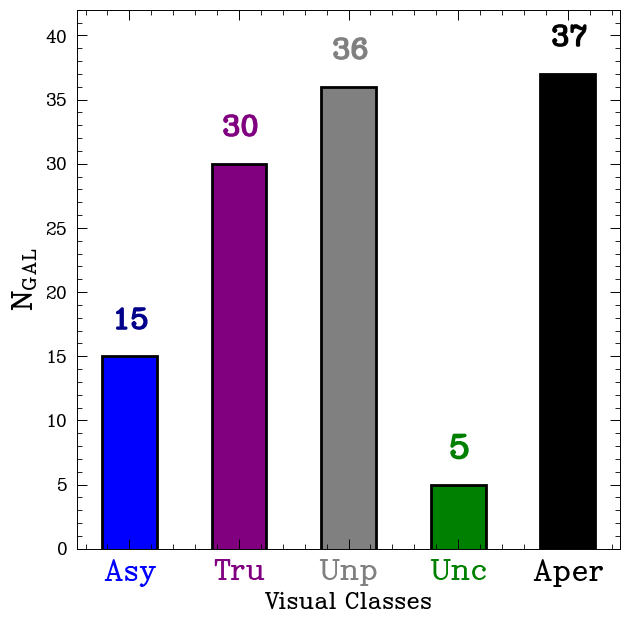}
    \caption{The number of galaxies per visual class. The colours are the same as described in Section~\ref{subsec:Visual Classification} --- unperturbed - grey, truncated - magenta, asymmetric - blue, unclear - green, and aperture - black. The numbers above each bin show the number of galaxies per class.}
    \label{fig:Visual_classes_histogram}
\end{figure}
\vspace{1em}

\noindent Figure~\ref{fig:Visual_classes_histogram} shows the distribution of the visual classes. The numbers above each bin indicate the number of galaxies in each class. Nearly two-thirds of the sample are classified as either unperturbed or aperture-affected, comprising 36 galaxies (\fraction{29.3}{3.7}{4.4}) and 37 galaxies (\fraction{30.1}{3.8}{4.4}), respectively, out of 123. Truncated galaxies make up 30/123 (\fraction{24.4}{3.4}{4.2}), asymmetric galaxies account for 15/123 (\fraction{12.2}{2.3}{3.6}) of the sample, and the minority is represented by unclear galaxies with 5/123 (\fraction{4.1}{1.1}{2.6}). This results in a total of \fraction{36.6}{4.1}{4.5} for RPS-affected galaxies, combining truncated and asymmetric classes (45/123). Associated uncertainties shown here and below are estimated following the approach given in \citet{Cameron2011}.

\begin{table}[!t]
    \centering
    \begin{NiceTabular}{|c|c|c|c|}[corners=NW]
        \Hline
        \rowcolor{gray!0}
        & \Block{1-1}{PASGs} & \Block{1-1}{SFGs} & \Block{1-1}{HDSGs} \\
        \Hline
        \rowcolors{gray!0}{}
        \textcolor{darkgray}{\textbf{Unperturbed}} & 1 & 34 & 1 \\
        \Hline
        \textcolor{magenta}{\textbf{Truncated}} & 11 & 17 & 2 \\
        \Hline
        \textcolor{blue}{\textbf{Asymmetric}} & 5 & 3 & 7 \\
        \Hline
        \textcolor{green}{\textbf{Unclear}} & 3 & 2 & 0 \\
        \Hline
        \textbf{Aperture} & 0 & 37 & 0 \\
        \Hline
    \end{NiceTabular}
    \caption{The breakdown of the visual classes as a function of spectral types described in Section~\ref{subsec:spectral classification maps}.}
    \label{tab:Breakdown of visual classes}
\end{table}

Moreover, we examine the distribution of visual classes as a function of spectral types. Table~\ref{tab:Breakdown of visual classes} presents the breakdown of classes. Nearly all unperturbed galaxies (34/36; \fraction{94.4}{6.5}{1.8}) are star-forming. More than half of truncated galaxies (\fraction{56.7}{9.1}{8.3}; 17/30) are also SFGs, while PASGs account for 11/30 (\fraction{36.7}{7.7}{9.3}). The majority of HDSGs (\fraction{70}{16.8}{10.0}) contribute to asymmetric galaxies, accounting for nearly the half of the class (\fraction{46.7}{11.6}{12.4}; 7/15), followed by PASGs with \fraction{33.3}{9.5}{13.4} (5/15) and SFGs with 3/15 (\fraction{20}{6.5}{13.6}). Aperture-affected sample only consists of star-forming galaxies, which is expected given that the majority of our sample (93/123) is star-forming.

\begin{figure}[!t]
    \centering
    \includegraphics[width=\textwidth]{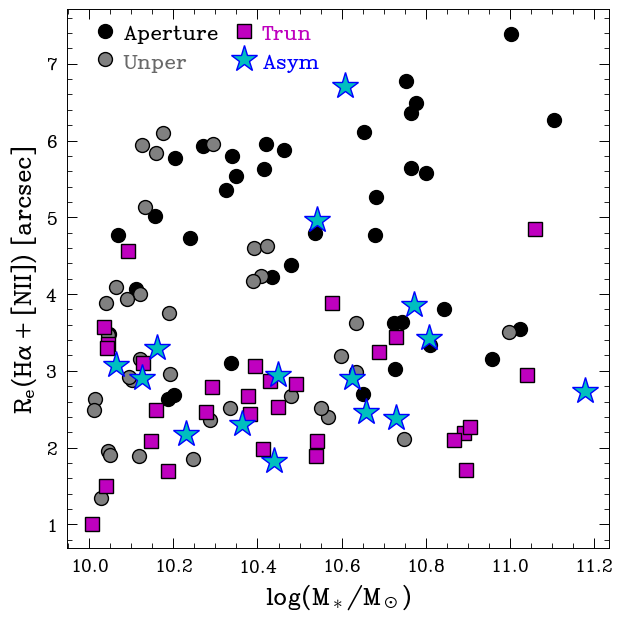}
    \caption{Distribution of stellar mass and half light radius of \ha + \nii flux, measured within the SAMI bundle, for the visual classes defined in Section~\ref{subsec:Visual Classification} --- unperturbed - grey points, truncated - magenta squares, asymmetric - cyan stars, and aperture - black points. Here, we exclude "Unclear" cases. The majority of galaxies with larger $\mathrm{R_{e}(H}\alpha \rm + [NII])$ values are primarily aperture-affected galaxies, quantitatively supporting the reasoning behind them, which is the emission filling the bundle.
    }
    \label{fig:Re emission plot}
\end{figure}

To understand the aperture-affected cases more quantitatively, we compare the stellar masses and the half-light radii of the total \ha+\nii flux, $\rm R_e$(\ha + \nii), which are measured using elliptical apertures. Figure~\ref{fig:Re emission plot} shows the distribution of mass and $\rm R_e$(\ha + \nii) across the samples. It is clearly seen that the majority of galaxies with larger $\rm R_e$(\ha + \nii) values are aperture-affected cases, while almost all unperturbed, truncated, and asymmetric galaxies exhibit lower $\rm R_e$(\ha + \nii) values, quantitatively agreeing that the ionised gas distribution of aperture cases is more extended. In addition, galaxies affected by the SAMI aperture appear to be more massive compared to their unperturbed counterparts. For the remainder of the paper, we exclude galaxies classified as either "Unclear" or "Aperture" from our analysis, which leaves us with 81 cluster galaxies as our final sample.

\subsection{Visual classification on quantitative space}\label{subsec:quantitative analysis}
\begin{figure*}[!t]
    \centering
    \includegraphics[width=\textwidth]{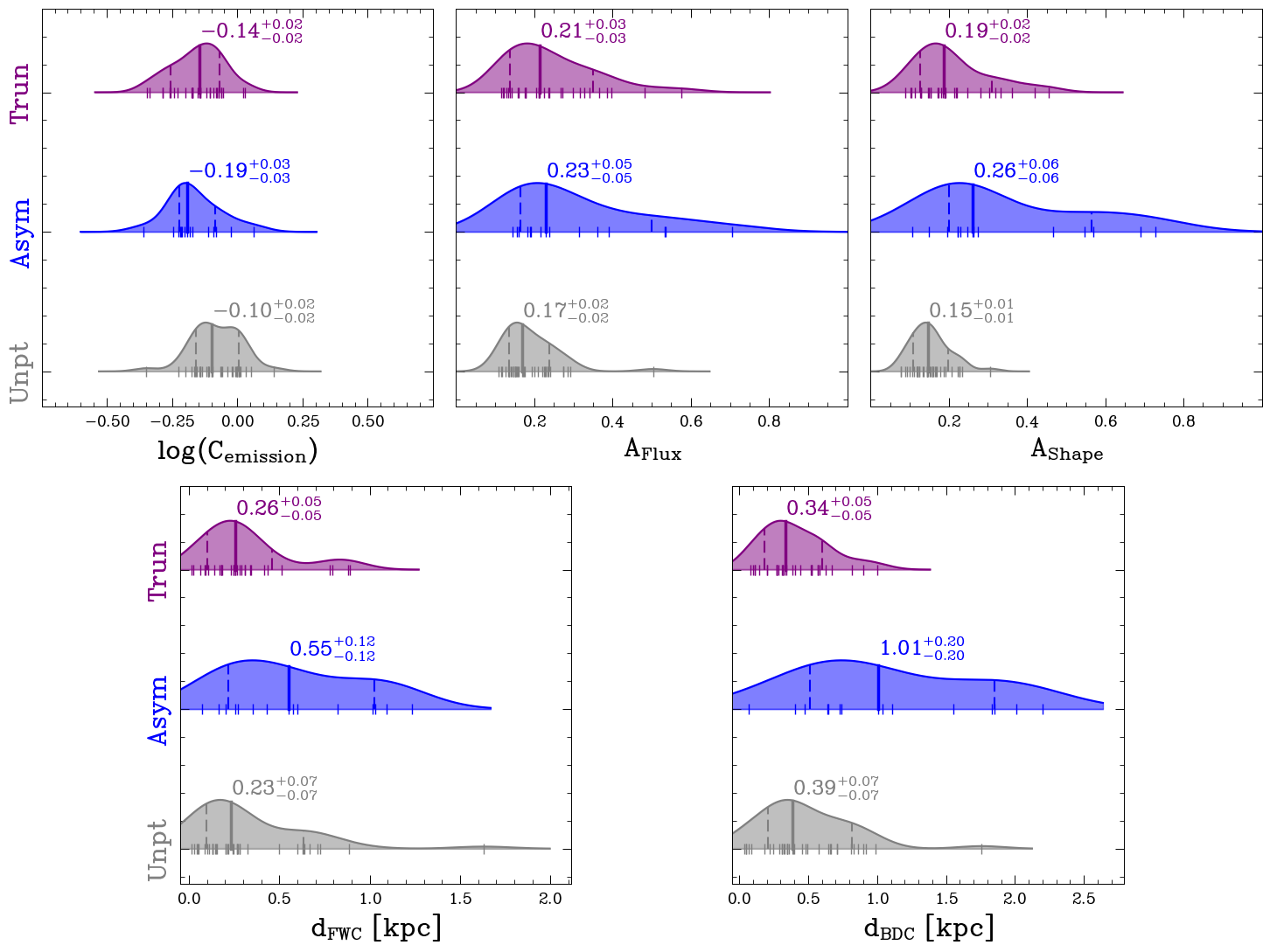}
    \caption{The probability distribution functions (PDFs) of non-parametric measures as a function of visual class. The colour code is the same as defined in Section~\ref{subsec:Visual Classification} --- unperturbed - grey, truncated - magenta, asymmetric - blue. The vertical lines for all panels mark the $16^{\rm th}, \ 50^{\rm th}, \text{ and, }84^{\rm th}$ percentiles, respectively. The medians (i.e. the $50^{\rm th}$ percentile) are also given next to each distribution, and the associated uncertainties are the standard errors, defined as $1.253\times\sigma/\sqrt{N}$. The stripes embedded within PDFs represent the individual values for each distribution. The top panel shows the distributions for concentration and asymmetry parameters (standard definition, $\rm A_{Flux}$, in the middle, shape asymmetry, $\rm A_{shape}$, in the right panel). Bottom panels show the offsets estimated via flux maps ($\rm d_{FWC}$; left) and binary detection maps ($\rm d_{BDC}$; right). These plots indicate that parameters, such as $\rm A_{shape} \text{ and } d_{BDC}$, are effective in separating visual classes quantitatively, especially those with asymmetric features. Both truncated and asymmetric galaxies have lower concentration values compared to their unperturbed counterparts. Moreover, especially the parameters defined using binary detection maps (e.g. $\rm A_{shape}$ and $\rm d_{BDC}$) are more robust at reproducing the visual classification.}
    %\textbf{, and aperture - black}
    \label{fig:Visual_shape_params}
\end{figure*}
Regardless of how high the expertise across the classifiers is, any kind of visual classification is prone to being subjective. Therefore, quantitative analysis methods based on non-parametric measures of the stellar light distribution have been developed to understand galaxy morphologies in a coherent and unbiased manner, such as CAS statistics \citep{Abraham1996, Bershady2000, Conselice2000, Conselice2003} and Gini-M$_{20}$ \citep{Lotz2004}. These methods have also been applied to nebular emission \citep{Nersesian2023} and exploited to identify ram-pressure-affected galaxies through different wavelengths \citep{McPartland2016, RP2020, Roberts2021b, Bellhouse2022, Krabbe2024}. Therefore, we similarly aim to identify the ram-pressure-affected galaxies in our sample quantitatively using different non-parametric measures such as concentration, asymmetry, and the offset between the ionised gas and the stellar continuum, as described below. 
\\
\\
\textbf{Concentration (C):} We adopt the concentration definition of \citet{Schaefer2017}, given by the ratio of the half-light radius determined in emission and continuum flux maps using elliptical apertures, as follows  
\begin{equation}
        C = \frac{r_{50}(H\alpha + [NII])}{r_{50}(Continuum)}
\end{equation}\label{eq:concentration}
\noindent This definition allows us to examine the emission relative to the continuum, aligning with the visual classification scheme. We estimate the concentration similar to the method outlined in \citet{Schaefer2017} and \citet{Owers2019}, except we consider the total \ha+\nii flux defined in Section \ref{subsec:ionized gas maps}. The interpretation of this parameter is that the higher the value, the more extended the ionised gas emission.  
    
\noindent \textbf{Asymmetry (A):} We employ the standard asymmetry expression \citep{Abraham1996, Conselice2000, Conselice2003} as follows  

    \begin{equation}
        A = \frac{\Sigma|I_0 - I_{180}|}{2\times\Sigma |I_0|}
    \end{equation}\label{eq:asymmetry}

\noindent with $I_0$ being the original emission line flux image and $I_{180}$ being the flux image rotated by 180$^{\circ}$. We fix the centre of rotation as the cube centre (i.e. galaxy centre). Due to its flux-weighted nature, the standard definition (hereafter $\rm A_{flux}$) is not sensitive to the faint/low-surface-brightness features, for instance, possible gas tails in our case. In order to account for these features, \citet{Pawlik2016} introduced a new asymmetry measure, namely "shape asymmetry" (hereafter $\rm A_{shape}$). It uses the same formulation defined above, yet instead of a flux image, a "binary detection" image, made of 1 and 0 based on the pixels included, is input into the calculation. In this way, pixels hosting the faint features, such as those found in ram-pressure stripped tails, are weighted equally with the rest. The following equation summarises the asymmetry definitions used here.
\vspace{0.3em}
\[
\mathrm{
A =
\begin{cases}
    \rm A_{\text{flux}}, & \text{if } \rm I_0 \text{ is flux image} \\
    \rm A_{\text{shape}}, & \text{if } \rm I_0 \text{ is binary detection image}
\end{cases}
}
\]
\vspace{0.3em}
\noindent \textbf{Offset (d):} We compute the centroid offset similarly to \citet{Liu2021}. As done in asymmetry, we determine two offsets between the cube centre, $\rm(x_c, y_c)$, and i) emission line flux-weighted centre (FWC) and ii) emission line binary detection centre (BDC), defined as $\rm (x_i, y_i)$ in the formula below, respectively. 
\vspace{0.2em}
\begin{equation}
   \rm d_i = \sqrt{(x_{c}-x_{i})^2 + (y_{c}-y_{i})^2} \ \ \ for \ i = \{FWC, BDC\}
\end{equation}\label{eq:offset}
\noindent Where $d_i$ returns the offset in pixels. Since a spaxel can cover a range of physical sizes in kpc, we multiply the offset by both the SAMI pixel scale and the physical size defined as in Section~\ref{subsec:SFR maps} to convert it to kpc.

Associated uncertainties are estimated via Monte Carlo realisations with $\rm N_{iteration}=100$. In each iteration, we perturb \ha and \nii line maps with noise sampled from a normal distribution of N(0, $\sigma_{flux}$), where $\sigma_{flux}$ is the flux error per spaxel. We then run perturbed maps through the detection procedure described in Section \ref{subsec:ionized gas maps} and measure the same quantities. The standard deviation of these measurements is assigned as the uncertainty per parameter.
\\
\\
Considering the combinations between flux and binary detection maps, we have five different non-parametric measurements in total. Figure~\ref{fig:Visual_shape_params} shows the distributions of these parameters as a function of the visual classes. Each visual class is coloured as described in Section~\ref{subsec:Visual Classification}. The top row presents concentration, flux-weighted asymmetry, and shape asymmetry parameters, while the offsets determined via flux-weighted and binary detection centroids are given in the bottom row. 
% The vertical lines denote the $16^{\rm th}, \ 50^{\rm th}, \text{ and, }84^{\rm th}$ percentiles, respectively, and the median of each distribution is shown next to it. The uncertainties of the median values are defined as $1.253\times\sigma/\sqrt{N}$ (i.e. standard errors).

\begin{table}[!h]
    \centering
    \begin{NiceTabular}{|c|c|c|c|}[corners=NW] % c|c|c|
        \Hline
        \rowcolor{gray!0}
        & \Block{1-1}{\textbf{Trun - Asym}} & \Block{1-1}{\textbf{Trun - Unper}} & \Block{1-1}{\textbf{Asym - Unper}}\\% & \Block{1-1}{T - U} & \Block{1-1}{A - U} & \Block{1-1}{N - U}\\
        \Hline
        \rowcolors{gray!0}{}
        $\rm \bf log(C_{emission})$ & 0.416 & \Block[fill=red!40,opacity=0.35]{1-1}{\textbf{0.007}} & \Block[fill=red!40,opacity=0.35]{1-1}{\textbf{0.002}}\\% & \textbf{0.040} & 0.071 & 0.301 \\
        \Hline
        $\rm \bf A_{flux}$ & 0.334 & 0.059 & \Block[fill=red!40,opacity=0.35]{1-1}{\textbf{0.008}}\\% & \textbf{0.035} & 0.171 & \textbf{0.001} \\
        \Hline
        $\rm \bf A_{shape}$ & \Block[fill=red!40,opacity=0.35]{1-1}{\textbf{0.005}} & \Block[fill=red!40,opacity=0.35]{1-1}{\textbf{0.011}} & \Block[fill=red!40,opacity=0.35]{1-1}{\textbf{0}}\\% & \textbf{0.005} & 0.173 & \textbf{0} \\
        \Hline
        $\rm \bf d_{FWC}$ & \Block[fill=red!40,opacity=0.35]{1-1}{\textbf{0.017}} & \Block{1-1}{0.814} & \Block[fill=red!40,opacity=0.35]{1-1}{\textbf{0.022}}\\% & 0.144 & 0.899 & 0.970 \\
        \Hline
        $\rm \bf d_{BDC}$ & \Block[fill=red!40,opacity=0.35]{1-1}{0} & \Block{1-1}{0.373} & \Block[fill=red!40,opacity=0.35]{1-1}{$\bf 2 \times 10^{-4}$}\\% & \textbf{0.043} & 0.313 & 0.730 \\
        \Hline
    \end{NiceTabular}
    \caption{p-values from the AD test of visual class pairwise comparisons of the shape parameters defined in Section~\ref{subsec:quantitative analysis}. Each row lists the parameters, while each row notes the visual class pair. Here, red shaded cells indicate $p<0.05$}
    \label{tab:p-values for shape parameters}
\end{table}

\vspace{1em}

\noindent Considering the concentration parameter distributions shown in the top left panel of Figure~\ref{fig:Visual_shape_params}, the truncated and asymmetric galaxies both exhibit systematically lower concentration values than those of unperturbed galaxies. The median values are -0.14 dex, -0.19 dex, and  -0.10 dex for truncated, asymmetric, and unperturbed galaxies, respectively. This similarity is primarily driven by the definition of the concentration parameter using fluxes. As already mentioned above, since asymmetric features, such as tails, have lower fluxes, this results in lower concentration values (i.e., more concentrated). Additionally, because the visual classification of asymmetric galaxies includes partially truncated emission, similar concentration values can be yielded. The distribution for unperturbed galaxies also extends towards low concentration values. Given that these unperturbed galaxies do not show any visual signatures of truncation, smaller values of concentration may be driven by extended emission in the presence of centrally concentrated flux, rather than physical truncation. We compare the concentration distributions pairwise employing the two-sample Anderson-Darling (AD) test \citep{ADtest}, which evaluates whether two samples originate from the same parent sample or not by taking the tails of the distributions into account. P-values for the pairwise comparisons of each parameter are given in Table~\ref{tab:p-values for shape parameters}. The AD test yields $\rm p\gg0.05$ only for the comparison of the concentration distributions between truncated and asymmetric galaxies, confirming the similarity quantitatively. In contrast, the test returns $\rm p\ll0.05$ for comparisons involving unperturbed galaxies, indicating that unperturbed and RPS-affected galaxies are not likely sampled from the same parent distribution.

\begin{figure*}[!t]
    \centering
    \resizebox{0.85\textwidth}{!}{
    \includegraphics[width=\linewidth]{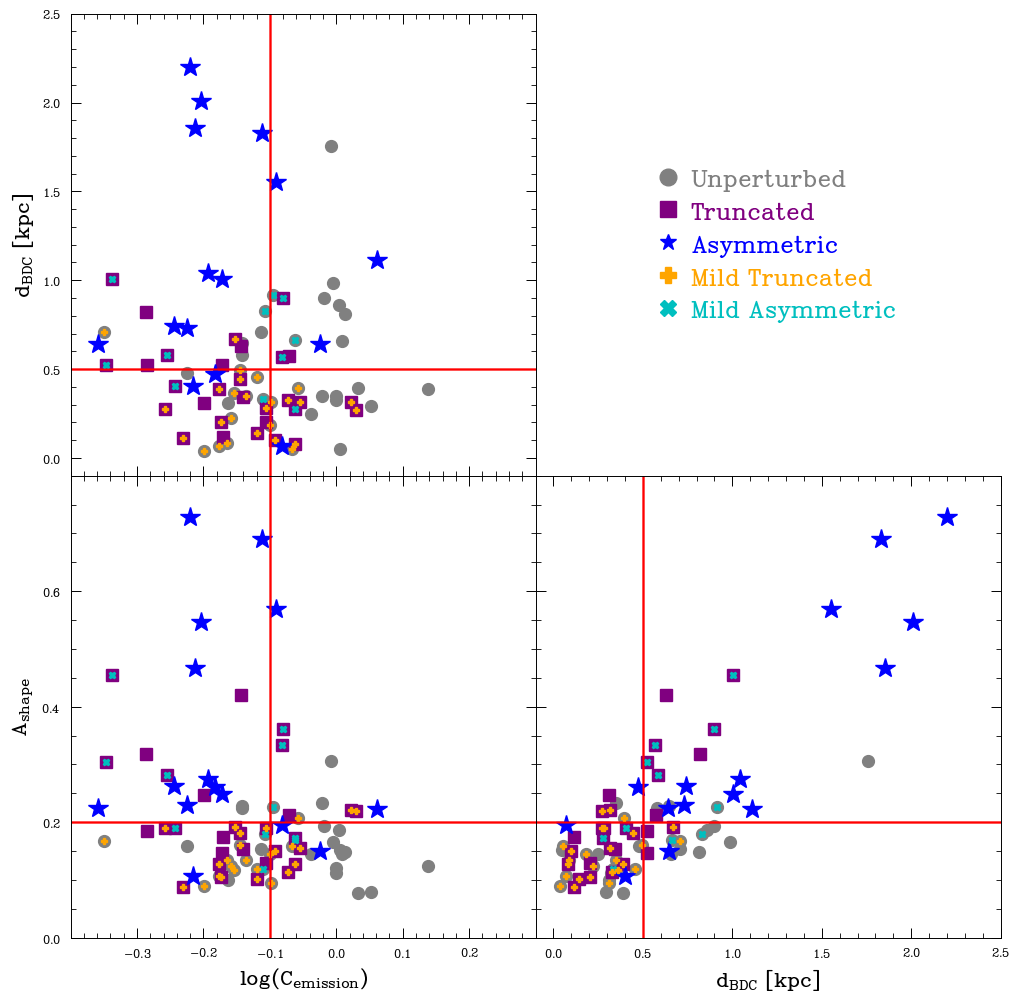}
    % The black crosses mark the aperture-affected galaxies.
    \caption{Corner plot for concentration, $\rm A_{shape}$, and $\bf d_{BDC}$ parameters. The grey points, magenta squares, and blue stars represent the galaxies classified as "unperturbed", "truncated", and "asymmetric", respectively. The orange pluses and cyan crosses are the galaxies showing mild truncation and mild asymmetry, respectively, based on the secondary comments. The red solid lines mark the rough demarcation lines between visual classes, as defined in Section~\ref{subsec:quantitative analysis}.}
    \label{fig:corner plot}
    }
\end{figure*}
\vspace{1em}
\noindent Focusing on the upper middle and right panels of Figure~\ref{fig:Visual_shape_params}, the flux weighted asymmetry, $\rm A_{flux}$, yields similar distributions across the visually identified classes, especially between truncated and asymmetric galaxies. On the other hand, the shape asymmetries, $\rm A_{shape}$, put forth more significant differences between classes, returning lower values for truncated and unperturbed galaxies, whereas asymmetric galaxies extend towards higher values. This is also quantitatively supported by the AD test. As shown in Table~\ref{tab:p-values for shape parameters}, only the asymmetric-unperturbed pair yields $\rm p < 0.05$ for $\rm A_{flux}$, showing that the flux weighted approach is unable to differentiate the asymmetric and truncated samples. In contrast, the comparisons of the distributions of shape asymmetries return p-values less than 0.05 for all pairs, indicating that $\rm A_{shape}$ is statistically more robust in capturing the asymmetric features compared to $\rm A_{flux}$. 
\\
\\
As shown in the bottom panel of Figure~\ref{fig:Visual_shape_params}, both $\rm d_{FWC}$ and $\rm d_{BDC}$ distributions are similar across visual classes. The majority of the truncated and unperturbed galaxies are clustered at the lower offset values (\s 0.3 kpc) with few extending towards higher values ($\gtrsim 1$ kpc). Notably, the asymmetric sample exhibits systematically higher offset values as well as a broad distribution in the offset. As we compare the distributions, the AD test returns $\rm p < 0.05$ for all pairwise comparisons of $\rm d_{FWC}$ and $\rm d_{BDC}$, except for the comparison between truncated and unperturbed samples. This indicates that the asymmetric galaxies are not likely drawn from the sample parent distribution, compared to the truncated and unperturbed galaxies.
%The distribution of $\rm d_{FWC}$ values for the unperturbed galaxies \textbf{extends} towards higher values. This may be due to features like spiral arms having higher flux densities. On the contrary, the asymmetric sample exhibits higher $\rm d_{BDC}$ values with respect to both truncated and unperturbed samples. The unperturbed galaxies are still spread over a large offset values. As we \textbf{compare the distributions}, the AD test returns $\rm p < 0.05$ for all pairwise comparisons of $\rm d_{FWC}$ and $\rm d_{BDC}$, except for the comparison of $\rm d_{FWC}$ between asymmetric and unperturbed samples.
\\
\\
Comparisons shown above reveal that the $\rm A_{shape}$ and $\rm d_{BDC}$ parameters --- derived from the binary detection map --- and concentration are more robust in segregating the RPS candidates from their unperturbed counterparts. This means that we can use these three parameters to demarcate RPS-affected galaxies. Figure~\ref{fig:corner plot} shows a corner plot for two-dimensional joint distributions between concentration, $\rm A_{shape}$, and $\rm d_{BDC}$ parameters. The grey points, magenta squares, and blue stars represent the visually classified unperturbed, truncated, and asymmetric galaxy samples, respectively. The black crosses, orange pluses, and cyan crosses show aperture-affected galaxies, and those exhibiting "mild truncation" and "mild asymmetry" based on the secondary comments, respectively. The red solid lines mark demarcation lines between visual classes, which are defined below. 
\\
\\
There are several noteworthy trends in the 2D parameter spaces shown in Figure~\ref{fig:corner plot}. The majority of the visually classified truncated galaxies reside in the lower left corner of all three plots, i.e., the quantitative morphologies align with them having concentrated ionised gas distributions that are symmetric about the centre of the galaxy. 5 out of 8 visually classified truncated galaxies that have $\rm A_{shape}> 0.25$ and $\rm d_{BDC} > 1$ were mostly also identified visually as having mild asymmetry, which gives further confidence in the quantitative measures. Furthermore, the majority of the visually identified asymmetric galaxies generally lie in regions of the three spaces that are distinct from the other two visual classes, e.g., they generally have both larger $\rm A_{shape}$ and $\rm d_{BDC}$ when compared to the other galaxies. This is expected because these two parameters are designed to quantify asymmetry and, therefore, provide confidence that the visual and quantitative schemes are yielding reliable results.
%Consistent with the 1D results presented in Figure~\ref{fig:Visual_shape_params}, the visually identified unperturbed galaxies primarily populate the higher $\rm log(C_{emission})$ and lower $\rm A_{shape}$ regions.
\\
\\
% We test \textbf{the null hypothesis, which states whether the two different samples are drawn from the same parent sample or not, between classes,} 
We compare the 2D distributions shown in Figure~\ref{fig:corner plot} between the samples to understand whether they are drawn from the same parent sample or not. For this purpose, we use the non-parametric multivariate two-sample kernel density estimation (KDE) test developed by \citet{Duong2012} and available within the \texttt{ks} library \citep{Duong2007} in \texttt{R}. Table~\ref{tab:2d KDE p-values for corner plot} summarises the p-values resulting from each pairwise comparison per parameter space. The KDE test returns $\rm p<0.05$ for the comparisons of $\rm log(C_{emission}) - d_{BDC}$, and $\rm log(C_{emission}) - A_{Shape}$ between the asymmetric and unperturbed galaxies, while it returns marginal significance, $\rm p = 0.057$, in $\rm log(C_{emission}) - d_{BDC}$ comparison for the truncated and unperturbed galaxies.
%The p-values mostly indicate marginal to statistically significant differences between classes for each parameter pair. The parameter spaces with $\rm log(C_{emission})$ and $\rm d_{BDC}$ return $\rm p<0.05$ for comparisons between asymmetric-unperturbed and truncated-unperturbed pairs, respectively. The comparisons between truncated and asymmetric samples yield $\rm p>0.05$, although the pairs with $\rm d_{BDC}$ return a marginal difference.
%\textbf{A smoothed cross-validation bandwidth is applied for smoothing the data to estimate the kernel densities, and is determined separately for each sample.}
\begin{table}[!h]
    \centering
    \begin{NiceTabular}{|c|c|c|c|}[corners=NW]
        \Hline
        \rowcolor{gray!0}
        & \Block{1-1}{Trun - Asym} & \Block{1-1}{Trun - Unper} & \Block{1-1}{Asym - Unper}\\
        \Hline
        \rowcolors{gray!0}{}
        $\rm \bf log(C_{emission})$ - $\rm \bf A_{shape}$ & 0.284 & 0.356 & \Block[fill=red!40,opacity=0.35]{1-1}{\textbf{0}} \\
        \Hline
        $\rm \bf log(C_{emission})$ - $\rm \bf d_{BDC}$ & 0.057 & \Block{1-1}{0.519} & \Block[fill=red!40,opacity=0.35]{1-1}{\textbf{0.028}} \\
        \Hline
        $\rm \bf A_{shape}$ - $\rm \bf d_{BDC}$ & 0.101 & \Block{1-1}{0.184} & \Block{1-1}{0.109} \\
        \Hline
    \end{NiceTabular}
    \caption{p-values from the 2D KDE test per pairwise comparison of each parameter space. Each row lists the parameters compared, while each column corresponds to a visual class pair. Red shaded cells mark pairs with p < 0.05.}
    \label{tab:2d KDE p-values for corner plot}
\end{table}

\noindent After testing different combinations of morphological parameters, we define the following demarcation lines between each sample as shown in Figure~\ref{fig:corner plot}: $\rm log(C_{emission}) \sim -0.1$ roughly separates the unperturbed and RPS-affected galaxies; $\rm A_{shape} \sim 0.2 \text{ and } d_{BDC} \sim 0.5$ kpc approximately divide the asymmetric sample from the truncated and unperturbed counterparts.
%It is worth noting that Figure~\ref{fig:Visual_shape_params} also points to a subsample of visually identified unperturbed galaxies exhibiting an extension to high $\rm A_{shape}$ and $\rm d_{BDC}$ values. This takes away from the robustness of those quantitative morphology measures in being able to identify asymmetric galaxies. However, as marked with black crosses in Figure~\ref{fig:corner plot}, these are entirely driven by the aperture-affected galaxies in the unperturbed sample, which suffer from the apertures of SAMI bundles containing the emission being offset from the centre of the galaxy. Having kept this matter in mind, w

\subsection{Comparison of the projected phase-space distributions}\label{subsec: PPS Distribution}
\begin{figure*}[!t]
    \centering
    \includegraphics[width=\linewidth]{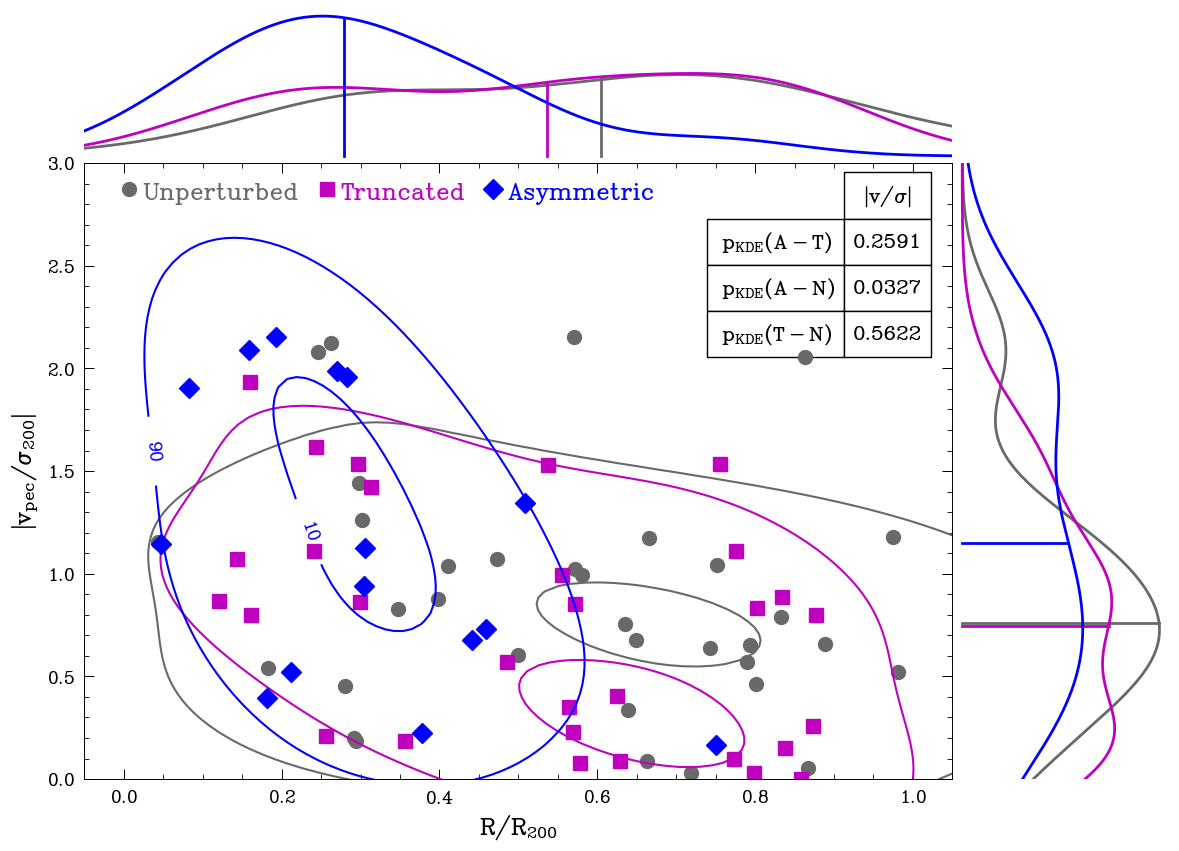}
    \caption{Projected phase-space distribution (\textit{Note: the unclear and aperture-affected sample is excluded}). The grey points, magenta squares, and blue diamonds represent the unperturbed, truncated, and asymmetric galaxies, respectively. The contours indicate $\bf 10^{th} \text{ and } 90^{th} $ percentiles of the 2D distribution of each sample, generated by \texttt{ks} library \citep{Duong2007} in \texttt{R}. The p-values from the 2D KDE test \citep{Duong2012}, comparing \rrtwo with $|v_{pec}/\sigma_{200}|$ for each pair, are given in the upper right. The top and right panels show the 1D KDEs of normalised cluster-centric distances and line-of-sight velocities, respectively, with solid lines indicating the bi-weigthed means for each class.}
    \label{fig:PPS}
\end{figure*}
% In the previous sections, we 
Analyses presented in the previous sections showed that the ionised gas morphologies differ in unperturbed, truncated, and asymmetric galaxies both visually and quantitatively. As the primary goal of this paper, we aim to understand whether these differences arise as a function of RPS activity (i.e., their accretion times). For this purpose, in this section, we examine the projected phase-space (PPS) distribution of each sample. It is well-established that the PPS diagram determined from the normalised projected cluster-centric distances, \rrtwo, and the normalised line-of-sight (LOS) velocity, $v_{pec}/\sigma_{200}$, is a useful tool to probe populations with different accretion times \citep{Haines2012, Haines2015, Noble2013, Oman2013, Rhee2017, Owers2019}. 
%Here, $v_{pec}$ is defined as $c\times(z_{galaxy} - z_{cluster})/(1+z_{cluster})$ with c being the speed of light; $\sigma_{200}$ is the cluster velocity dispersion defined within the characteristic cluster radius, \rtwo, which encloses 200 times the critical density, $\rho_{crit}$, for the given redshift.
\\
\\
Figure~\ref{fig:PPS} shows the distribution of each visual class in the PPS diagram, using absolute values of $v_{pec}/\sigma_{200}$, as defined in Section~\ref{subsec:SAMI-GS}. The grey points, magenta squares, and blue diamonds represent the unperturbed, truncated, and asymmetric galaxies, respectively. The same coloured contours are generated by \texttt{ks} library \citep{Duong2007} in \texttt{R}, and show the $\rm 10^{th}, \text{ and } 90^{th}$ percentiles of each distribution, respectively. The upper and right panels display 1D KDEs of the sample distributions for each axis, with the solid lines indicating the bi-weighted means per axis and sample, respectively. Furthermore, Table~\ref{tab:PPS properties} presents the means and standard deviation of individual distributions of normalised cluster-centric distances and absolute LOS velocities per visual class, determined using a bi-weight estimator \citep{Beers1990}. The uncertainties are estimated through bootstrapping as follows. We generate 100 mock distributions (i.e., $\rm N_{iteration} = 100$) by randomly resampling the original parameter distribution, and compute the mean and standard deviation in each iteration using the same procedure as for the real data. Uncertainties are then estimated using the $\rm 16^{th}$ and $\rm 84^{th}$ percentiles of the bootstrapped parameter distributions.

\begin{table}
    \centering
    \resizebox{0.85\textwidth}{!}{
    \begin{NiceTabular}{|c|c|c|c|c|}[corners=NW] %c|c|
        \Hline
        \rowcolor{gray!0}
        & \Block{1-2}{$\rm R/R_{200}$} & & \Block{1-2}{$|v_{pec}/\sigma_{200}|$} &  \\%\Block{1-2}{$|\mathrm{v}/\sigma|$} & \\
        \Hline
        & \Block{1-1}{\textbf{BW Mean}} & \Block{1-1}{\textbf{BW STDEV}} & \Block{1-1}{\textbf{BW Mean}} & \Block{1-1}{\textbf{BW STDEV}} \\%& \Block{1-1}{$\mu$} & \Block{1-1}{$\sigma$} \\
        \Hline
        \rowcolors{gray!0}{}
        \textbf{\textcolor{darkgray}{Unper}} & $\bf 0.61_{-0.05}^{+0.04}$ & $\bf 0.29_{-0.03}^{+0.03}$ & $\bf 0.76_{-0.06}^{+0.08}$ & $\bf 0.53_{-0.09}^{+0.11}$ \\%& $0.79_{-0.08}^{+0.11}$ & $0.62_{-0.07}^{+0.07}$\\
        \Hline
        \textbf{\textcolor{magenta}{Trun}} & $0.54_{-0.05}^{+0.05}$ & $0.26_{-0.02}^{+0.02}$ & $0.75_{-0.19}^{+0.10}$ & $0.57_{-0.05}^{+0.05}$ \\%& $0.75_{-0.15}^{+0.15}$ & $0.57_{-0.03}^{+0.03}$ \\
        \Hline
        \textbf{\textcolor{blue}{Asym}} & $0.28_{-0.06}^{+0.04}$ & $0.17_{-0.03}^{+0.04}$ & $1.15_{-0.22}^{+0.18}$ & $0.71_{-0.07}^{+0.09}$ \\%& $1.15_{+-.28}^{+0.21}$ & $0.71_{+-.13}^{+0.10}$ \\
        \Hline
    \end{NiceTabular}
    }
    \caption{The bi-weighted (BW) means and standard deviations \citep{Beers1990} of normalised cluster-centric distances and LOS velocities in absolute for each sample. Associated uncertainties are estimated through bootstrapping.}
    \label{tab:PPS properties}
\end{table}
\vspace{2mm}
\noindent Figure~\ref{fig:PPS} and Table~\ref{tab:PPS properties} both show very clearly that the PPS distribution of asymmetric galaxies is strikingly different compared to those of truncated and unperturbed galaxies. The asymmetric galaxies are concentrated at lower cluster-centric distances (\rrtwo \lsim 0.5) with the smallest standard deviation. In comparison, the truncated and unperturbed galaxies appear to be located at higher cluster-centric distances, yet exhibit almost uniform distributions across the plane. The 1D two-sample AD test returns $\rm p \ll 0.05$ for the pairs with the asymmetric sample, and $\rm p = 0.154$ for the unperturbed-truncated pair, confirming the difference in the radial distribution, quantitatively. 
\\
\\
Considering the distribution of $|v_{pec}/\sigma_{200}|$, both the truncated and unperturbed samples exhibit lower dispersions, as indicated by \texttt{BW STDEV} in Table~\ref{tab:PPS properties}, with respect to their asymmetric counterparts. The AD test returns p-values between 0.3-0.4 for $|v_{pec}/\sigma_{200}|$ comparison between pairs including unperturbed galaxies, whereas truncated - asymmetric pair yields a p-value of 0.045, indicating that they present a statistically significant difference in $|v_{pec}/\sigma_{200}|$ distribution. When non-absolute LOS velocities are compared (i.e., $v_{pec}/\sigma_{200}$), the same test returns p-values higher than 0.05 for all pair combinations, with a marginal significance of $\rm p = 0.064$ for the asymmetric-unperturbed pair.
\\
\\
Lastly, to test the significance of the difference in PPS distributions across the samples, using the absolute LOS velocities (i.e. $|v_{pec}/\sigma_{200}|$), we employ the non-parametric multivariate two-sample KDE test, which has been adopted by several studies \citep{Lopes2017, deCarvalho2017, Owers2019, Costa2024, Cakir2025}. As shown in the upper right of Figure~\ref{fig:PPS}, the KDE test yields p = 0.033 only for the asymmetric-unperturbed pair. When considering the non-absolute LOS velocities (i.e. $v_{pec}/\sigma_{200}$), p-values between 0.2 and 0.5 are returned. This leads us to the conclusion that the PPS distribution of the asymmetric sample is different to the unperturbed and truncated counterparts, although with marginal statistical significance.

\subsection{Star formation activity}\label{sec:SF activity}
The difference in PPS found in Section~\ref{subsec: PPS Distribution}  indicates that the asymmetric galaxies may be caught at a different accretion phase when compared with the unperturbed and truncated galaxies. The larger velocity and smaller cluster-centric distance, coupled with the asymmetric ionised gas morphology, further indicate that asymmetric galaxies may be observed at an evolutionary phase closer to peak ram-pressure compared to the truncated and unperturbed galaxies. To understand the impact of RPS on the star-forming properties, in this section, we investigate star formation activity across the asymmetric, truncated, and unperturbed samples. We first investigate the integrated star formation activity in Section~\ref{subsec:global SF activity}. We then investigate the resolved star formation activity through two diagnostics: resolved star formation rate - mass surface density plane, $\Sigma_{\mathrm{SFR}} - \Sigma_\ast$, and the specific star formation rates (sSFRs) as a function of the galactocentric distances ($\rm r/r_{eff}$).

\subsubsection{Integrated star-forming properties}\label{subsec:global SF activity}
\noindent For the purpose of investigating global star-forming properties, we determine integrated SFRs by summing the SFR maps defined in Section~\ref{subsec:SFR maps}, using only SF spaxels, following a method similar to previous studies \citep{Gonzalez-Delgado2016, Medling2018, Sanchez2019, Mun2024}. When summing the per-spaxel SFRs, we do not apply any aperture correction; this means that, as noted by \citet{Medling2018}, the integrated SFRs given here are not true global SFRs. However, given the similar selection of the cluster and GAMA control samples, we expect that relative comparisons of the integrated star-forming properties should not be affected by the lack of aperture correction. We also note that the per-spaxel spectroscopic classification outlined in Section~\ref{subsec:spectral classification maps} differs from that adopted by \citet{Medling2018, Sanchez2019} in terms of diagnostics used and the SNR cut applied for the emission lines. Therefore, SFR estimates for the common galaxies may differ between these studies.

%Before diving into the local star formation activity, we investigate the global activity through the integrated star formation rate - mass, SFR-M, plane. For this purpose, we use the stellar masses from \citet{Bryant2015, Owers2017}, determined using Equation~\ref{eq:stellar mass}. Similar to previous studies \citep{Gonzalez-Delgado2016, Medling2018, Sanchez2019, Mun2024}, the integrated SFR for each galaxy is estimated by summing the SFR maps defined in Section~\ref{subsec:SFR maps}, resulting in 623 galaxies with available total SFRs. For the remaining 271 galaxies without SF spaxels, we set the total SFR to $10^{-5} \ \mathrm{M}_\odot\ \rm yr^{-1}$. When summing the per-spaxel SFRs, we do not apply any aperture correction, similarly to \citealt{Medling2018, Sanchez2019}, acknowledging the caveat raised by \citet{Medling2018} that the integrated SFRs given here are not true global SFRs. We note that the per-spaxel spectroscopic classification outlined in Section~\ref{subsec:spectral classification maps} differs from that adopted by \citet{Medling2018, Sanchez2019} in terms of diagnostics used and the SNR cut applied for the emission lines. Therefore, SFR estimates for the common galaxies may differ between these studies.}}

\begin{figure}[!t]
    \centering
    \includegraphics[width=\textwidth]{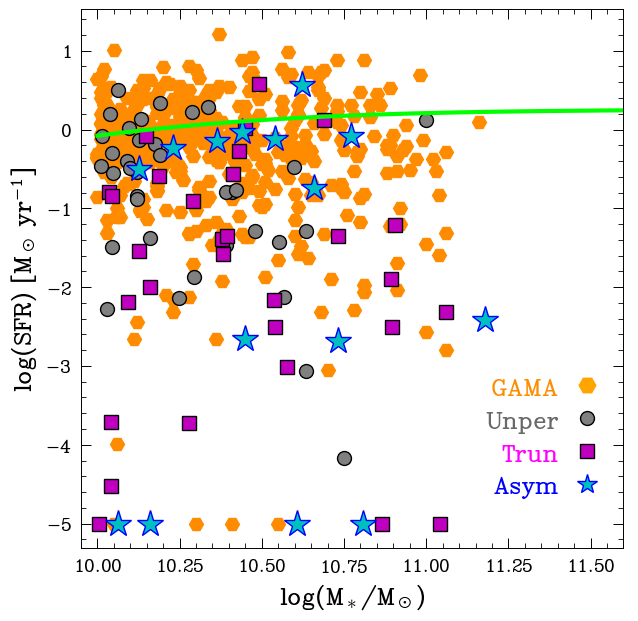}
    \caption{Integrated star formation rate - stellar mass plane. Integrated star formation rates are measured by summing star formation rates of star-forming spaxels within the SAMI aperture, while stellar mass is estimated through ($g-i$) colours, as defined in \citet{Bryant2015, Owers2017}. The grey points, magenta squares, and cyan stars show the visually identified unperturbed, truncated, and asymmetric galaxies. The orange hexagons represent the GAMA sample; the darker the bins, the higher the number of galaxies in the bin. Here, we set log(SFR) values to -5 if no star-forming spaxels are detected. The solid lime curved line is the SFMS best fit defined by \citet{Fraser-McKelvie2021} for the SAMI and MANGA galaxies.}
    \label{fig:integrated SFR-M}
\end{figure}

Figure~\ref{fig:integrated SFR-M} shows the integrated SFR-stellar mass distribution for our sample. The grey points, magenta squares, and cyan stars represent visually identified unperturbed, truncated and asymmetric galaxies, while the orange hexagons are the GAMA control galaxies. A small number of GAMA (4), asymmetric (4), and truncated (3) galaxies have no "star-forming" spaxels, and we set their SFR to $10^{-5} \ \mathrm{M}_\odot\ \rm yr^{-1}$. The solid lime line shows the best fit to the SFMS defined by \citet{Fraser-McKelvie2021} and determined using SAMI and MANGA galaxies. The majority of the GAMA control sample (357/462) and the unperturbed cluster galaxies (19/36) lie within 1$\sigma$ (0.76 dex) of the \citet{Fraser-McKelvie2021} SFMS best fit. On the other hand, both the truncated and asymmetric galaxies are systematically below the SFMS fit of \citet{Fraser-McKelvie2021}, with 10/15 and 26/30 of the asymmetric and truncated galaxies, respectively, found more than 0.5$\sigma$ below the SFMS. These findings indicate that global star formation in the asymmetric and truncated galaxies is suppressed relative to the unperturbed and GAMA control galaxies. 
%The green solid line is the best fit in the form of a curve, whereas the blue dashed line is a linear best fit.}}

%\textcolor{red}{\textbf{The overall galaxy distribution shows two main sequences. Nearly half of the sample (443/894) follows the SFMS, while a clear population without ongoing star formation (271/894) lie at the bottom edge, whose is fixed to -5. When considering visual classes, most unperturbed (23/36) and asymmetric samples (8/15) lie on the SFMS, showing star formation levels comparable to the bulk of the GAMA control sample. In contrast, the majority of the truncated galaxies (20/30) sit below the main sequence along with a few unperturbed and asymmetric galaxies, indicating suppressed star formation.}}

\subsubsection{Resolved Star formation rate - Stellar Mass Relation}\label{subsubsec: SFR-M}%\label{subsec:resolved SF activity}
While we found in Section~\ref{subsec:global SF activity} that the global SFR is suppressed for asymmetric and truncated galaxies, that analysis does not allow us to ascertain the impacts of ram-pressure on the local star-forming properties of galaxies in our sample. To that end, we now use the resolved star-formation---stellar-mass distribution to determine if star-formation is locally suppressed or enhanced, and how the resolved star-forming properties relate to those on a global scale.

%\textcolor{red}{\textbf{The global star formation activity investigated in the previous section suggests that star formation is suppressed in most truncated galaxies compared to their unperturbed and asymmetric counterparts, which yield stars enough to be located on the SFMS. The next step is to analyse the spatial distribution of star formation activity to understand the origin of the global activity shown above.}}

\begin{figure*}[!t]
    \centering
    \includegraphics[width=\textwidth]{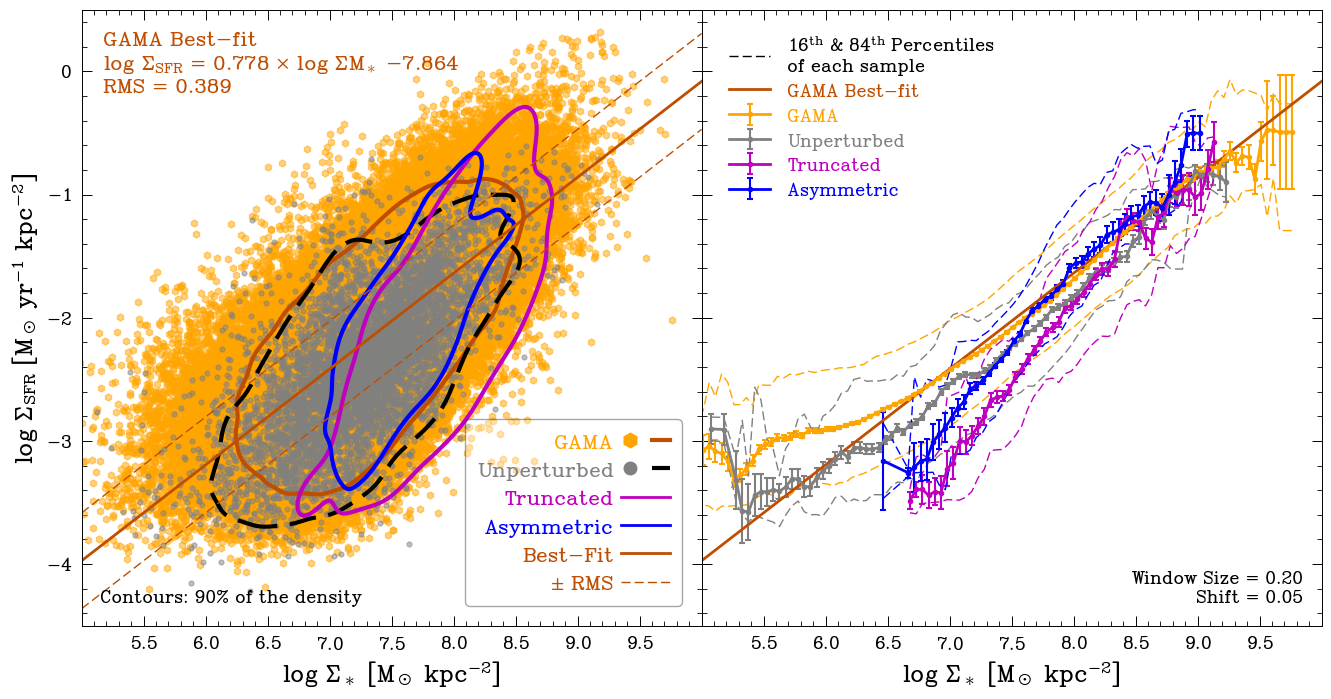}
    \caption{The spatially resolved star formation rate - mass plane ($\Sigma_{\mathrm{SFR}} - \Sigma_{\ast}$) for the whole sample (i.e. cluster + GAMA galaxies) using spaxels classified as "SF" by \citet{Owers2019}. \textbf{Left panel:} The orange hexagons represent the distribution for GAMA spaxels, while the grey points are spaxels from unperturbed galaxies. The dark orange, black, magenta and blue contours enclose 90\% of the 2D density, estimated by \texttt{seaborn} using \textit{Scott's rule of thumb}, for the GAMA, unperturbed, truncated, and asymmetric galaxies, respectively. The dark orange solid line marks the best-fit estimated by \texttt{ltsfit} \citep{Cappellari2013} for the GAMA galaxies, with the dashed lines indicating the root mean square error of the fit. The best-fit parameters are shown in the upper left of the panel, and in Table~\ref{tab:SFRM bestfit params}. \textbf{Right panel:} The solid lines show the running medians for the same distributions shown in the left panel, with the same colour scheme. The medians are defined within 0.2 dex bins and shifted 0.05 dex (i.e. window size and shift) in $\Sigma_{\ast}$. The error bars are the standard errors on the median (i.e. $1.253 \times \sigma_{\mathrm{bin}}/\sqrt{\mathrm{N_{bin}}}$) and the dashed lines represent $\rm 16^{th} \text{ and } 84^{th}$ percentiles of the distributions per bin. The dark orange solid line is the best-fit of the GAMA sample. }
    \label{fig:resolved SFR-M}
\end{figure*}

\vspace{1em}

We defined resolved star formation rate and stellar mass surface density maps using the $\Sigma_{\rm SFR}$ and $\Sigma_{\ast}$ defined in Section~\ref{subsec:SFR maps} and ~\ref{subsec:stellar mass maps}. The left panel of Figure~\ref{fig:resolved SFR-M} shows the $\log\Sigma_{\mathrm{SFR}} - \log \Sigma_\ast$ relation for all "star-forming" spaxels in the GAMA control (orange hexagons) and unperturbed (grey points) samples. The dark orange,  black, magenta, and blue contours mark the loci enclosing 90\% of 2D density, generated using the \texttt{seaborn} library \citep{Seaborn}, for the GAMA control, unperturbed, truncated, and asymmetric samples, respectively. The dark orange solid line shows the best-fit linear fit to the $\log\Sigma_{\mathrm{SFR}} - \log \Sigma_\ast$ relation of  star-forming spaxels for the GAMA control sample, determined using \texttt{ltsfit} \citep{Cappellari2013}. The dashed orange line shows the root-mean-square error on the fit. Table~\ref{tab:SFRM bestfit params} presents the best-fit parameters for the GAMA control sample, as well as the parameters determined from fits to the truncated, asymmetric, and unperturbed samples.

Despite the large scatter, the left panel of Figure~\ref{fig:resolved SFR-M} reveals that $\Sigma_{\rm SFR}$ and $\Sigma_\ast$ show a clear correlation in which spaxels with larger $\Sigma_\ast$ exhibit higher star formation rates. This holds independently of the sample to which the spaxels belong, although the truncated and asymmetric samples appear to host a steeper relation when compared with the GAMA control and unperturbed samples. These differences in slope are confirmed by the parameters from the best-fit relations shown in Table~\ref{tab:SFRM bestfit params}, where the truncated and asymmetric samples have slopes $\simeq 1.4-1.5$, while the GAMA control and unperturbed samples have slopes $\simeq 0.8-0.9$.

The right panel of Figure~\ref{fig:resolved SFR-M} shows the median log$\Sigma_{\rm SFR}$ values, estimated through a sliding window approach (i.e. running median), with a window of full width of 0.2 dex and a step size of 0.05 dex along $\mathrm{log } \ \Sigma_{\ast}$. The uncertainties per bin are defined as the standard error on the median --- $1.253 \times \sigma_{\mathrm{bin}}/\sqrt{\mathrm{N_{bin}}}$. The $\rm 16^{th} \text{ and } 84^{th}$ percentiles of each bin are also shown with the dashed lines for each sample. The dark orange solid and dashed lines show the same best-fit and error as that shown in the left panel for the GAMA sample. For log $\Sigma_\ast>8\ \mathrm{M}_\odot \ \mathrm{kpc}^{-2}$, the median profiles show comparable trends for the GAMA, unperturbed,  asymmetric, and truncated samples. In contrast, for lower masses, log $\Sigma_\ast < 8\ \mathrm{M}_\odot \ \mathrm{kpc}^{-2}$, the asymmetric and truncated galaxies diverge from the unperturbed and GAMA samples, exhibiting lower $\Sigma_{\rm SFR}$ values, which drives the steeper slopes seen in Table~\ref{tab:SFRM bestfit params}. Furthermore, the truncated galaxies have systematically lower log $\Sigma_{\rm SFR}$ values than the asymmetric sample. This is most evident, at log $\Sigma_\ast \lesssim 8.5\ \mathrm{M}_\odot \ \mathrm{kpc}^{-2}$, where the offset is \s0.2 dex.

%\subsubsection{Star formation rate - Mass Plane}\label{subsubsec: SFR-M}
%\textbf{In Section~\ref{subsec:SFR maps} and ~\ref{subsec:stellar mass maps}, we defined star formation rate and stellar mass surface density maps. The procedure outlined in Section~\ref{subsec:stellar mass maps} failed to produce stellar mass maps for one "unperturbed" and one "truncated" galaxy because one of the $grz$ magnitudes being missing. From the remaining sample, we define the resolved star-forming main sequence (rSFMS; $\Sigma_{\mathrm{SFR}} - \Sigma_\ast$) for each subsample, including only spaxels classified as "star-forming", independent of the global classification of their host galaxy. On the other hand, since we consider the total emission from both photoionisation and non-star-forming ionisation, some galaxies do not contribute the rSFMS relation. Having said that, in order to understand the differences between subsample, we fit a linear relation in logarithmic space using \texttt{ltsfit} \citep{Cappellari2013}. The number of galaxies and spaxels used and the best-fitting coefficients per sample are shown in Table~\ref{tab:SFRM bestfit params}.}

\begin{table}
    \centering
    \resizebox{\textwidth}{!}{
    \begin{NiceTabular}{cccccc}[corners=NW] %c|c|
        \Hline
        \Hline
        \rowcolor{gray!0}
        \texttt{Sample} & $\rm N_{GAL}$ & $\rm N_{SF-spx}$ & \texttt{Slope} & \texttt{Intercept} & \texttt{RMS} \\
        \Hline
        \textbf{\textcolor{orange}{GAMA}} & 458 & 158074 & 0.778 $\pm$ 0.002 & -7.864 $\pm$ 0.014 & 0.389 \\
        \textbf{\textcolor{darkgray}{Unper}} & 35 & 6382 & 0.897 $\pm$ 0.009 & -8.901 $\pm$ 0.066 & 0.399 \\
        \textbf{\textcolor{magenta}{Trun}} & 25 & 1253 & 1.452 $\pm$ 0.031 & -13.482 $\pm$ 0.250 & 0.461 \\
        \textbf{\textcolor{blue}{Asym}} & 11 & 1714 & 1.540 $\pm$ 0.028 & -13.797 $\pm$ 0.212 & 0.360 \\
        \Hline
        \Hline
    \end{NiceTabular}
    }
    \caption{Best-fit coefficients (i.e., slope and intercept) of the spatially resolved $\Sigma_{\mathrm{SFR}}-\Sigma_\ast$ relation for each sample. $\rm N_{GAL}$ and $\rm N_{SF-spx}$ denote the number of galaxies and star-forming spaxels used to define the best-fit. RMS is the root mean square of residuals between best-fit and individual points.}
    \label{tab:SFRM bestfit params}
\end{table}

\subsubsection{Radial Trends in Specific Star Formation Rate (sSFR)}\label{subsubsec: radial sSFR}

\begin{figure*}[!t]
    \centering
    \includegraphics[width=\textwidth]{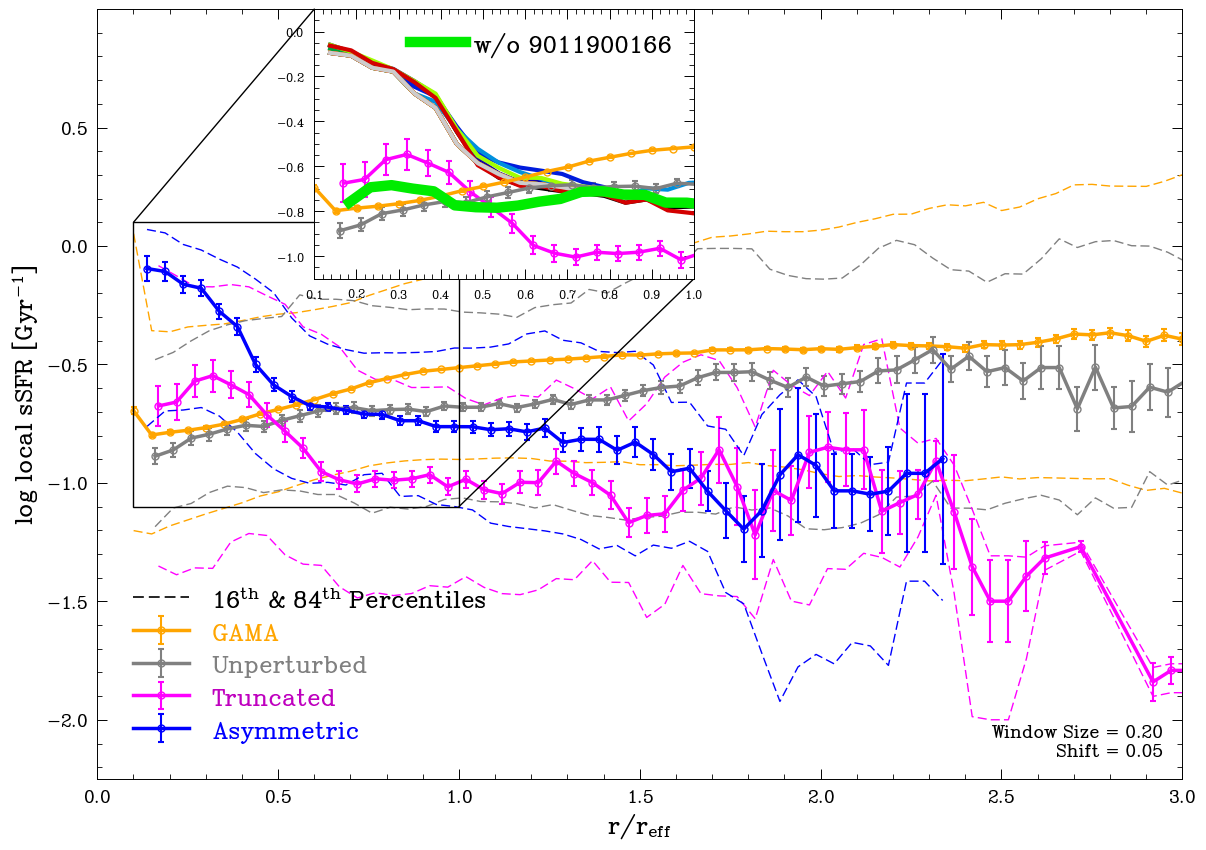}
    \caption{Radial profile of specific star formation rates for each sample. The medians are determined with the step size of 0.2 and shifted by steps of 0.05 in $\rm r/r_{eff}$. The colours are the same as Figure~\ref{fig:resolved SFR-M}. The error bars and the dashed lines show the standard errors on medians, and the $\rm 16^{th} \text{ and } 84^{th}$ percentiles of the distribution per bin. The inset zooms in on the central regions (i.e. $\rm r<r_{eff}$). We overlay the radial sSFR profiles of the asymmetric galaxies derived by a jackknife approach on the main samples. Here, each colour from the \texttt{nipy\_spectral} colour map (between black and light grey) represents the galaxy excluded for that iteration (i.e. the profile without that galaxy). While nearly all profiles show similarities with the main asymmetric profile, only the lime profile (without \texttt{9011900166}) exhibits much lower sSFRs, comparable with the other samples.}
    \label{fig:radial sSFR profiles}
\end{figure*}

In Section~\ref{subsubsec: SFR-M}, we showed that there are differences between the resolved SFMS of unperturbed and ram-pressure-affected galaxies. As the next step, to rule out the mass dependency on the star-formation activity, we investigate the radial sSFR (i.e. SFR per unit mass) profiles. For this purpose, we first measure the distance of each spaxel with respect to the galaxy centre. The ellipticities and position angles from the input catalogue \citep{Bryant2015, Owers2017} are used to account for projection effects. We then normalise the distances by the effective radius, $\rm r_{eff}$, of each galaxy to enable like-for-like comparisons.
\\
\\
Figure~\ref{fig:radial sSFR profiles} shows the radial sSFR profile of each sample. The orange, grey, magenta, and blue solid lines represent the medians for the GAMA, unperturbed, truncated, and asymmetric samples. Similar to that in Figure~\ref{fig:resolved SFR-M}, a sliding window approach is used to determine the medians of log sSFR binned in normalised galactocentric distance, their associated uncertainties (error bars), and the $\rm 16^{th} \text{ and } 84^{th}$ percentiles (dashed lines). The inset plot zooms in on the central $\rm r_{eff}$.  The unperturbed galaxies show a comparable median radial profile to the GAMA control sample, with a small offset ($\lesssim0.2$ dex). Both the unperturbed and GAMA samples show a slight decline in log sSFR, \s 0.25 dex, towards their centres. On the other hand, the asymmetric and truncated galaxies show the opposite trend, with a declining sSFR profile as the radial distance increases.
% \vspace{1mm}
\\
\\
\noindent Notably, the asymmetric galaxies show somewhat enhanced central star formation activity compared to the other samples. For $\rm r/r_{eff}<0.5$, the difference in log sSFR between the asymmetric sample and the rest reaches $\sim 0.7$ dex. To assess whether the enhancement seen in the asymmetric galaxies is a population-wide trend or driven by a small number of galaxies, we use jackknife resampling to determine the median sSFR profiles for a subsample of asymmetric galaxies. In each iteration, we exclude a galaxy and then estimate the median. As shown with the lime profile in the inset plot, when \texttt{9011900166} is omitted, the level of central sSFRs becomes comparable with the remaining samples. Therefore, this enhancement is primarily due to this galaxy, instead of a representative trend for the asymmetric sample. Still, even after removing \texttt{9011900166}, asymmetric galaxies maintain higher sSFRs in the outskirts (i.e. $\rm r/r_{eff}>0.5$) than their truncated counterparts, and this trend seems to disappear for $\rm r/r_{eff} > 1.5$, where data gets noisier for both samples. 
\\
\\
In light of these findings, we can draw a conclusion that star formation in ram-pressure-affected galaxies is suppressed in the outskirts with respect to unperturbed counterparts (as well as the control sample), while central sSFRs remain comparable.

\section{Discussion}\label{sec: Discussion}
This study aims to investigate how cluster environments influence their constituent galaxies through gas stripping by ram pressure, which requires identifying galaxies with such indicators. For this purpose, in Section~\ref{subsec:Visual Classification}, we introduced a visual classification scheme based on the ionised gas morphologies of cluster galaxies, defining subsamples such as "\textcolor{darkgray}{Unperturbed}", "\textcolor{blue}{Asymmetric}", "\textcolor{magenta}{Truncated}". We also quantitatively explored these visual classes on the morphological parameter spaces --- concentration, asymmetry and offset between galaxy and ionised gas centres. We investigated the projected phase space distribution across these subsamples in Section~\ref{subsec: PPS Distribution}. Asymmetric galaxies are found to occupy a narrow region, close to the cluster centre and have a large velocity dispersion, while both truncated and unperturbed galaxies are widely distributed across PPS, being commonly found at larger cluster-centric distances. Section~\ref{sec:SF activity} presented an analysis of both the global and resolved star formation activity of the visual classes. RPS candidates (i.e., truncated and asymmetric galaxies) exhibit both globally suppressed star formation as well as locally suppressed star formation activity in the outskirts compared to their unperturbed counterparts. This might be attributed to an evolutionary sequence tightly linked to the RPS stage. In this section, we discuss the differences in the identification methods and compare our interpretation with the previous literature. 

\subsection{Identifying the candidates}\label{subsec:Discussion - Identification}

\subsubsection{Comparison in visual identification}\label{subsubsec:Discussion - Identification-Visual}
In order to highlight the differences in the selection of RPS candidates, we need a sample of galaxies that are common across different studies. We achieve this by cross-matching our initial cluster sample (before excluding aperture and unclear classes; N=125) with the sample of jellyfish candidates identified in \citet{Poggianti2016}, based on optical imaging, due to a partial overlap in cluster samples. This sample of jellyfish candidates has also been observed partially in other wavelengths, such as in the ultraviolet \citep{George2024, George2025} and radio continuum \citep{Roberts2021a}. More recently, the GASP team also released the full sample of ram-pressure-stripped galaxies observed with MUSE \citep{Poggianti2025}, which will enable us to compare the ionised gas component.
\\
\\
The cross-match reveals seven galaxies in common with \citet{Poggianti2016}, whose IDs are given in the Table~\ref{X-match with GASP}. Our classification results in "truncated" for 2 galaxies (i.e. JO10 and JO182), and "aperture" for the rest (i.e. JO7, JO46, JO49, JO200, and JO201). If we consider the JClass parameter from \citet{Poggianti2016}, JO49, JO200, and JO201 have $\rm JClass \geq 3$, indicating moderate to extreme stripping features. Moreover, strong features such as tails have also been detected in multi-wavelength observations, specifically in the UV for JO200 and JO201 \citep{George2024, George2025}, and in the radio continuum for JO46 and JO49 \citep{Roberts2021a}. The full GASP sample provides ionised gas morphologies (i.e., \ha) for only four galaxies \citep{Poggianti2025} --- JO10, JO49, JO200, and JO201. Their classification of ionised gas ---referred to as JType (see \citealt{Poggianti2025})--- indicates that JO49, JO200, and JO201 also show strong to extreme stripping (i.e., $\rm JType = 1{-}2$), while JO10 is classified as truncated (i.e. $\rm JType = 3$).

\begin{table*}[!t]
    \centering
    \resizebox{0.75\textwidth}{!}{
    \renewcommand{\arraystretch}{.8}
    \begin{tabular}{cccccccc}
        \hline
        \hline
        $\rm ID_{SAMI}$ & Our class & $\rm ID_{JO}$ & \texttt{JClass} & \texttt{JType} & UV Tail & Radio Tail \\
        \hline
        9011900367 & \textbf{Aperture}  & JO7 & 2 &  &  & \\
        9011900441 & \textcolor{magenta}{Truncated}  & JO10 & 1 & 3.0 &  & \\
        9016800167 & \textbf{Aperture}  & JO46 & 2 &  &  & \checkmark \\
        9016800093 & \textbf{Aperture}  & JO49 & 3 & 1.0 &  & \checkmark \\
        9388000046 & \textcolor{magenta}{Truncated}  & JO182 & 2 &  &  & \\
        9008500268 & \textbf{Aperture}  & JO200 & 3 & 1.0 & \checkmark & \\
        9008500074 & \textbf{Aperture}  & JO201 & 5 & 2.0 & \checkmark & \\
        \hline
        \hline
    \end{tabular}}
    \caption{The comparison between galaxies common between this study and \citet{Poggianti2016} and \citet{Poggianti2025}. $\rm ID_{SAMI}$ in the first columns is the SAMI ID of the galaxy; the second column shows the visual classes that assigned in this study; the third column presents the IDs from the jellysfish candidate catalogue of \citet{Poggianti2016}; jellyfish class (\texttt{JClass}) and jellyfish type (\texttt{JType}) are the visual classifications based on the wide field B-band imaging and \ha flux distribution, defined by \citet{Poggianti2016, Poggianti2025}. UV and radio information are adopted from \citet{George2024, George2025} and \citet{Roberts2021a}.}
    \label{X-match with GASP}
\end{table*}

\noindent The comparison above shows that only JO10 yields a consistent result. For galaxies exhibiting strong jellyfish features, particularly in UV and optical imaging (including \ha, as a tracer of young stellar populations), we could not detect any such features. As shown in Table~\ref{X-match with GASP}, the primary reason behind this is that JO49, JO200, and JO201 are affected by the SAMI aperture. Considering the SAMI field of view (FoV), which is 7.5 arcseconds in radius, JO49, JO200, and JO201 have r-band effective radii of \s7, 10, and 9 arcseconds \citep{Owers2019}, respectively, showing that SAMI bundles cover only the centres of these galaxies. Compared to the SAMI's, the MUSE's FoV in wide-field mode is 20 times larger ($1^\prime\times 1^\prime$). Additionally, wide-field UV and optical imaging and radio continuum surveys tend to cover the whole galaxy.
\\
\\
When the comparison is reversed, no overlap is found between our asymmetric sample and those from previous studies. This is not primarily driven by such an aperture effect as discussed above. Instead, it is likely because asymmetric galaxies identified in this study do not exhibit the pronounced stripping signatures, such as star-forming tails, typical of extreme jellyfish examples. Indeed, a similar discussion has also been conducted in \citet{Poggianti2025}. In their cluster control sample, three out of four star-forming galaxies, which are selected to be undisturbed in optical broadband imaging, actually show some degree of stripping within the \ha distribution. Moreover, RPS may not always trigger star formation \citep{Fossati2016, Pedrini2022}. In fact, the tail-like features identified within our asymmetric sample are not star-forming, supporting other mechanisms such as shocks and turbulence, rather than photoionisation. In an upcoming follow-up paper, a subsample of asymmetric galaxies identified here will be investigated in greater detail (Quattropani et al., in prep). Overall, in the case of galaxies at the early phase of stripping or with a non-star-forming tail, optical imaging might fail in identifying these objects.

\subsubsection{Morphological parameters}\label{subsubsec:Discussion - Identification-Quantitative}
In Section~\ref{subsec:quantitative analysis}, we quantitatively analysed the ionised gas morphologies. We found that the morphological parameters --- concentration, shape asymmetry, and spatial offset between galaxy and actual (i.e. not flux-weighted) emission centre--- are able to separate the majority of visual classes. While the parametrisation used in this study is not identical to all of the other studies, comparable findings have been reported in earlier works. 
\\
\\
\citet{Roberts2021a} identified 95 "jellyfish" galaxies through visual classification of LoTSS radio continuum (\s144 MHz) images. They similarly measured the shape asymmetry parameter \citep[$\rm A_{shape}$;][]{Pawlik2016}  for the full sample, finding that the majority of jellyfish galaxies (\s85\%) have $\rm A_{shape} > 0.3$, whereas this threshold excludes \s70\% of the general LoTSS cluster population. This is consistent with our findings, where the majority of asymmetric galaxies (\s80\%) identified in this study have $\rm A_{shape} > 0.2$, with relatively low contamination from other subsamples (i.e., $\lesssim 35\%$). Based on optical morphology analyses, RPS candidates are found to be outliers in the standard diagnostic parameter spaces such as concentration-asymmetry, Gini-$\rm M_{20}$ \citep{McPartland2016, RP2020, Krabbe2024}. Furthermore, \citet{Krabbe2024} reported a positive correlation between JClasses and morphological parameters for $\rm JClass\ge3$ --- especially, stronger in asymmetry. Although we cannot directly compare our results with theirs, except for \citet{Roberts2021a}, they qualitatively support our findings that asymmetric galaxies occupy a distinct region in our diagnostic spaces.
\\
\\
For the SAMI cluster sample, \citet{Owers2019} measured \ha flux concentration using the recipe in \citet{Schaefer2017}. They only considered spaxels whose emission is classified as intermediate (INT), star-forming (SF), or weak star-forming (wSF). They found that galaxies with strong \hd absorption signatures (i.e., HDSGs) have much lower concentration (i.e., C$_{H\alpha, \rm cont} <1$) compared to cluster SFGs, indicating central star formation. As we refer back to Table~\ref{tab:Breakdown of visual classes}, the same sample of HDSGs is classified as either truncated (N=2) or asymmetric (N=7) --- ram-pressure affected candidates. Different from their approach, we combine \ha and \nii from both SF and nSF spaxels. Still, as shown in Section~\ref{subsec:quantitative analysis}, we also see a consistent trend in truncated and asymmetric galaxies, returning smaller concentration values.

\subsection{Evolutionary context}\label{subsubsec:Discussion - Evolution}
One of the main findings of this study emerges from the projected phase-space analysis presented in Section~\ref{subsec: PPS Distribution}. We found that asymmetric galaxies cumulate in a narrow cluster-centric distance band (i.e. $\rm 0.1\lesssim R/R_{200} \lesssim0.6$), while exhibiting a larger velocity dispersion (i.e. $\sigma(|v_{pec}|)_\mathrm{Asym} = 1.15^{+0.18}_{-0.22}\  \sigma_{200}$) compared to the truncated and unperturbed samples. These findings are in line with conclusions drawn from earlier studies of RPS candidates \citep{Smith2010, Yoon2017,  Jaffe2018, Jung2018, Owers2019, RP2020, Roberts2021a, Roberts2021b, Bellhouse2022, Salinas2024, Poggianti2025}, in which galaxies severely affected by RPS are mostly found near the cluster centre, and have higher relative velocities. 
\\
\\
In addition to observational evidence, simulations are useful in segregating populations based on accretion times, which can be easily tracked in both real three-dimensional and projected phase spaces. Indeed, \citet{Owers2019} examined the PPS distribution of HDSGs ---which make up the majority of the asymmetric sample--- comparing the simulated orbital libraries of cluster galaxies from \citet{Oman2013} and \citet{Oman2016}. Their analysis revealed that HDSGs present a similar PPS with a population of infalling galaxies that have crossed $\rm 0.5R_{200,3D}$ within the last 1 Gyr. If we expand this to include the full asymmetric sample, we can draw the same conclusion about them being consistent with a recent infalling population, also in line with the findings of \citet{Rhee2017}. Moreover, given that the majority of the asymmetric galaxies are HDSGs, this also provides support for the RPS scenario outlined in \citet{Owers2019} being the primary mechanism responsible for the quenching of those galaxies. On the other hand, the broad PPS distributions of truncated and unperturbed galaxies make it more challenging to estimate their accretion timescales compared to asymmetric galaxies. Nonetheless, the contours shown in the lower left panels of Figure 16 in \citet{Owers2019} for the crossing time between 0.8 and 1.2 Gyr, as well as the distribution of intermediate infallers defined in \citet[see their Figures 6,7,8]{Rhee2017}, qualitatively coincide with the PPS distribution of the truncated galaxies, suggesting that they are probably in the later stage of evolution.
\\
\\
That said, the analyses of the resolved star-forming properties in Section~\ref{subsubsec: SFR-M} provide additional evidence to support our interpretation of there being an evolutionary link with RPS activity. We found that resolved star-forming main sequences ($\Sigma_{\mathrm{SFR}}$ vs $\Sigma_\ast$) introduced different trends between the RPS candidates and their unperturbed counterparts. The slope of rSFMS becomes steeper for the RPS candidates (i.e. \s1.5) compared to the rest (i.e., \s0.8). For the high-mass end (i.e. $\mathrm{log}\ \Sigma_\ast \gtrsim 8\ \mathrm{M}_\odot \ \mathrm{kpc}^{-2}$), the subsamples yield comparable $\Sigma_{\mathrm{SFR}}$ values, while RPS candidates have lower star formation activity for the $\mathrm{log}\ \Sigma_\ast \lesssim 8\ \mathrm{M}_\odot \ \mathrm{kpc}^{-2}$ region, possibly indicating suppression, which is stronger for the truncated sample. Given that the slopes of both the GAMA (field) and unperturbed galaxies are similar and also consistent with the range reported by earlier studies \citep[0.6-1;][]{Cano-Diaz2016, Abdurro'uf2017, Hsieh2017, Mun2024}, the suppression signal can, therefore, be considered robust.

These findings, however, contradict those presented in \citet{Vulcani2020}. Using a sample of 40 local cluster galaxies undergoing ram-pressure stripping selected from the GASP survey, \citet{Vulcani2020} investigated the rSFMS and compared their results with a control sample of 30 galaxies drawn from both cluster and field environments. The rSFMS of the stripping sample shows systematically enhanced star formation activity at any given $\Sigma_\ast$, independent of the degree of stripping. Their truncated galaxy sample exhibits a very narrow rSFMS and does not significantly differ from the control sample within the uncertainties. The best-fit slopes of both the stripping and control samples are nearly identical (\s1.6) and closely overlap with those of our RPS candidates. This suggests that the GASP control sample used in \citet{Vulcani2020} may not be representative of a truly unperturbed population. As discussed in Section~\ref{subsubsec:Discussion - Identification-Visual}, \citet{Poggianti2025} reported that three out of four optically unperturbed, star-forming cluster galaxies show evidence of stripping in their ionised gas distributions. When expanded to include the full GASP cluster control sample \citep[N = 17; see Table 2 of][]{Vulcani2019}, the majority (13/17) exhibit mild to strong evidence of stripping, with JTypes in the range 0.5–1 \citep[from Tables 2 and 3 of][]{Poggianti2025}. 

On the other hand, our results are in line with \citet{Zhu2024}. Through a suite of galaxy-scale wind tunnel simulations, \citet{Zhu2024} investigated the impact of ram-pressure stripping on a low-mass disk galaxy ($\mathrm{M}_\ast = 10^{9.7} \mathrm{\ M}_\odot$). Their suite consists of three different wind (W) tunnel runs (12W, 13W, and 14W) for different halo masses ---Milky Way-like ($\mathrm{M}_\ast = 10^{12} \mathrm{\ M}_\odot$), group-like ($\mathrm{M}_\ast = 10^{13} \mathrm{\ M}_\odot$), and cluster-like ($\mathrm{M}_\ast = 10^{14} \mathrm{\ M}_\odot$)--- and an isolation run (iso case). Compared to the isolated case, 13W and 14W runs yield a much steeper relation between $\Sigma_{\mathrm{SFR}}$ and $\Sigma_\ast$, due to truncation in the low-mass elements of star-forming disk ($\Sigma_\ast \lesssim 10^{6.5} \text{ and } 10^{7} \ \mathrm{M}_\odot \ \mathrm{kpc}^{-2}$ for 13W, and 14W, respectively), and the enhanced star formation activity for the high-mass end ($\Sigma_\ast \gtrsim 10^{7.1} \text{ and } 10^{7.6} \ \mathrm{M}_\odot \ \mathrm{kpc}^{-2}$ for 13W, and 14W, respectively). The truncation reported at the low-mass end clearly supports our observational trend.
\\
\\
As rSFMSs are translated into radial star formation activity profiles given in Section~\ref{subsubsec: radial sSFR}, the difference between various samples becomes more evident. The median sSFR profile of unperturbed galaxies presents a slight increase towards the galaxy outskirts, also found for GAMA field galaxies. This aligns with the results presented in \citet{Abdurro'uf2017}. They studied the spatially resolved properties of 93 local ($\rm 0.01< z < 0.02$) massive ($\mathrm{log\ M_\ast/M_\odot}>10.5$) spiral galaxies, by pixel-to-pixel spectral energy distribution (SED) fitting method using the Galaxy Evolution Explorer (GALEX) far- and near- ultraviolet (FUV and NUV), and Sloan Digital Sky Survey (SDSS) $ugriz$ bands. The average radial sSFR profile of their sample yields a similar increasing trend in sSFR with increasing distance, which can be explained by the presence of a central bulge or a bar, or active galactic nuclei \citep[aka "inside-out" quenching; ][]{Gavazzi2015, Gonzalez-Delgado2016, Abdurro'uf2017, Ellison2018, Medling2018, Lin2019, Bluck2020b}.  

In contrast to unperturbed and GAMA galaxies, RPS candidates exhibit radially decreasing sSFR profiles, with significantly stronger suppression in the outskirts. This suppression is most pronounced in truncated galaxies, consistent with an outside-in quenching scenario. In the central regions (i.e., $\rm r/r_{eff} < 0.5$), star formation activity remains comparable to that of unperturbed galaxies, with RPS candidates showing slightly enhanced central sSFRs. However, the caveat here is that, as shown in Figure~\ref{fig:radial sSFR profiles}, the large difference in the central sSFRs that results from omitting \texttt{9011900166}, this central enhancement for the RPS sample is likely driven by the small sample size. Yet, similar trends were reported by \citet{Koopmann2004b} for Virgo spirals, where the majority (27 out of 52) were found to have truncated H$\alpha$ disks. Furthermore, these truncated galaxies, along with spirals displaying H$\alpha$ asymmetries (e.g., edge-enhanced or extra-planar emission), showed a range of star formation activity—typically normal to mildly enhanced in their central regions.
% We already mentioned in Section~\ref{subsubsec: radial sSFR} that the central elevation observed for the asymmetric sample is driven by 9011900166.
\\
\\
By combining the results from the projected phase-space analysis and the resolved star-forming properties, we infer that visually identified truncated galaxies are at a more advanced evolutionary stage than their asymmetric counterparts, corresponding to a post-RPS phase, whereas unperturbed galaxies have not yet been affected by ram-pressure stripping. Supporting evidence was also found by Pak et al. (in prep), where galaxies with spatially truncated star-forming regions \textit{(or central SF)} found near the cluster centre appear to have quenched more recently (i.e., undergone RPS later) than those at higher cluster-centric distances. Moreover, since more than half of the truncated galaxies remain star-forming, this suggests that they have survived during the first pericentric passage. This finding contrasts with the results of \citet{Oman2016} and \citet{Lotz2019}, who report that core passage can fully quench star formation in satellite galaxies for the first infall. However, \citet{Lotz2019} also highlight a mass dependence in the rate of survivors, in which galaxies with $\mathrm{M}_\ast > 1.5\times 10^{10} \mathrm{M}_\odot$ are more likely to retain star formation through their first pericentric passage. Given the stellar mass range probed in this study ($log(M_\ast/M_\odot) \geq 10$), what \citet{Lotz2019} found supports the presence of truncated galaxies with central star-formation in the SAMI cluster sample.

\section{Summary and Conclusion}\label{sec: Conclusion}
Cluster environments regulate star formation activity, notably through gas removal by ram-pressure. To unveil how this occurs, we need a well-defined sample of galaxies exhibiting signatures of ongoing or recent ram pressure stripping. In this study, using spatially resolved data from the SAMI Galaxy Survey, we identified galaxies which may be affected by ram pressure through a visual classification scheme based on their ionised gas morphologies. We selected a subsample of cluster galaxies with reliable emission, and classified them into visual classes---unperturbed, truncated, asymmetric---and studied their properties such as non-parametric morphological parameters, the projected phase-space distribution, and resolved star formation activities. Below, we highlight the key findings presented in this study,

\begin{enumerate}
    \item Quantitative analysis on the ionised gas morphologies shows that diagnostic spaces constructed from the concentration of ionised gas relative to the stellar continuum, the shape asymmetry, and the spatial offset between the galaxy centre and the non-flux-weighted centre of the ionised gas distribution are able to separate galaxies, aligning with their visual classes.
    
    \vspace{1em}
    
    \item The projected phase space distributions present differences between the asymmetric sample and the other visual classes. Asymmetric galaxies are coherently located at small cluster-centric distances (i.e. $\rm 0.1\lesssim R/R_{200} \lesssim0.6$), and have a larger line-of-sight velocity dispersion ($\sigma(|v_{pec}|)_\mathrm{Asym} = 0.71^{+0.09}_{-0.07} \ \sigma_{200}$) relative to both truncated and unperturbed samples. Comparison with \citet{Owers2019} via \hd-strong galaxies (galaxies with recently quenched star formation), which comprise the majority of the asymmetric galaxies, indicates that the asymmetric sample is consistent with an infalling population having crossed $\rm 0.5R_{200,3D}$ in the last 1 Gyr. 
    
    \vspace{1em}
    
    \item The resolved star-forming main sequence (rSFMS), $\Sigma_{\mathrm{SFR}} - \Sigma_\ast$, of the ram-pressure-affected galaxies ---asymmetric and truncated sample--- is much steeper, compared to the unperturbed counterparts and the field sample. For $\mathrm{log} \ \Sigma_\ast\gtrsim8 \ \mathrm{M}_\odot \rm\ kpc^{-2} $, rSFMSs are comparable across the samples, while ram-pressure-affected galaxies yield lower $\Sigma_{\mathrm{SFR}}$ values for the lower mass end (i.e., $\mathrm{log} \ \Sigma_\ast\lesssim8 \ \mathrm{M}_\odot \rm\ kpc^{-2} $), more prominent for truncated galaxies.
    
    \vspace{1em}
    
    \item Radial sSFR profiles exhibit different trends between subsamples. The unperturbed and field galaxies similarly show an increasing star formation activity towards the galaxy outskirts, favouring inside-out quenching. In contrast, star formation activity in RPS candidates is suppressed in the outskirts (i.e., $\rm r>0.5 r_{e}$), being more significant for the truncated galaxies, supporting outside-in quenching. When central (i.e., $\rm r<0.5 r_{e}$) star formation is considered, all samples yield comparable star formation activity, indicating that no enhanced central activity within RPS candidates.
    % , hinting at an elevated activity for RPS candidates, in particular for the asymmetric sample.
\end{enumerate}

\noindent These results suggest an evolutionary sequence among galaxies with different ionised gas morphologies. Unperturbed galaxies likely represent a pre-RPS population, having either only recently entered the cluster environment or entered long ago, yet having circular orbits that prevent interaction between the high-density ICM region. Asymmetric galaxies exhibit characteristics of a population undergoing ram-pressure stripping, while truncated galaxies are a post-RPS population outbounding towards the cluster outskirts after the core-passage, a potential backsplash population. 
\\
\\
While ram-pressure stripping is the primary focus of this study, the potential contribution of other mechanisms, such as preprocessing \citep{Mahajan2012, Haines2015}, cannot be excluded. Addressing this requires extending similar analyses to a larger sample of galaxies out to cluster outskirts. This is where the next-generation Hector Galaxy Survey (\citealt{Bryant2024, Oh2025}; Bryant et al., in prep) comes into play. The Hector Galaxy Survey will provide spatially resolved spectroscopy for a larger cluster sample, including more massive halos ($\mathrm{log \ M_{200}/M}_{\odot} > 14.5$) and more diverse in dynamical activity, out to 2\rtwo, where both the first infallers and the backsplash galaxies can be found \citep{Balogh2000, Gill2005}. This will enable us to study the environment-driven evolution of galaxies by providing a complete picture out to the cluster outskirts.

\begin{acknowledgement}
We thank the anonymous referee for their constructive
feedback that has helped to improve the clarity of the manuscript. 

OÇ and GQ acknowledge the financial support from the Australian Government Research Training Program Scholarship (RTP; \textcolor{largeblue}{https://doi.org/10.82133/C42F-K220}) to conduct this research. This research was partially supported by the Australian Research Council Centre of Excellence for All Sky Astrophysics in 3 Dimensions (ASTRO 3D), through project number CE170100013. This work was supported by the Korea Astronomy and Space Science Institute under the R\&D program (Project No. 2025-1-831-01) supervised by the Ministry of Science and ICT (MSIT). MP acknowledges support from the National Research Foundation of Korea (NRF) grant funded by the Korean government (MSIT) (No. 2022R1A2C1004025). JJB acknowledges support of an Australian Research Council Future Fellowship (FT180100231). AR recognises the support from the Australian Research Council Centre of Excellence in Optical Microcombs for Breakthrough Science (project number CE230100006), funded by the Australian Government. S.K.Y. acknowledges support from the Korean National Research Foundation (RS-2025-00514475; RS-2022-NR070872).  

%%%%%%%%%%%%%%%%%%%%%%%%%%%%%%%%%%%%%%%%%
%%%%%%%%%%%%%% SAMI Survey %%%%%%%%%%%%%%
%%%%%%%%%%%%%%%%%%%%%%%%%%%%%%%%%%%%%%%%%
The SAMI Galaxy Survey is based on observations made at the Anglo-Australian Telescope. The Sydney-AAO Multi-object Integral field spectrograph (SAMI) was developed jointly by the University of Sydney and the Australian Astronomical Observatory. The SAMI input catalogue is based on data taken from the Sloan Digital Sky Survey, the GAMA Survey and the VST ATLAS Survey. The SAMI Galaxy Survey is supported by the Australian Research Council Centre of Excellence for All Sky Astrophysics in 3 Dimensions (ASTRO3D), through project number CE170100013, the Australian Research Council Centre of Excellence for All-sky Astrophysics (CAASTRO), through project number CE110001020, and other participating institutions. The SAMI Galaxy Survey website is \textcolor{largeblue}{http://sami-survey.org/}.

The SAMI Galaxy Survey made extensive use of the Anglo-Australian Telescope (AAT). We acknowledge the traditional custodians of the land on which the AAT stands, the Gamilaraay people, and pay our respects to elders past and present, and all the technical and observing support of the staff at the AAT throughout the SAMI Galaxy Survey.

%%%%%%%%%%%%%%%%%%%%%%%%%%%%%%%%%%%%%%%%%
%%%%%%%%%%%%%% GAMA Survey %%%%%%%%%%%%%%
%%%%%%%%%%%%%%%%%%%%%%%%%%%%%%%%%%%%%%%%%
GAMA is a joint European-Australasian project based around a spectroscopic campaign using the Anglo-Australian Telescope. The GAMA input catalogue is based on data taken from the Sloan Digital Sky Survey and the UKIRT Infrared Deep Sky Survey. Complementary imaging of the GAMA regions is being obtained by a number of independent survey programmes, including GALEX MIS, VST KiDS, VISTA VIKING, WISE, Herschel-ATLAS, GMRT and ASKAP, providing UV to radio coverage. GAMA is funded by the STFC (UK), the ARC (Australia), the AAO, and the participating institutions. The GAMA website is \textcolor{largeblue}{https://www.gama-survey.org/}.

%%%%%%%%%%%%%%%%%%%%%%%%%%%%%%%%%%%%%%%%½%%
%%%%%%%%%%%%%% Legacy Survey %%%%%%%%%%%%%%
%%%%%%%%%%%%%%%%%%%%%%%%%%%%%%%%%%%%%%%%½½%
The Legacy Surveys consist of three individual and complementary projects: the Dark Energy Camera Legacy Survey (DECaLS; Proposal ID \#2014B-0404; PIs: David Schlegel and Arjun Dey), the Beijing-Arizona Sky Survey (BASS; NOAO Prop. ID \#2015A-0801; PIs: Zhou Xu and Xiaohui Fan), and the Mayall z-band Legacy Survey (MzLS; Prop. ID \#2016A-0453; PI: Arjun Dey). DECaLS, BASS and MzLS together include data obtained, respectively, at the Blanco telescope, Cerro Tololo Inter-American Observatory, NSF’s NOIRLab; the Bok telescope, Steward Observatory, University of Arizona; and the Mayall telescope, Kitt Peak National Observatory, NOIRLab. Pipeline processing and analyses of the data were supported by NOIRLab and the Lawrence Berkeley National Laboratory (LBNL). The Legacy Surveys project is honoured to be permitted to conduct astronomical research on Iolkam Du’ag (Kitt Peak), a mountain with particular significance to the Tohono O’odham Nation.

NOIRLab is operated by the Association of Universities for Research in Astronomy (AURA) under a cooperative agreement with the National Science Foundation. LBNL is managed by the Regents of the University of California under contract to the U.S. Department of Energy.

This project used data obtained with the Dark Energy Camera (DECam), which was constructed by the Dark Energy Survey (DES) collaboration. Funding for the DES Projects has been provided by the U.S. Department of Energy, the U.S. National Science Foundation, the Ministry of Science and Education of Spain, the Science and Technology Facilities Council of the United Kingdom, the Higher Education Funding Council for England, the National Center for Supercomputing Applications at the University of Illinois at Urbana-Champaign, the Kavli Institute of Cosmological Physics at the University of Chicago, Center for Cosmology and Astro-Particle Physics at the Ohio State University, the Mitchell Institute for Fundamental Physics and Astronomy at Texas A\&M University, Financiadora de Estudos e Projetos, Fundacao Carlos Chagas Filho de Amparo, Financiadora de Estudos e Projetos, Fundacao Carlos Chagas Filho de Amparo a Pesquisa do Estado do Rio de Janeiro, Conselho Nacional de Desenvolvimento Cientifico e Tecnologico and the Ministerio da Ciencia, Tecnologia e Inovacao, the Deutsche Forschungsgemeinschaft and the Collaborating Institutions in the Dark Energy Survey. The Collaborating Institutions are Argonne National Laboratory, the University of California at Santa Cruz, the University of Cambridge, Centro de Investigaciones Energeticas, Medioambientales y Tecnologicas-Madrid, the University of Chicago, University College London, the DES-Brazil Consortium, the University of Edinburgh, the Eidgenossische Technische Hochschule (ETH) Zurich, Fermi National Accelerator Laboratory, the University of Illinois at Urbana-Champaign, the Institut de Ciencies de l’Espai (IEEC/CSIC), the Institut de Fisica d’Altes Energies, Lawrence Berkeley National Laboratory, the Ludwig Maximilians Universitat Munchen and the associated Excellence Cluster Universe, the University of Michigan, NSF’s NOIRLab, the University of Nottingham, the Ohio State University, the University of Pennsylvania, the University of Portsmouth, SLAC National Accelerator Laboratory, Stanford University, the University of Sussex, and Texas A\&M University.

BASS is a key project of the Telescope Access Program (TAP), which has been funded by the National Astronomical Observatories of China, the Chinese Academy of Sciences (the Strategic Priority Research Program “The Emergence of Cosmological Structures” Grant \# XDB09000000), and the Special Fund for Astronomy from the Ministry of Finance. The BASS is also supported by the External Cooperation Program of the Chinese Academy of Sciences (Grant \# 114A11KYSB20160057), and the Chinese National Natural Science Foundation (Grant \# 12120101003, \# 11433005).

The Legacy Survey team makes use of data products from the Near-Earth Object Wide-field Infrared Survey Explorer (NEOWISE), which is a project of the Jet Propulsion Laboratory/California Institute of Technology. NEOWISE is funded by the National Aeronautics and Space Administration.

The Legacy Surveys imaging of the DESI footprint is supported by the Director, Office of Science, Office of High Energy Physics of the U.S. Department of Energy under Contract No. DE-AC02-05CH1123, by the National Energy Research Scientific Computing Center, a DOE Office of Science User Facility under the same contract; and by the U.S. National Science Foundation, Division of Astronomical Sciences under Contract No. AST-0950945 to NOAO.

This study made extensive use of \texttt{Python} \citep{Python3} libraries --- \textsc{\texttt{Astropy}} \citep{Astropy2013, Astropy2018, Astropy2022}, \textsc{\texttt{Numpy}} \citep{Numpy}, \textsc{\texttt{Scipy}} \citep{Scipy}, \textsc{\texttt{Matplotlib}} \citep{Matplotlib}, \textsc{\texttt{Smplotlib}} \citep{Smplotlib}, \textsc{\texttt{Pandas}} \citep{Pandas}, \textsc{\texttt{Seaborn}} \citep{Seaborn}, and \textsc{\texttt{Jupyter}} \citep{Jupyter}; and \textsc{\texttt{R}} \citep{R}.

\end{acknowledgement}

% PASA uses footnotes, not endnotes. \endnote in this template will behave like \footnote; and \printendnotes will not output anything.
% \printendnotes

\bibliography{Bibliography}

\appendix
\onecolumn
% \counterwithin{figure}{section}
\section{A Mosaic of Visually Identified Asymmetric Galaxy Sample}
\begin{figure*}[!ht]
    \centering
    \begin{tabular}{cc}
        \includegraphics[width=0.5\textwidth]{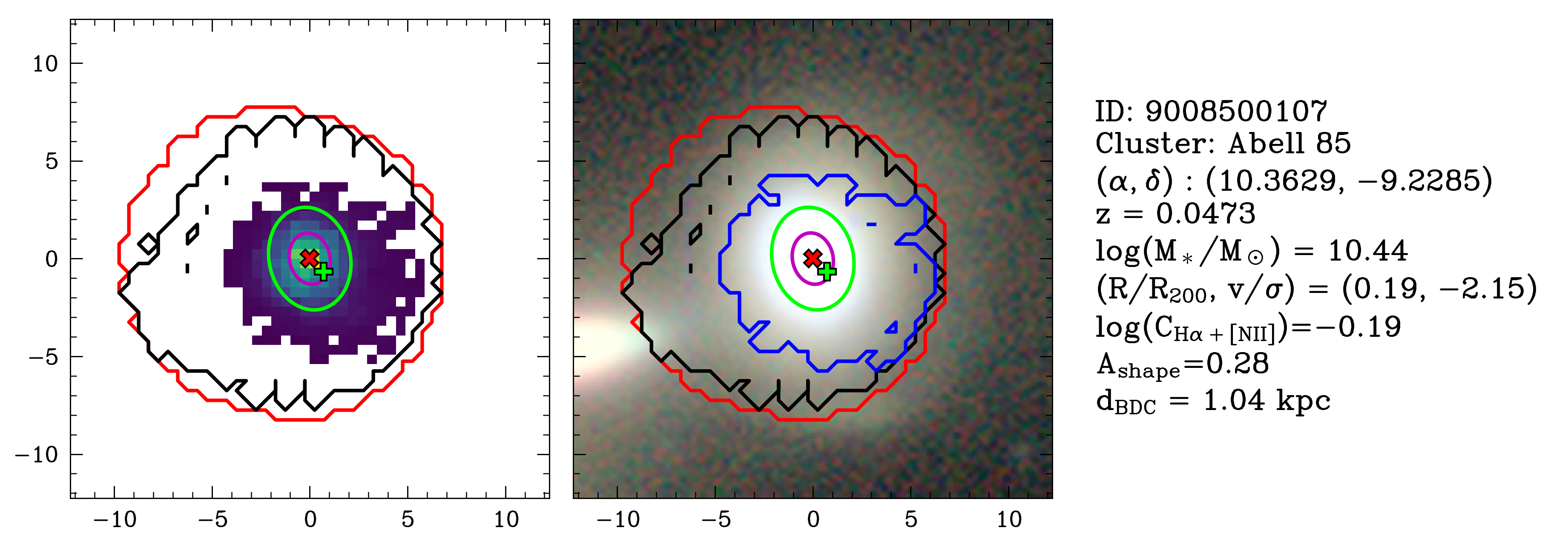} &
        \includegraphics[width=0.5\textwidth]{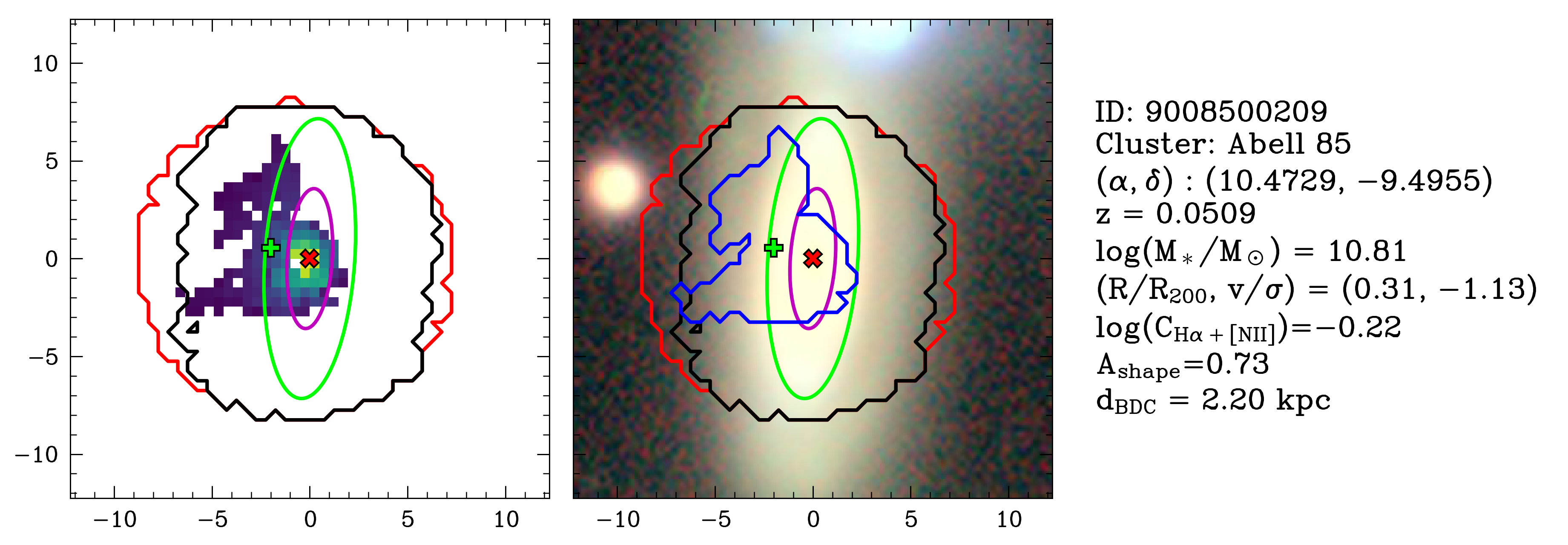} \\
        \includegraphics[width=0.5\textwidth]{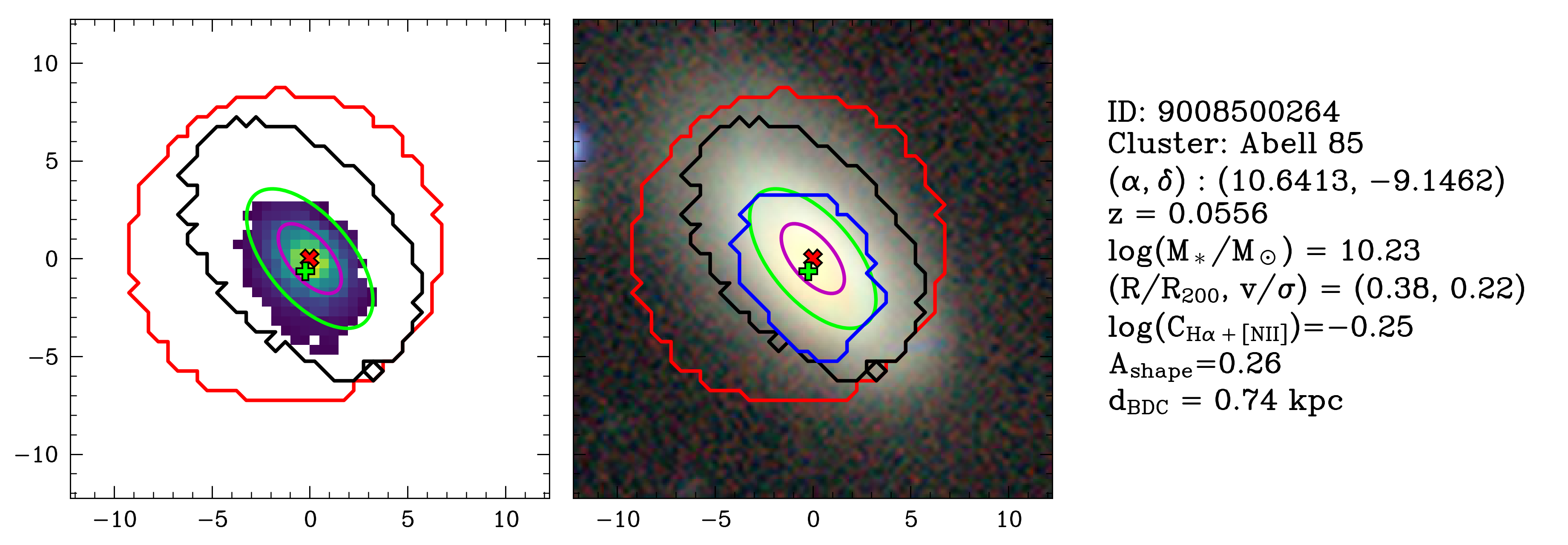} & 
        \includegraphics[width=0.5\textwidth]{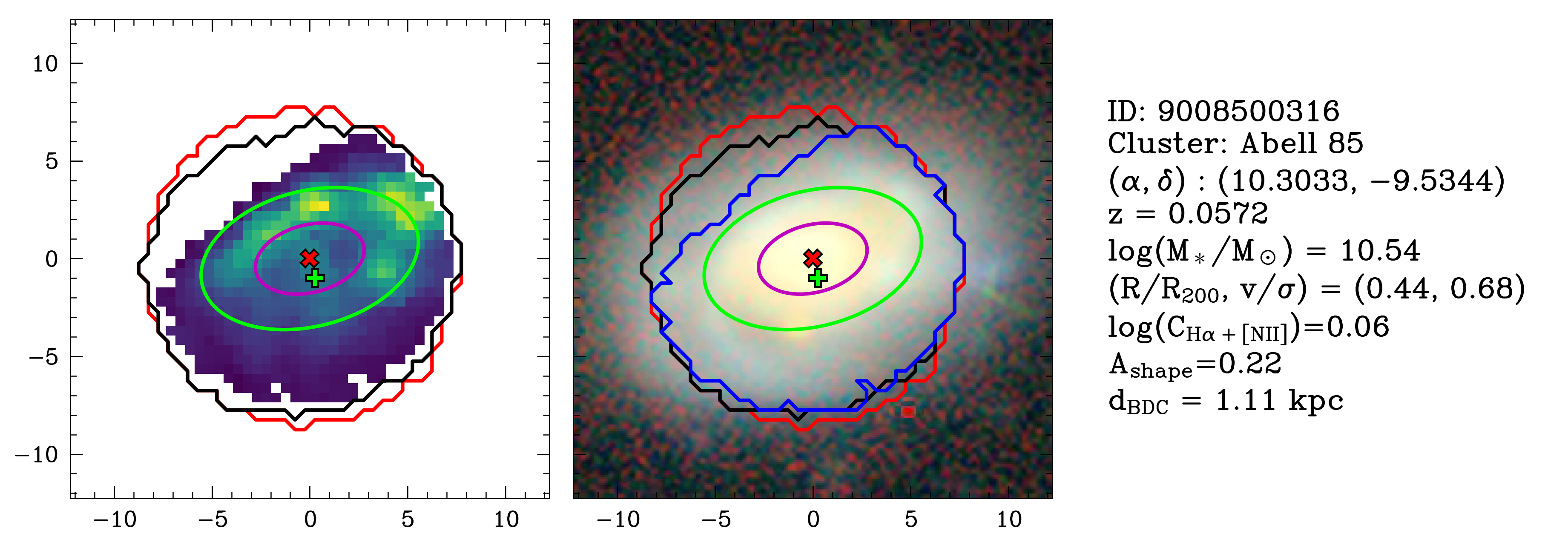} \\
        \includegraphics[width=0.5\textwidth]{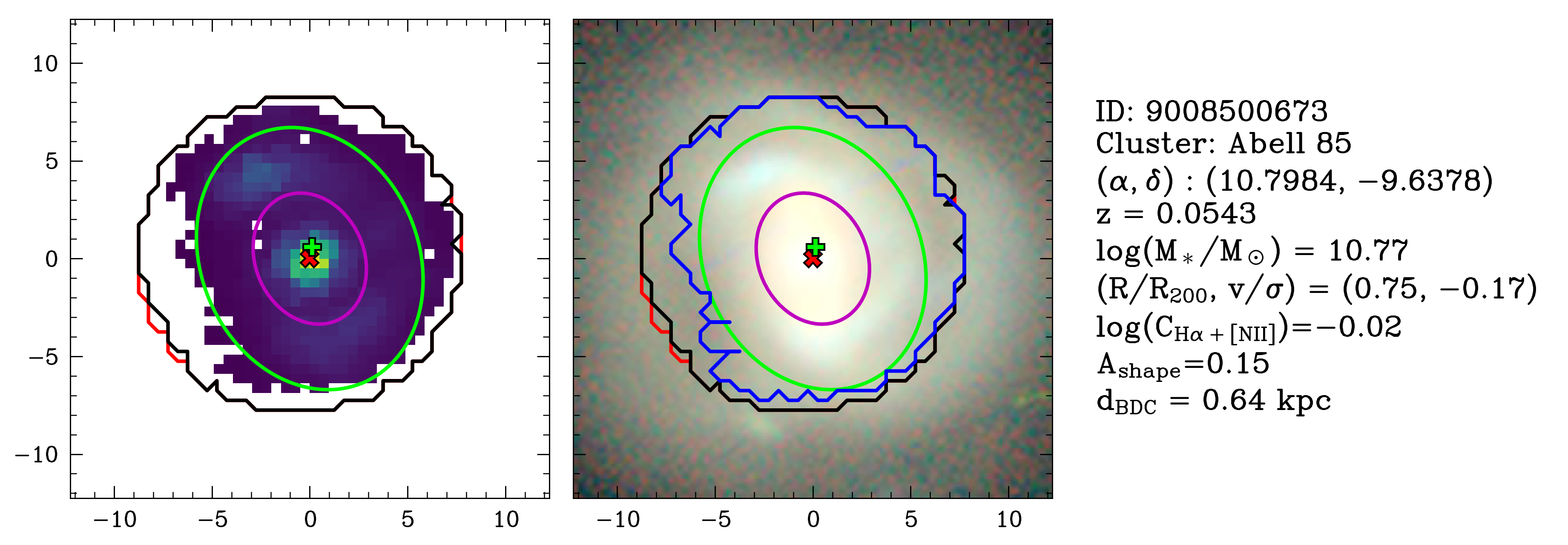} &
        \includegraphics[width=0.5\textwidth]{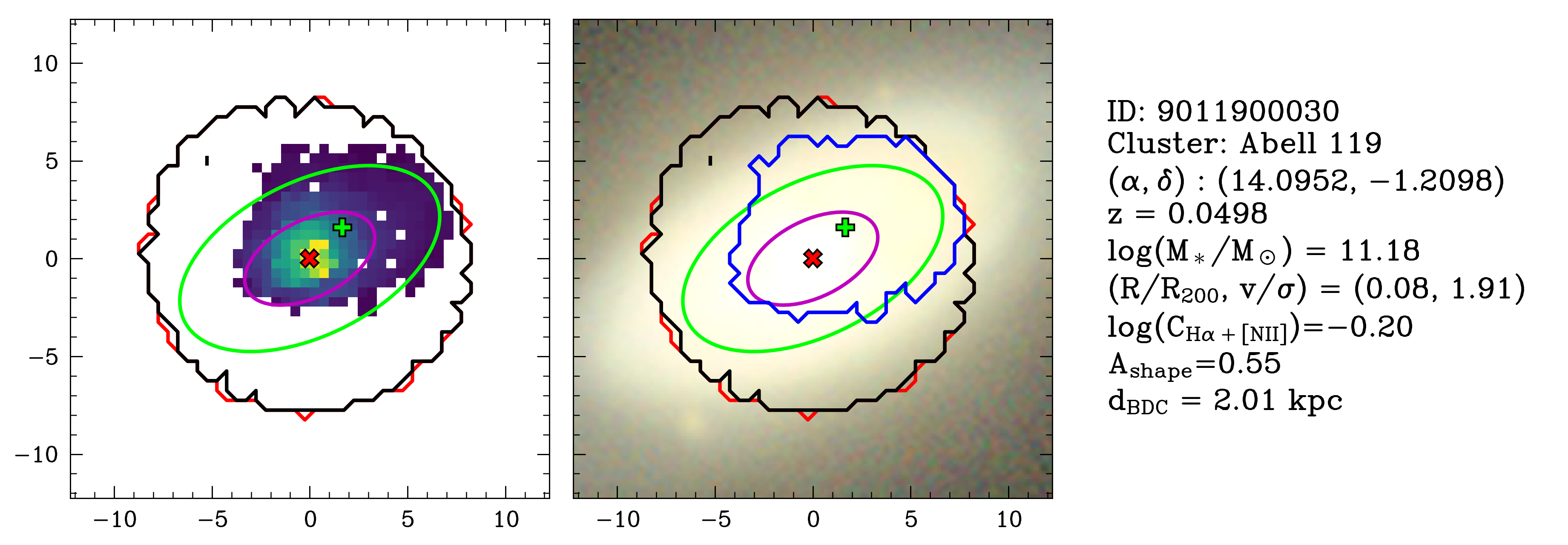} \\
        \includegraphics[width=0.5\textwidth]{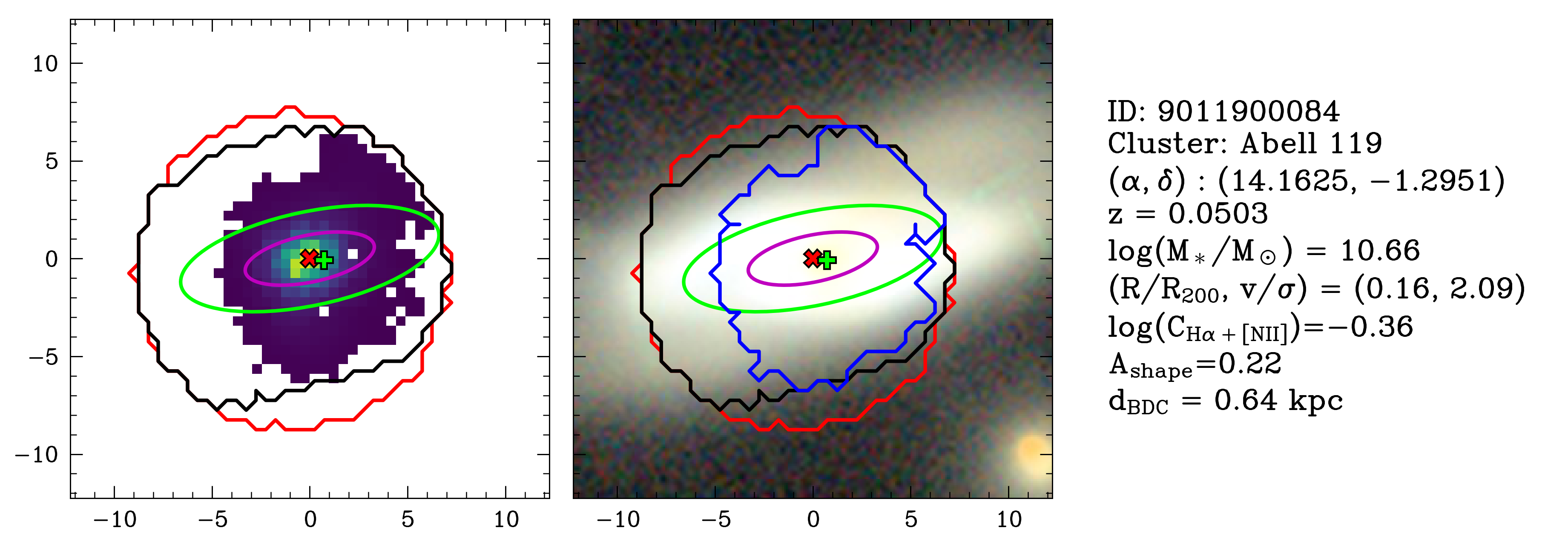} &
        \includegraphics[width=0.5\textwidth]{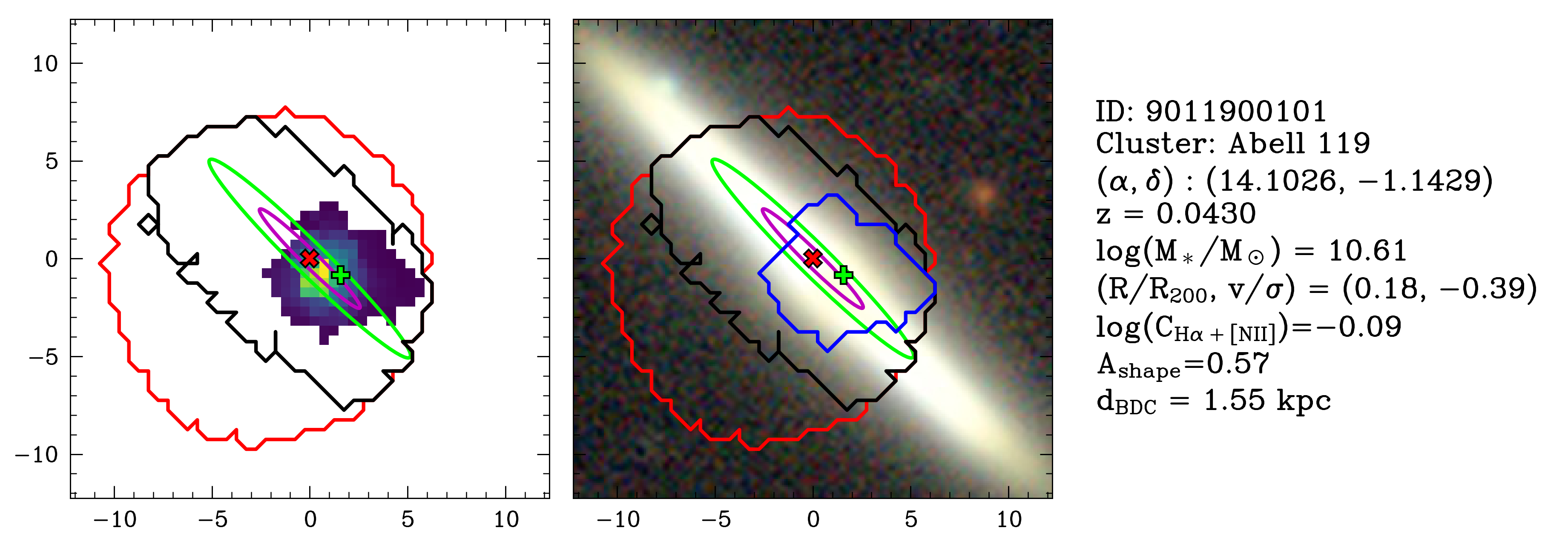} \\
        \includegraphics[width=0.5\textwidth]{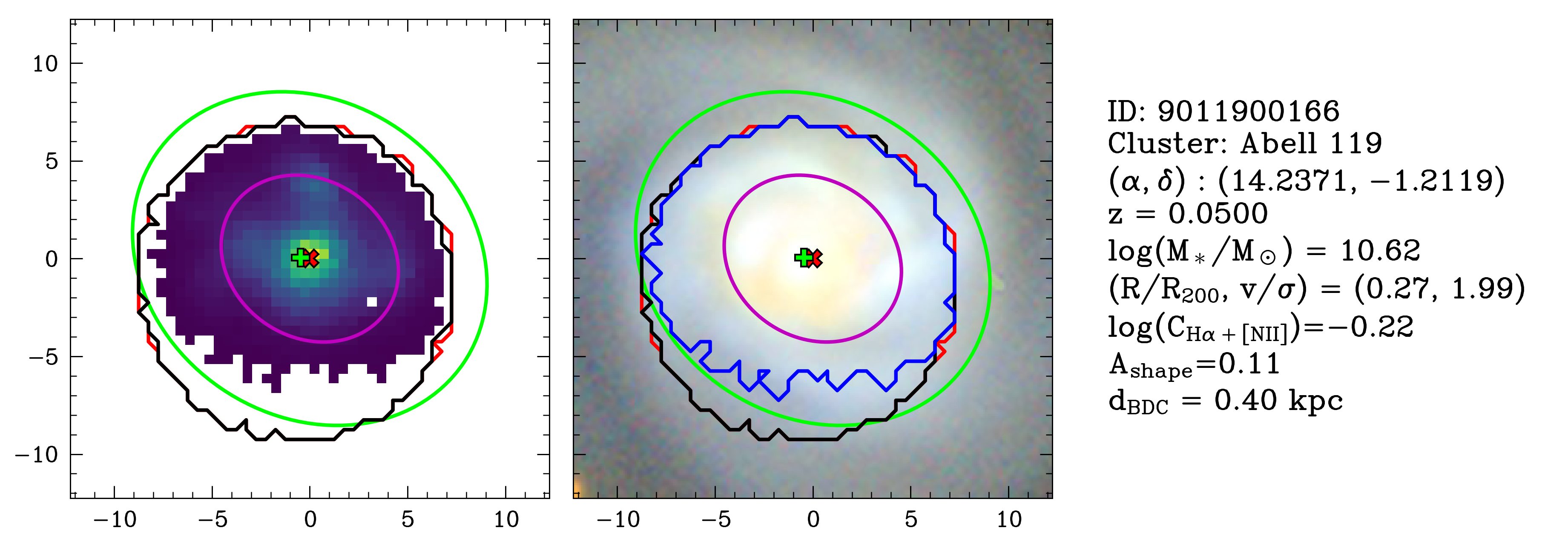} &
        \includegraphics[width=0.5\textwidth]{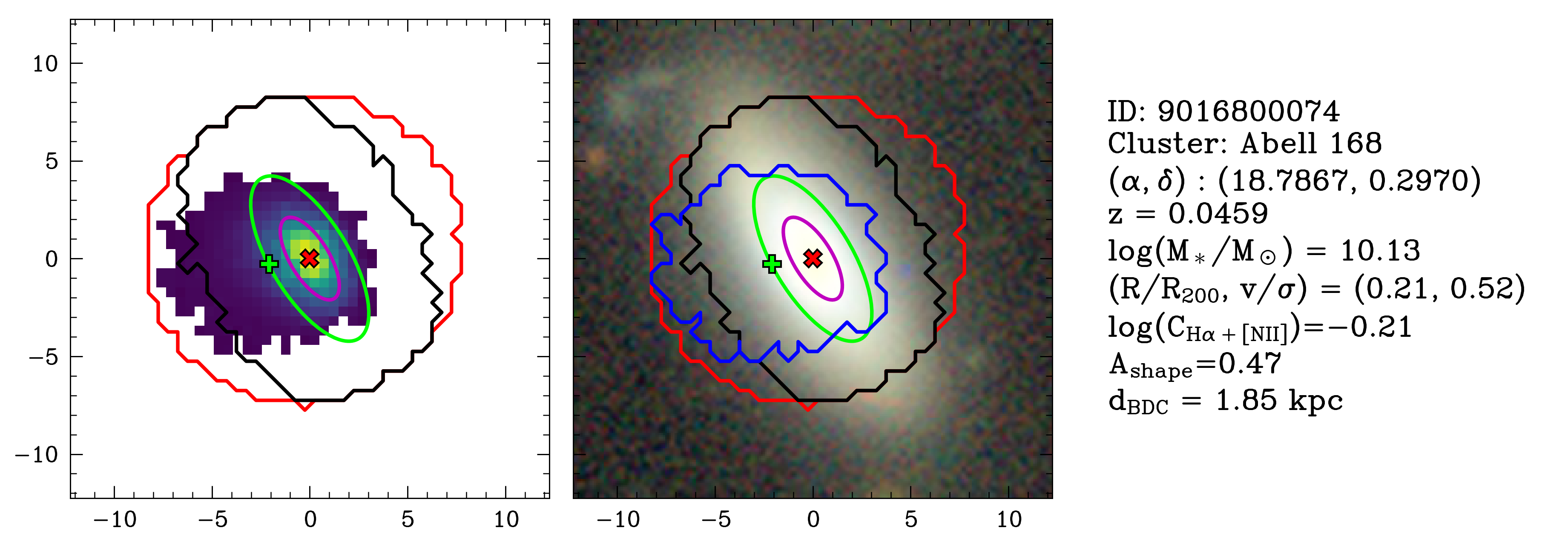} \\
        \includegraphics[width=0.5\textwidth]{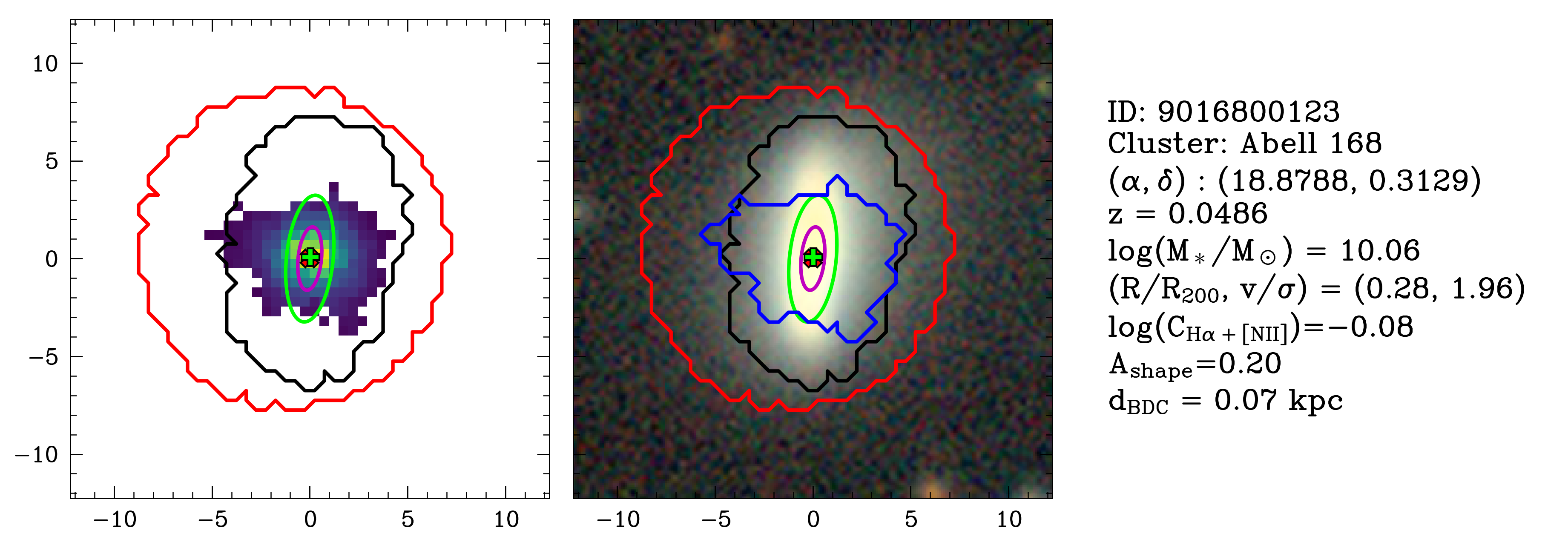} &
        \includegraphics[width=0.5\textwidth]{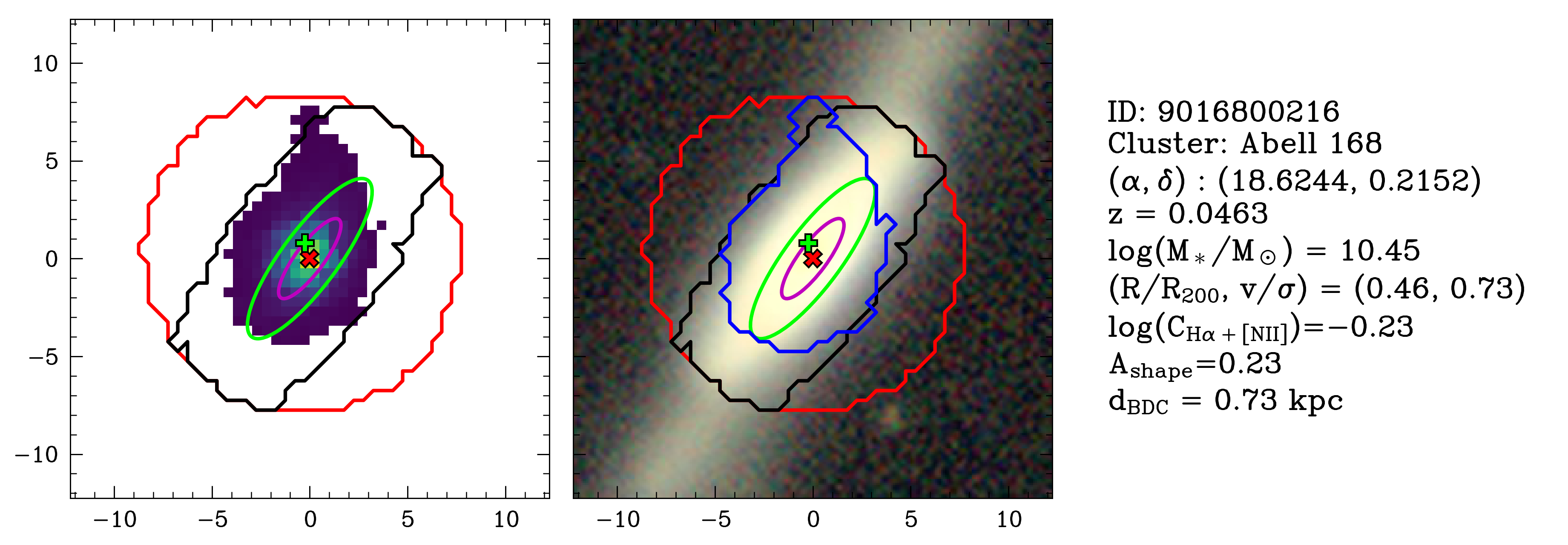} 
    \end{tabular}
    \caption{Full compilation of visually identified asymmetric galaxies in this study. For each galaxy, three panels are shown. In the left panel, the ionised gas distribution is mapped using the \texttt{viridis} colourmap. The black contour traces the stellar continuum around \ha\ and \nii\ at $\mathrm{SNR}=2$. The red contour marks the SAMI bundle edge, derived from the median of the red spectrum. The red cross and lime plus sign indicate the galaxy centre and the non–flux-weighted (binary detection) centre of the ionised gas distribution, respectively. The middle panel shows the composite $griz$  image from Legacy Survey DR10 and the composite $gri$  image from KiDS DR5 \citep{Wright2024} for \texttt{9388000124}. The contours and symbols match those in the left panel, except here the emission is overlaid as blue contour rather than displayed as a colourmap. In the right panel, we present galaxy properties including stellar mass, normalised cluster-centric distance and line-of-sight (LOS) velocity, concentration, shape asymmetry, and the offset between the stellar and ionised gas centres. The magenta and lime ellipses correspond to 0.5 and 1 $\rm R_e$, respectively. The unit of the x and y axes is arcsecond. North is towards the up, and East is towards the left.}
    \label{fig:Asymmetric Galaxies part1}
\end{figure*}

\begin{figure*}[!h]
    \centering
    \begin{tabular}{cc}
        \includegraphics[width=0.5\textwidth]{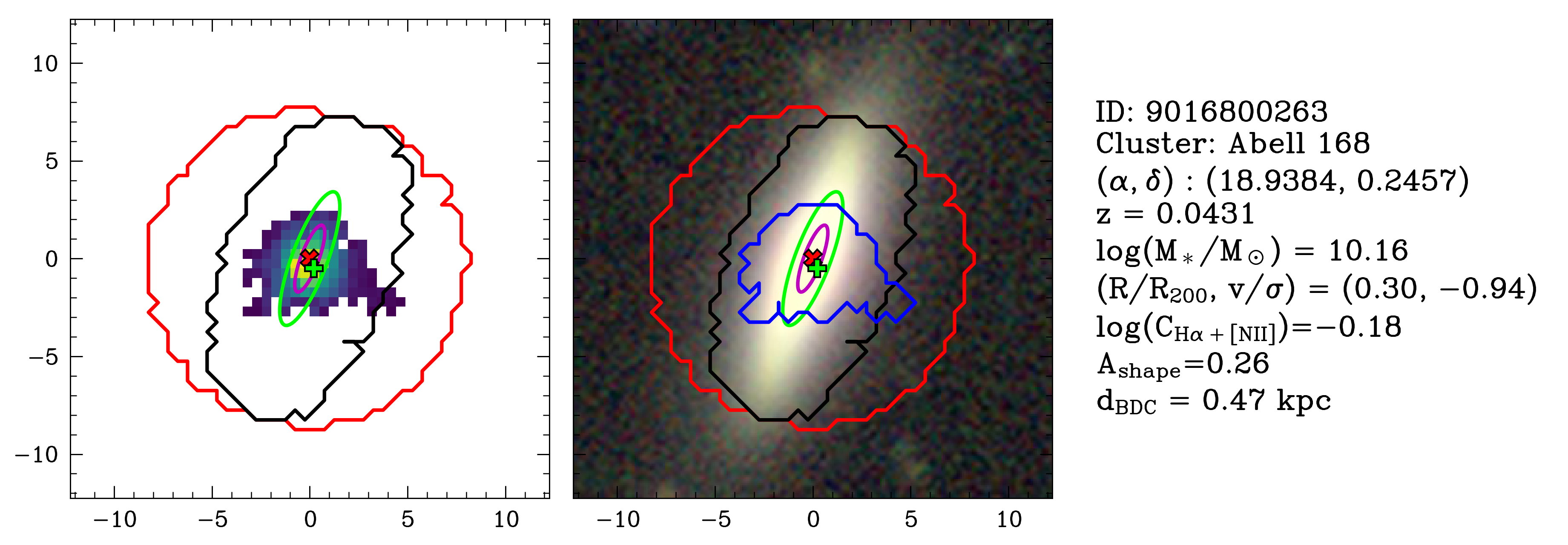} &
        \includegraphics[width=0.5\textwidth]{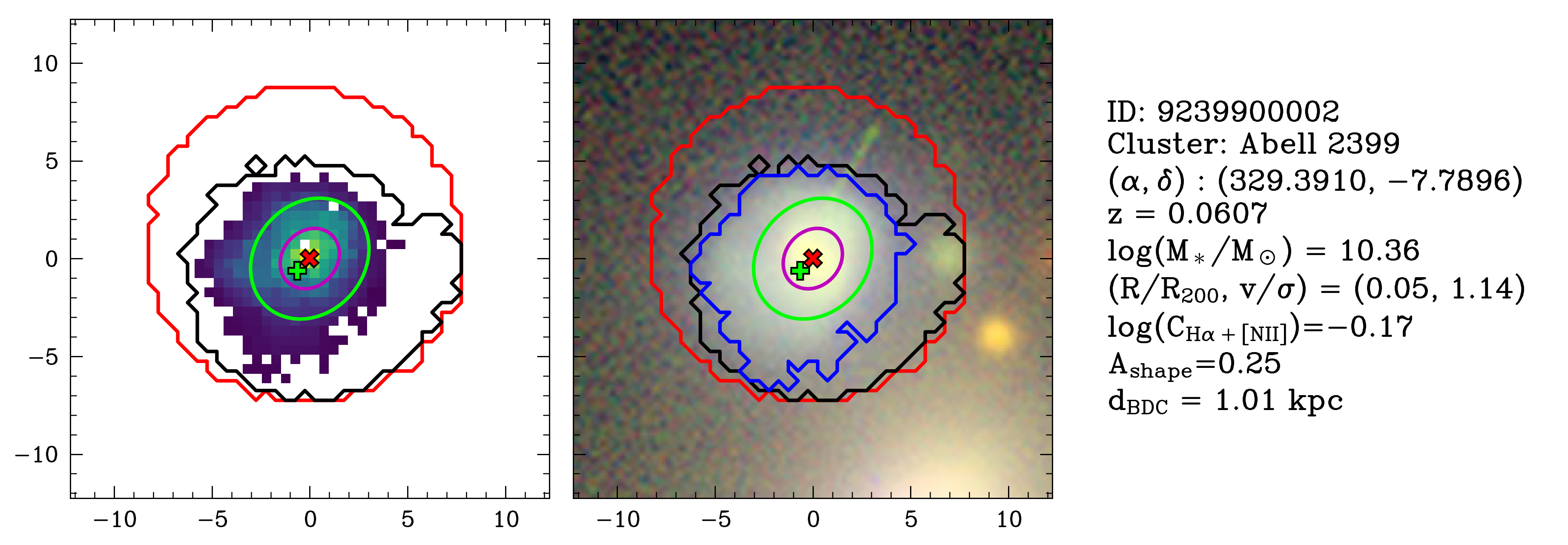}\\
        \includegraphics[width=0.5\textwidth]{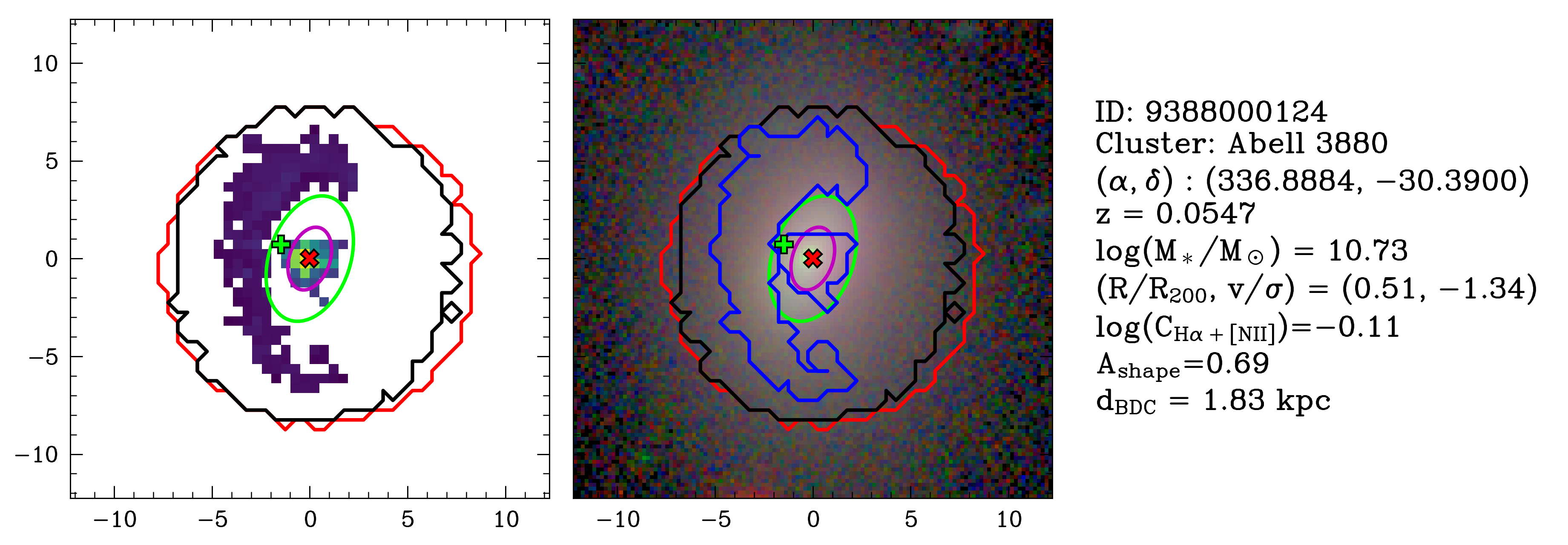}
    \end{tabular}
    \caption{(Continued.)}
    \label{fig:Asymmetric Galaxies part2}
\end{figure*}

\FloatBarrier

\section{A Subsample of Visually Identified Truncated Galaxies}
\begin{figure*}[!h]
    \centering
    \begin{tabular}{cc}
        \includegraphics[width=0.5\textwidth]{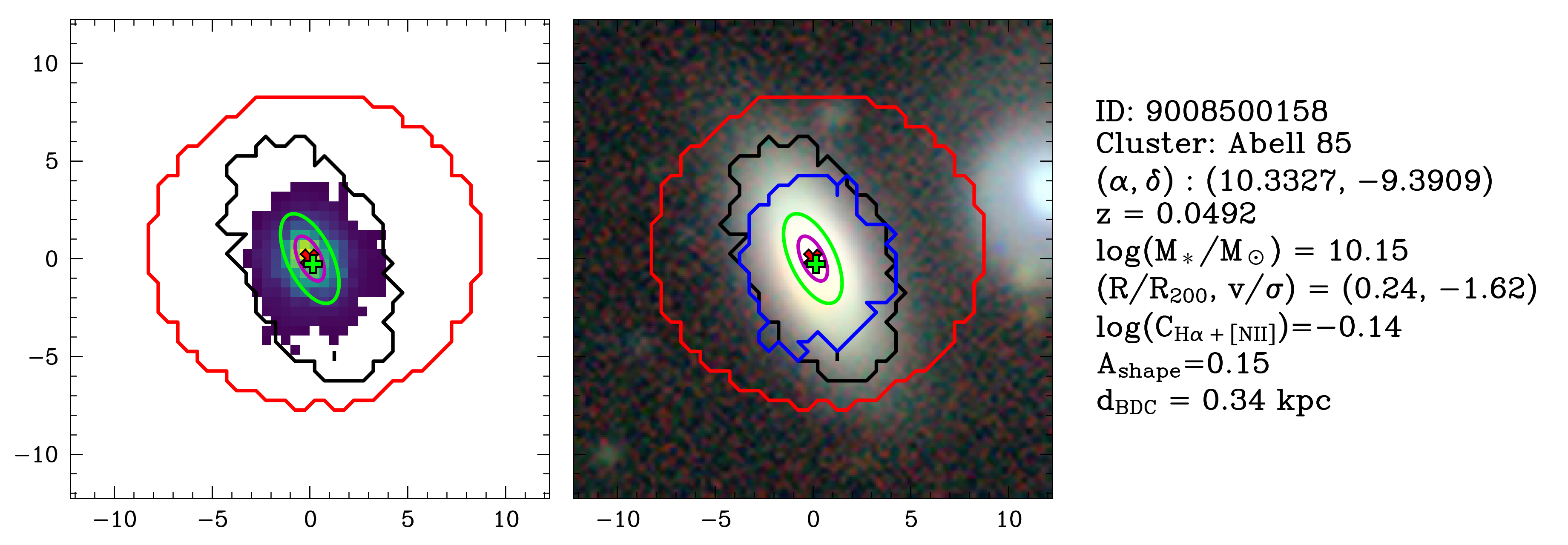} & 
        \includegraphics[width=0.5\textwidth]{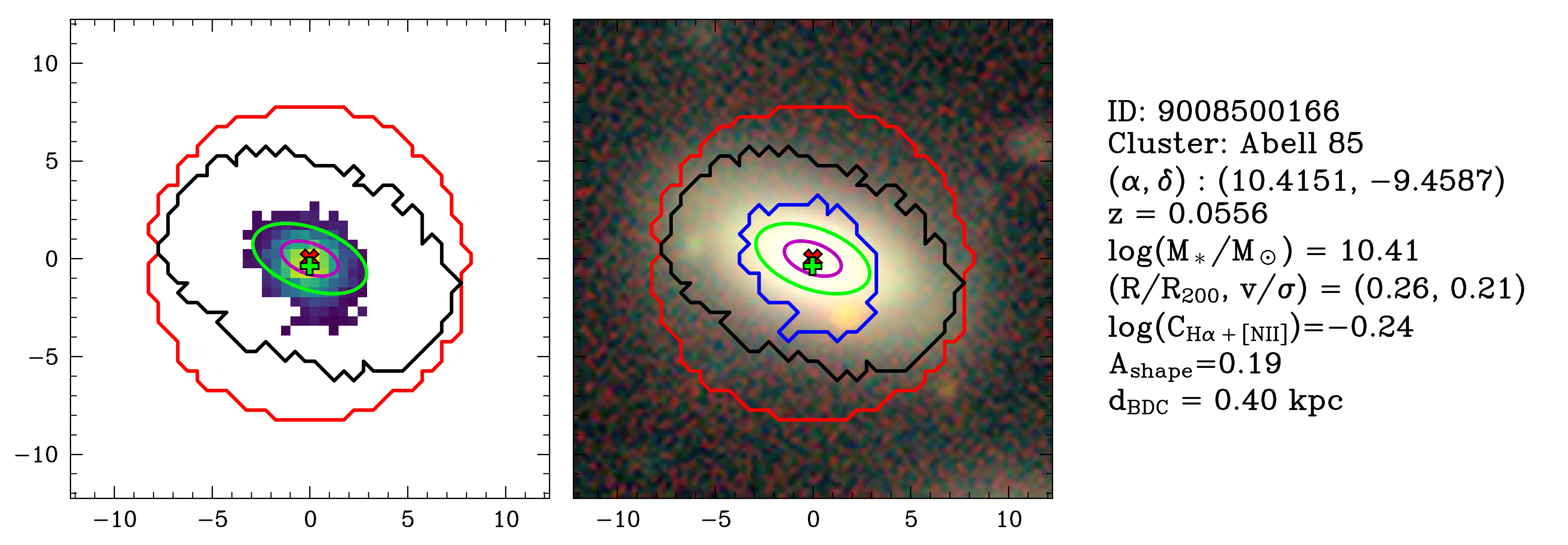} \\
        \includegraphics[width=0.5\textwidth]{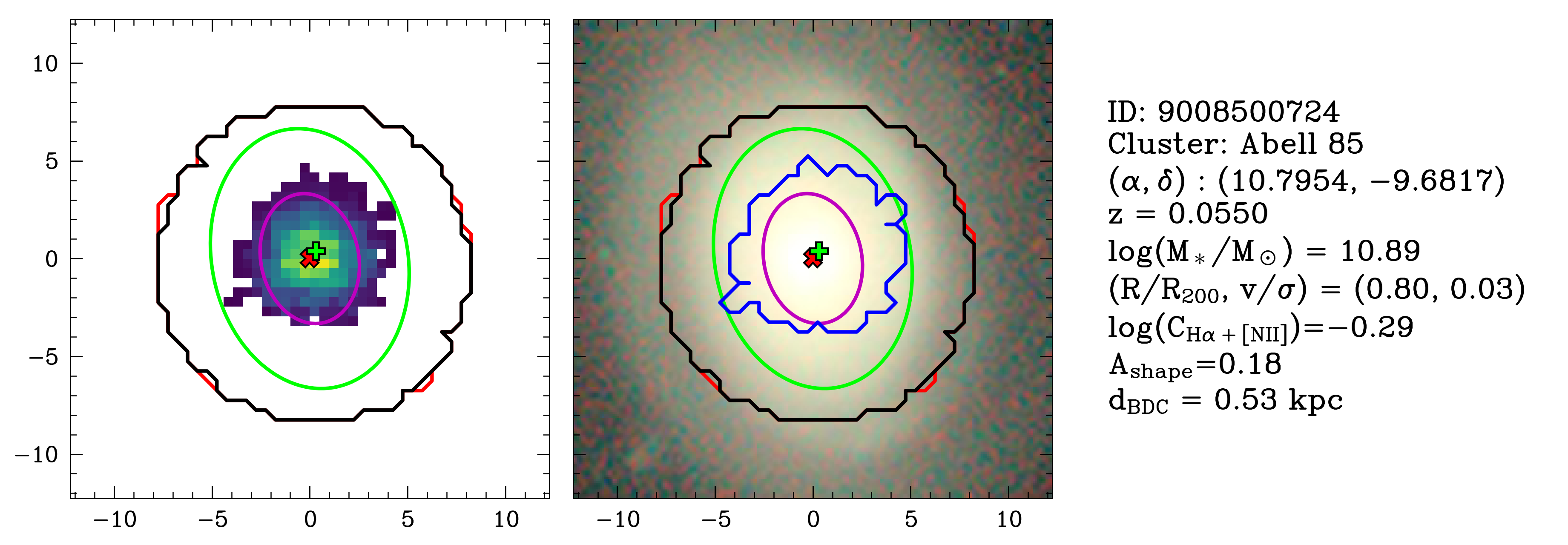} &
        \includegraphics[width=0.5\textwidth]{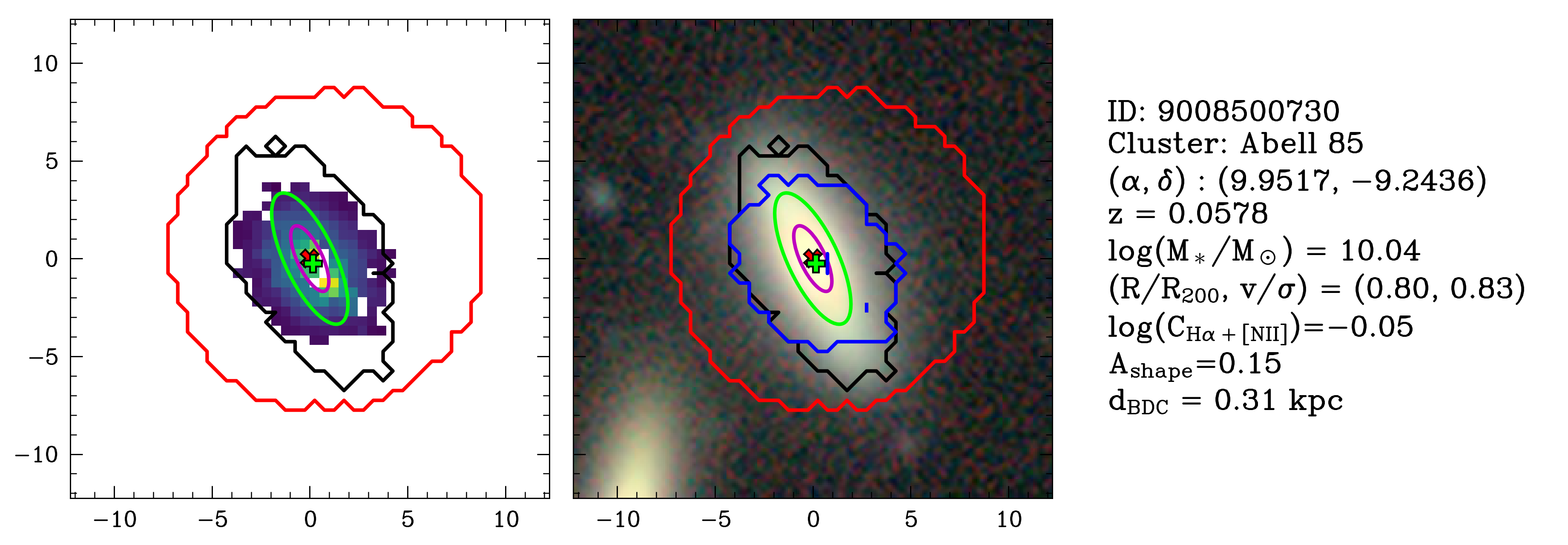} \\
        \includegraphics[width=0.5\textwidth]{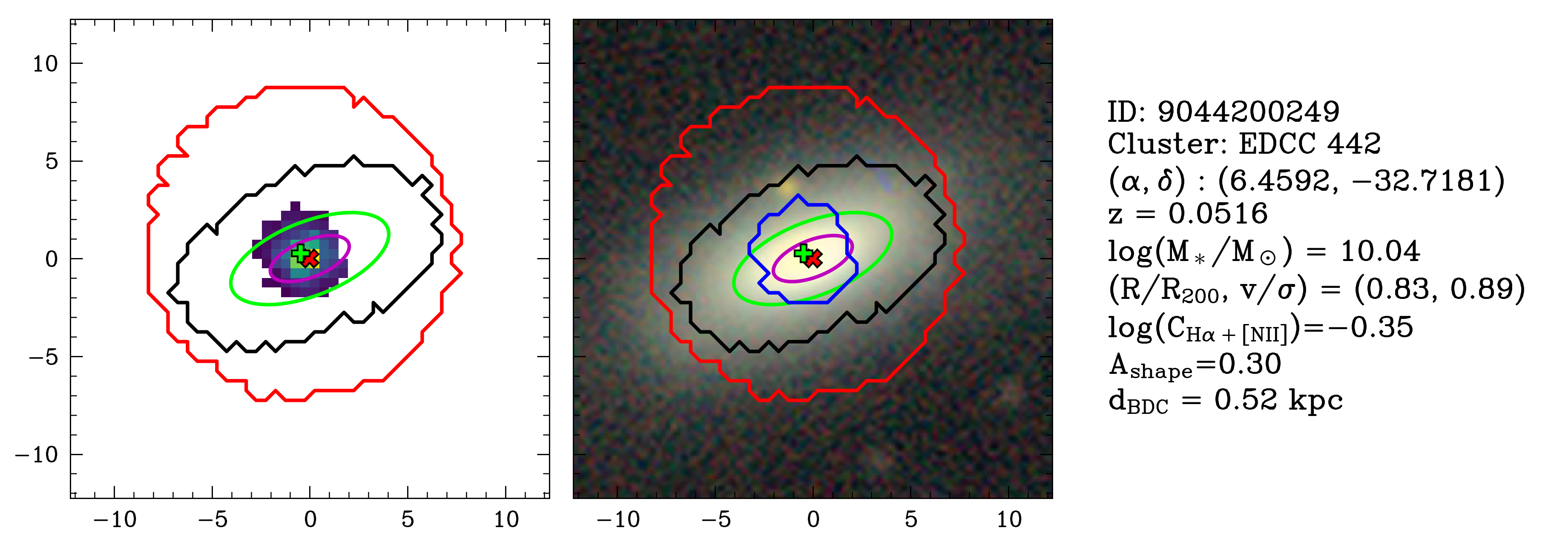} &
        \includegraphics[width=0.5\textwidth]{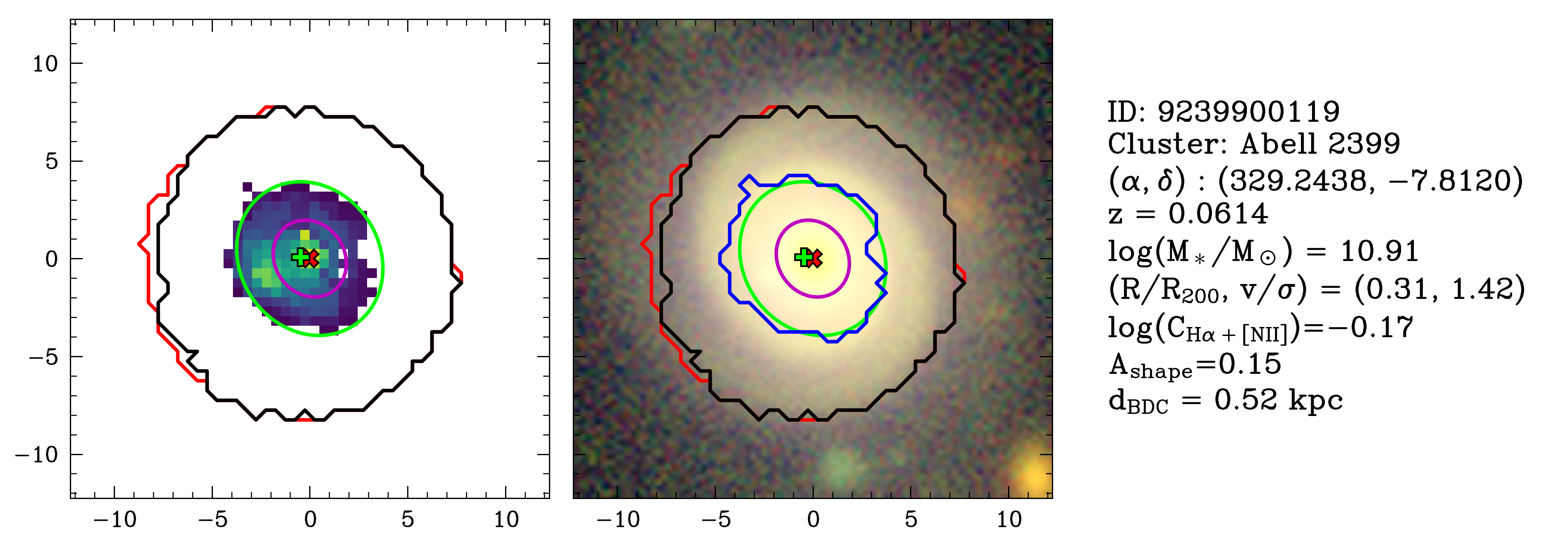} \\
        \includegraphics[width=0.5\textwidth]{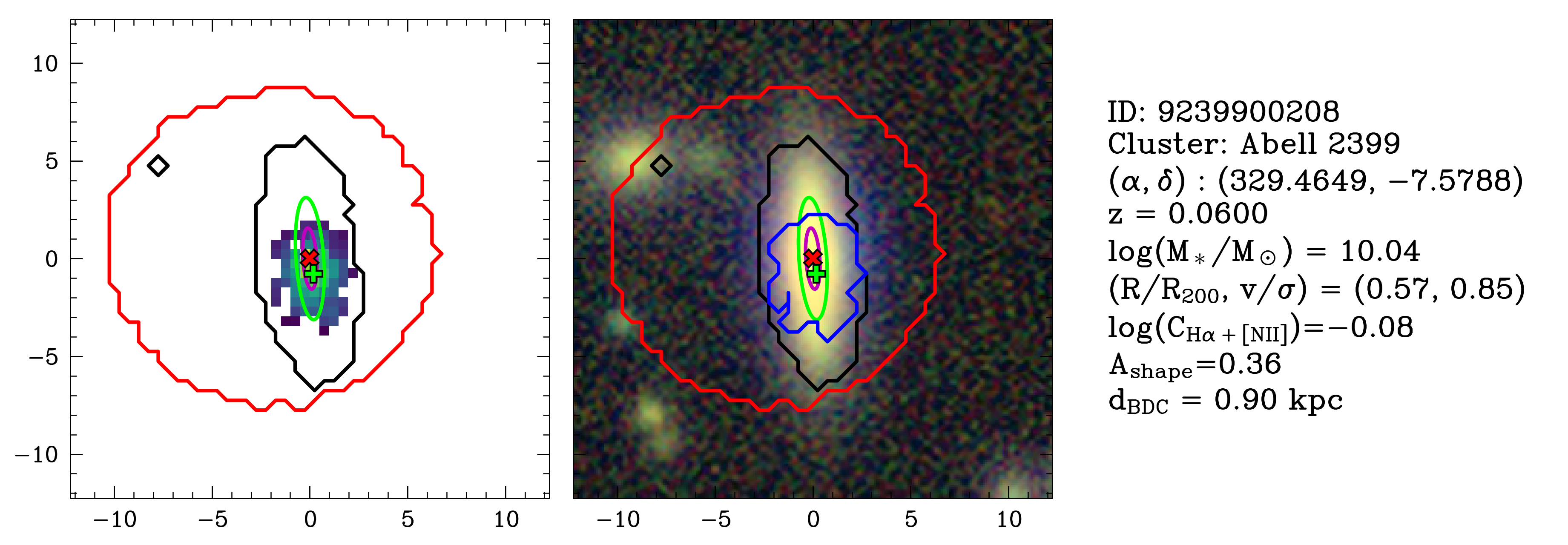} &
        \includegraphics[width=0.5\textwidth]{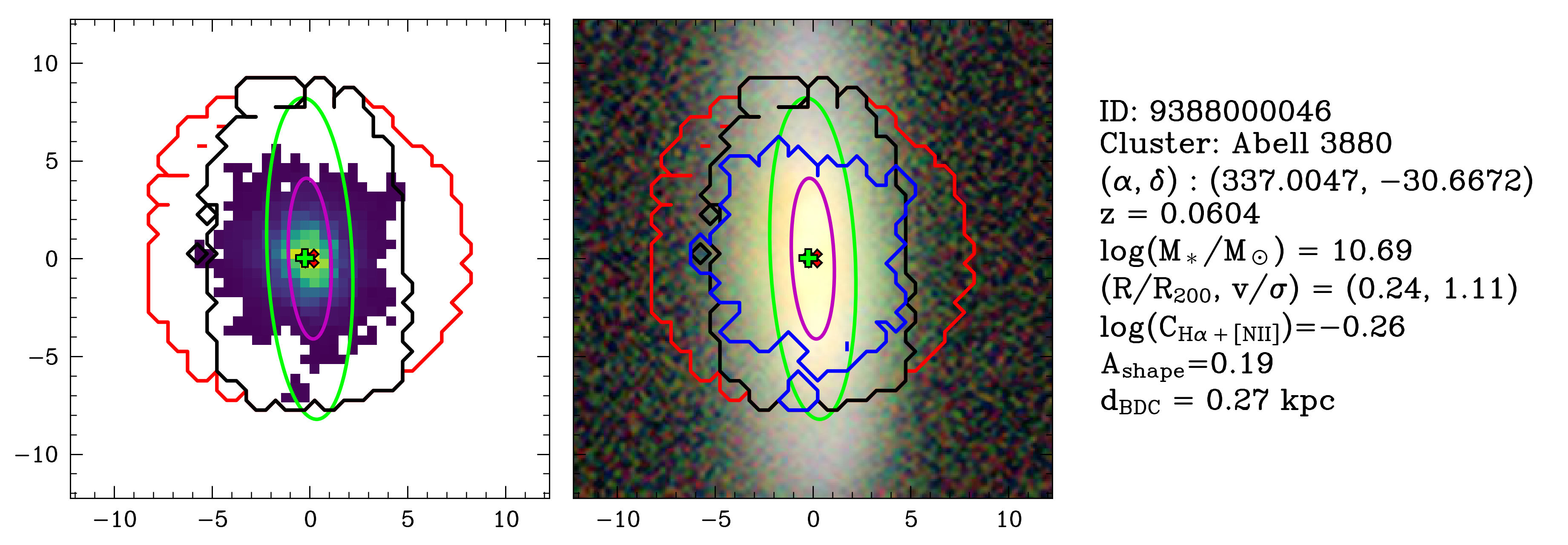} \\
    \end{tabular}
    \caption{Same as Figure~\ref{fig:Asymmetric Galaxies part1}, but for a subsample of visually identified truncated sample.}
    \label{fig:Truncated Galaxy Examples}
\end{figure*}

\FloatBarrier

\clearpage
\section{Examples for Aperture-affected cases}

\begin{figure*}[!h]
    \centering
    \begin{tabular}{cc}
        \includegraphics[width=0.5\textwidth]{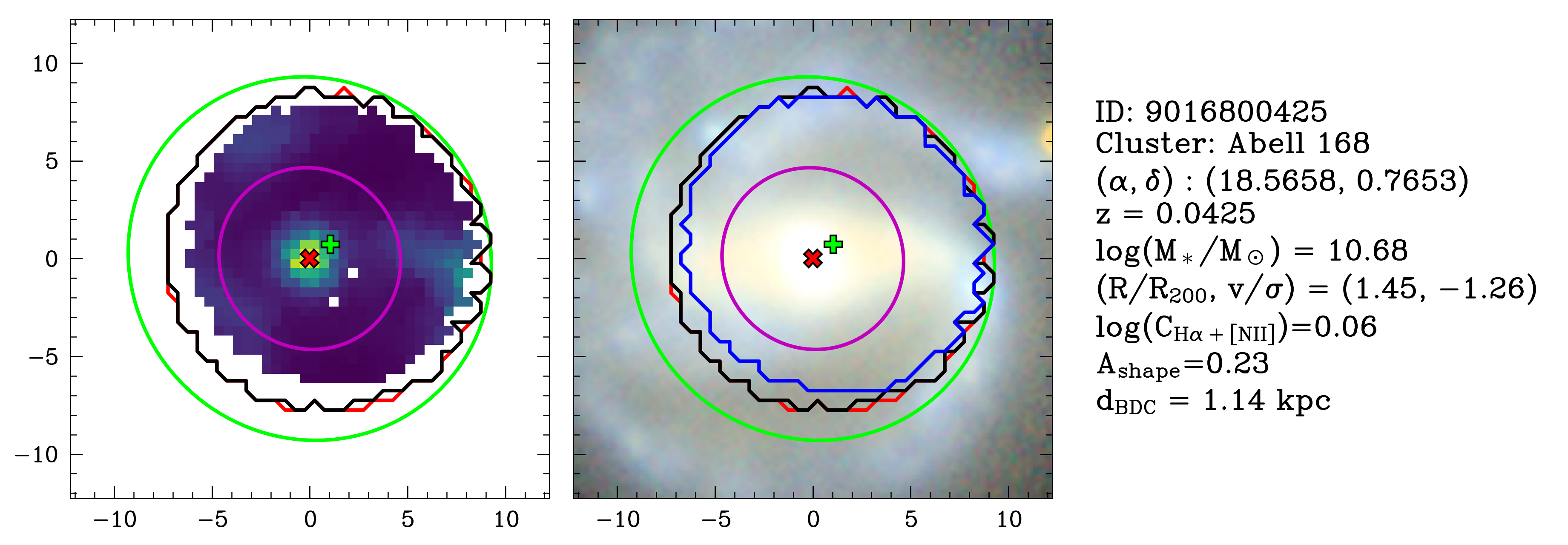} &
        \includegraphics[width=0.5\textwidth]{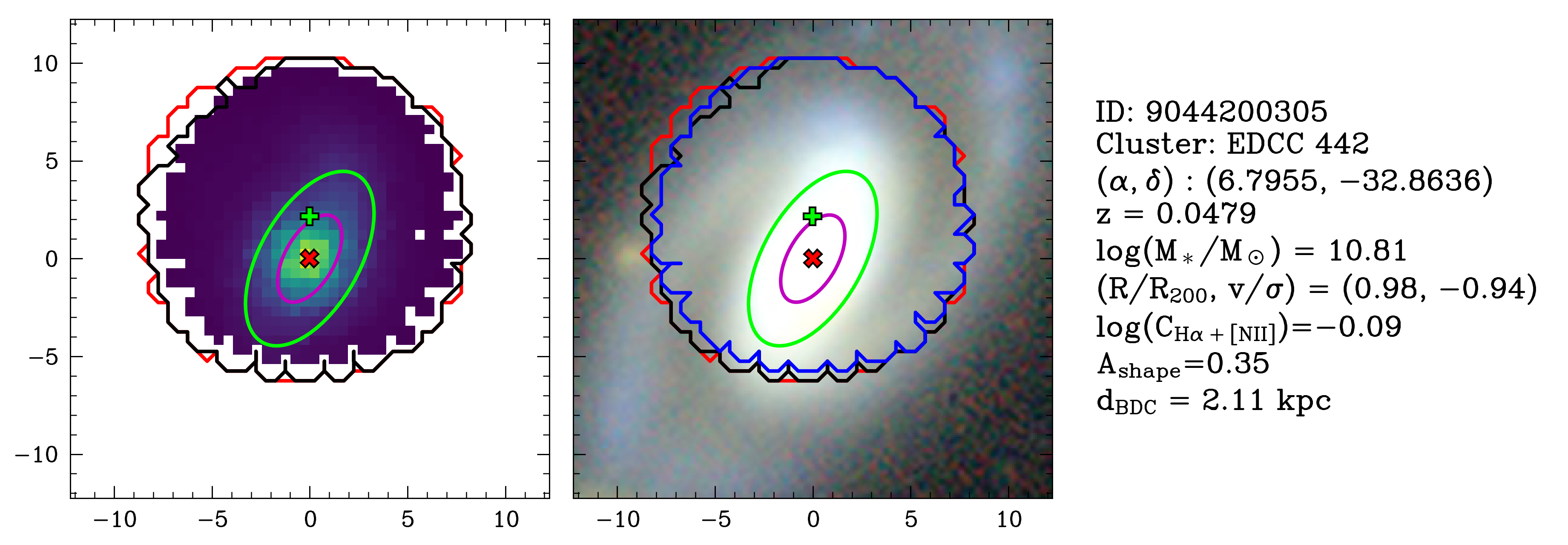} \\
        \includegraphics[width=0.5\textwidth]{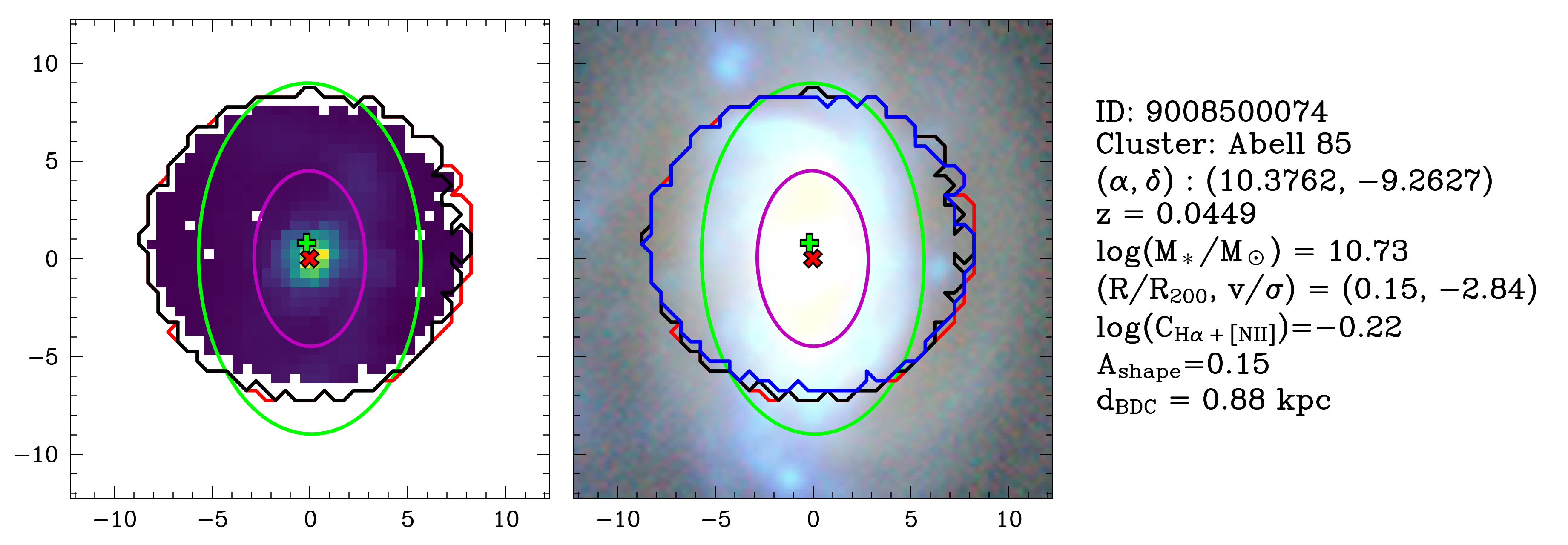} &
        \includegraphics[width=0.5\textwidth]{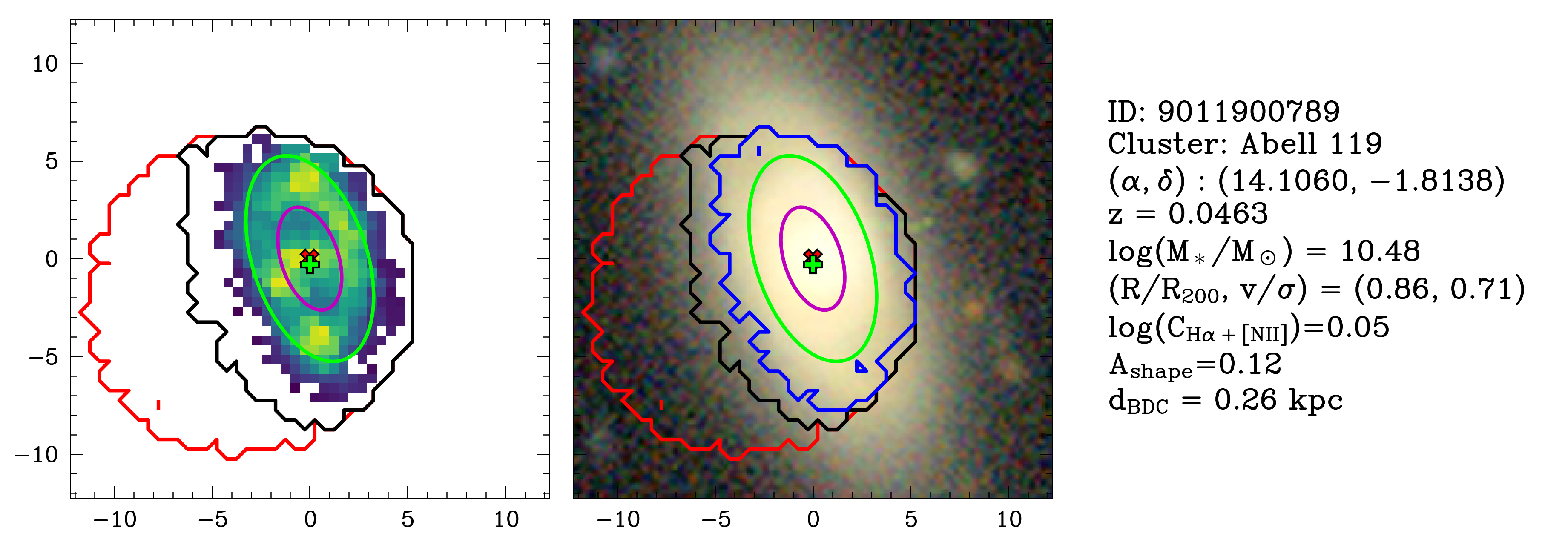}
    \end{tabular}
    \caption{Same as Figure~\ref{fig:Asymmetric Galaxies part1}, but for a handful of aperture-affected examples.}
    \label{fig:Aperture-affected examples}
\end{figure*}

\end{document}